\documentclass[12pt, a4paper]{article}
\usepackage[applemac]{inputenc}
\usepackage{mltex}
\usepackage{amsmath}
\usepackage{amssymb}
\usepackage{setspace}
\usepackage{latexsym}
\usepackage{fancyhdr}

\usepackage[pdftex]{graphicx}

\graphicspath{{figures/}}

\newcommand{\ji}{\mathrm{j}}

\newcommand{\di}{\displaystyle}

\begin{document}
\title{A proposal for a minimal model of free reeds}
\author{L.  Millot, Cl. Baumann\\
l.millot@ens-louis-lumiere.fr}
\date{} 

\maketitle
\begin{abstract}
In this paper we propose a minimal model for free reeds  taking into account the significant phenomena. This free reed model may be used to build models of free reed instruments which permit numerical simulations. Several definitions for the section by which the airflow passes through the reed are reviewed and a new one is proposed which takes into account the entire escape area under the reed and the reed thickness. To derive this section, it is necessary to distinguish the neutral section (the only section of the reed which always keeps its length constant while moving) from the upstream or downstream sections. A minimal configuration is chosen to permit the instabilities of both (-,+) and (+,-) reeds on the basis of a linear analysis of instabilities conditions. This configuration is used to illustrate,  with temporal simulations, the minimal model for both kinds of reeds and to discuss the model assumptions. Some clues are given about the influence, on the playing frequency and on the dynamic of the sound, of two main parameters of the geometrical model: the size of the volume and the level of the excitation. It is shown that the playing frequency of a (+,-) reed can vary in a large range according to the  size of the volume upstream of the reed; that the playing frequency is nearly independent of the excitation but that the dynamic of the sound increases with the excitation level. Some clues are also proposed to determine the nature of the bifurcation for free reeds: it seems that free reeds may present inverse bifurcations. The influence of the reed thickness is also studied for configurations where the reed length or the reed width vary to keep the mass constant. This study shows that the reed thickness can have a great influence on the sound magnitude, the playing frequency and the magnitude of the reed displacement which justifies its introduction in the reed model.

This article has been published in Acta Acustica united with Acustica, Vol. 93 (2007), p. 122-144.

\end{abstract}

\section{Introduction}
\paragraph{}Since the first works of Helmholtz \cite{Helmholtz:54}, much attention has been paid to the study of reeds and reeds instruments. One can find a great deal of works related to striking reed auto-oscillation \cite{Wilson:74, Nederveen:98, Saneyoshi:87, Grand:96, Kergomard:99, Dalmont:00, Kergomard:00, Ollivier:04, Ollivier:05} and synthesis of the clarinet or saxophone \cite{Schumacher:78, Schumacher:81, McIntyre:83, Gazengel:95} for instance. But there are far less works related to free reeds dynamics \cite{Koopman:97, Bahnson:98, Cottingham:99a, Cottingham:99b, Cottingham:05} and most of the articles dealing with modeling of free reeds are only concerned with frequency-domain approach and the threshold of reed auto-oscillation \cite{Singhal:76, Fletcher:79a, Fletcher:79c, Fletcher:79b, Johnston:87, Fletcher:93, Cuesta:95, Fletcher:99a, Tarnopolsky:00}. To have an idea of most of the latest  available information on free-reed modeling, the reader is invited to search in \cite{Fletcher:00}. We adopt Fletcher's convention \cite{Fletcher:93, Fletcher:00} to designate both kinds of reed: (-,+) for an blown closed reed and (+,-) for a blown open reed. 

\paragraph{}A first attempt to obtain a temporal model for a free-reed is due to St-Hilaire, Wilson and  Beavers \cite{Hilaire:71}. In this paper, a three-dimensional far-field flow is matched with a two-dimensional near field one. Since this first attempt, Van Hassel and Hirschberg \cite{Hassel:95} have shown that it would not be more difficult to solve the three-dimensional problem rather than using a 2D/3D matching which is rather arbitrary. Some new research on the 3D modeling seem to be on the way \cite{Misdariis:00, Hassel:01}.

\paragraph{}To derive a satisfactory model for the free reed  we already have some key information about the behavior of the free reeds. Indeed, it has been shown \cite{Cottingham:99b, Millot:99, Millot:01}  that the reed motion is almost perfectly sinusoidal during playing because of the non-striking nature of the reeds.  Another experiment made by Misdariis, Ricot and Causs{\'e} \cite{Misdariis:00, Ricot:99} using a laser vibrometer confirmed the sinusoidal character of free reed motion.

\paragraph{}The present research is a proposal of a  model which may be used to obtain time si\-mulation of the instrument useful for sound synthesis and comparison with measurements. While the model is globally one-dimensional, the description of the flow through the reed takes into account the side and front  escape areas under both reeds.  In section 2, we first point out some ambiguities concerning the description of the reed motion. Then we complete the free reed model by discussing the definition of the total volume flow which passes through the reed, by taking into account the local longitudinal and transverse  deformations of the reed. In section~3, we first introduce a minimal loading configuration which may permit the instabilities of both kinds of reed: (-,+) and (+,-) reeds. The conditions for the instability of both  kinds of reed are derived according to a linear ana\-lysis. We then illustrate the model for both kinds of reed and discuss the validity of the assumptions. Then, some preliminary results are given about the influence of some main control parameters: the excitation, the size of the volume upstream of the reed and the reed thickness. Some clues about the nature of the bifurcation in the case of free reeds are also given.  In the Appendix, one can find the details for the derivation of the motion of the different reed sections, for the pumped volume flow, for the calculation of the instability condition and some keypoints about the algorithm we used to perform numerical simulations.

\paragraph{}We must emphasize that this study has not been confronted with experiments for the moment in the case of the configurations we used: only numerical simulations are used to illustrate the model we propose. But this  free reed model has been successfully used for the study of  chromatic playing on a diatonic harmonica, which will be presented elsewhere.

\section{Reed minimal model \label{sec:reedmotion}}
\subsection{Description of the reed motion}
\begin{figure}[h!tbp]
    \centering
    \includegraphics[width=7.5cm]{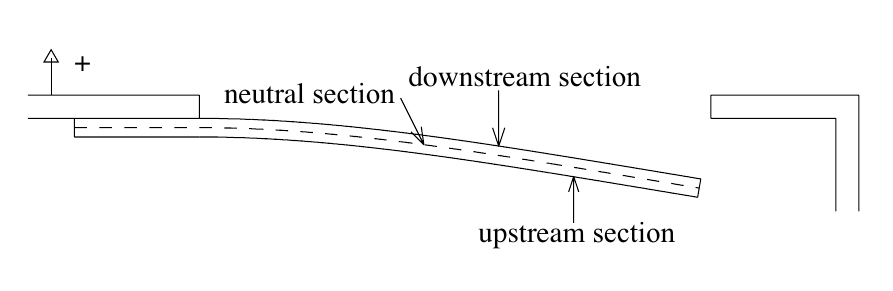}
    \includegraphics[width=7.5cm]{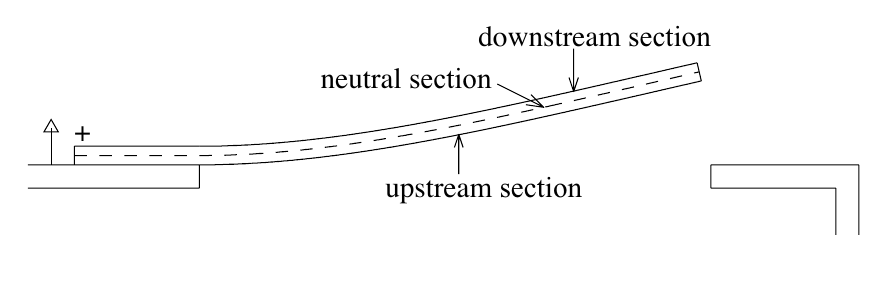}
   \caption{Definition of upstream and downstream sections for the two types of reeds which will be studied: (-,+) reed on the left and (+,-) reed on the right. The upstream section is related to the face which is in front of the inside of the instrument while the downstream section faces the outside of the instrument. The introduction of these two sections will be useful to define the output sections for the airflow.}
    \label{f1}
\end{figure}

\paragraph{}As demonstrated by Millot \cite{Millot:99, Millot:01} using strain gauges, Ricot  \cite{Ricot:99} or  Cottingham \cite{Cottingham:99b} using a laser vibrometer, the reed motion is sinusoidal for a free reed: each reed is moving on the first transverse eigenmode of a free-clamped beam, the Euler-Bernoulli assumption for the beams. So, to derive our model, we only consider pure flexion motion of the reed on its first transverse mode, which means that the cross sections of the reed are assumed to stay normal to the neutral section, the only section which does not suffer from compression or extension.  In the following, we call respectively  upstream and downstream sections, the faces of the reeds inside and outside of the instruments as illustrated by Figure \ref{f1}. We then assume that the target face is the upstream one for the (-,+) reed and the downstream one for the (+,-) reed, using the conventions proposed by Fletcher \cite{Fletcher:79a}. 

\paragraph{}Table 1 gives all the conventions used for the model.

\begin{table}[h!tp]
\begin{center}
\begin{tabular}{ll}
symbol & related quantity\\
\hline
$L_r$ & length of the reed\\
$W_r$ & width of the reed\\
$e_r$ & thickness of the reed\\
$e_s$ & thickness of the support\\
$s$ & normalized distance from the clamped end of the reed\\
$h_{min}$ & clearance gap existing between the reed and the support\\
$h_{n}$ & transverse displacement of the neutral section at the free tip of the reed\\
$h_{up}$ & transverse displacement of the upstream section at the free tip of the reed\\
$h_{do}$ & transverse displacement of the downstream section at the free tip of the reed\\
$h_{n,0}$ & mean value for $h_{n}$ during the oscillation\\
$h_{n,00}$ & rest value for $h_{n}$ when the reed is unblown\\
$h_{n,000}$ & value of $h_{n}$ for which the reed is flat \\
$h_{n}(s)$ & transverse displacement of the neutral section at position $s.L_r$\\
$h_{up}(s)$ & transverse displacement of the upstream section at position $s.L_r$\\
$h_{do}(s)$ & transverse displacement of the downstream section at position $s.L_r$\\ 
$\psi(s)$ & normalized modal reed factor for the first transverse eigenmode\\
$\zeta$ & equilibrium displacement at the free tip: $\zeta=h_{n} -h_{n,00}$\\
$\Delta p$ & pressure difference between the upstream and downstream of the reed\\
\end{tabular}
\end{center}
\caption{Definition of the symbols introduced for the reed description.}
\end{table}

\paragraph{}The reed motion can be described by considering an equivalent mass-spring-damper oscillator with a single degree of freedom, the neutral section transverse displacement at the free tip of the reed, which will be noted $h_{n}$ in the following. In almost all the former works dealing with the dynamics of reeds \cite{Backus:63,Wilson:74, Thompson:79, Fletcher:79a, Schumacher:81, Sommerfeldt:88, Hirschberg:90, Fletcher:93, Hirschberg:94, Gazengel:95, Millot:99, Ricot:99, Tarnopolsky:00}, the equivalent reed motion equation, which is normally only valid for the neutral section, is used for the upstream or downstream section following the assumption that this is a fair approach as the reeds are quite thin. For the derivation of our model, we can not afford such a simplification as we need to distinguish neutral, upstream and downstream sections and because we think that the reed thickness can not be neglected.

\paragraph{}The respective expressions for the transverse local displacement at  the distance $s.L_r$ from the clamped end ($s\in[0,1]$), of the neutral, upstream and downstream sections are respectively for a (-,+) reed:
\begin{equation}
	h_{n}(s)=h_{n,000} + \big( h_{n} -h_{n,000} \big)\psi(s),
\end{equation} 

\begin{equation}
	h_{up}(s)=h_{n}(s) -\frac{e_{r}}{2}\frac{L_r}{\sqrt{L^2_r+( h_{n}-h_{n,000} )^2\psi'(s)^2}},
\end{equation} 

\begin{equation}
	h_{do}(s)=h_{n}(s) +\frac{e_{r}}{2}\frac{L_r}{\sqrt{L^2_r+( h_{n}-h_{n,000} )^2\psi'(s)^2}},
\end{equation} 
where $\di h_{n,000}=-e_s-\frac{e_{r}}{2}$ represents the transverse position of the neutral section when the reed is flat and $\psi(s)$ is the modal reed factor at the distance $s.L_r$ from the clamped end. It is then obvious that the upstream, downstream and neutral transverse displacements differences are greater as much as $e_r$, $L_r$ and $h_{n}$ increase. 

\paragraph{}We consider that the equivalent reed motion equation is a function of the neutral section equilibrium displacement at the free tip of the reed, $\zeta=h_{n} -h_{n,00}$, where $h_{n,00}$ is the neutral section transverse position at rest (we assume that the reed adopts a first eigenmode profile at rest). This motion equation is then given by:
\begin{equation}
 \frac{d^2 \zeta}{dt^2} + Q^{-1}\omega_0\frac{d \zeta}{dt}+\omega^2_0 \zeta = \mu \Delta p,
	\label{eq:Reed}
\end{equation} 
where $\Delta p$ is the pressure difference between the upstream and the downstream of the reed  (see Appendix for the analytical calculation of parameter $\mu$). 

\paragraph{}As the reed motion equation is defined, we now need to consider the description of the volume flow passing through the reed.

\subsection{Airflow through the reed}
\paragraph{}Following Thompson \cite{Thompson:79}, Hirschberg \cite{Hirschberg:94} and Gazengel \cite{Gazengel:95}, the first component of the output flow to  consider, noted $u_p$, is  the volume flow induced by the motion of the reed (the "pumped" flow) whose expression is:
\begin{equation}
	u_p(t)=S_r \frac{d \zeta}{dt}.
	\label{eq:Up}
\end{equation} 
As explained in the Appendix, the effective section introduced by the calculation of $u_p$ is equal to the effective section associated with the over-pressure $\Delta p$ in the reed motion equation.

\paragraph{}The second contribution to the output flow is the flow passing under the reed. This flow is noted $u_t$ and its expression is:
\begin{equation}
	u_t(t)=\alpha .S_u(h_n).v_j,
	\label{eq:Ut}
\end{equation} 
where $\alpha$ is the \textit{vena contracta} coefficient, $v_j$ the velocity inside the free jet and $S_u$ the useful section whose definition will be discussed in the next section. We also assume that the jet velocity is the same in the whole jet even if the jet sheds all around the reed. This assumption may be a point of improvement for the proposed model but, in this case, a three dimensional description of the flow through the reed may be needed which is out of purpose here. As the reed scales are tiny compared to the wavelengths, we think that considering the same velocity and over-pressure all over the reed is a fair first approach.

\paragraph{}To complete the description of the airflow through the reed, we need to introduce a relation between the jet velocity $v_j$ and the over-pressure $\Delta p$. This relation is a Bernoulli equation for which we assume an incompressible, irrotational (because of the free jet formation) and quasi-stationary airflow. Assuming the volumic kinetic energy at the upstream of the reed to be considerably smaller than the downstream one in the free jet, when blowing, the Bernoulli equation between upstream and downstream of the reed is then:
\begin{equation}
	\Delta p=\frac{1}{2}\rho_0 v_j^2,
	\label{eq:Bern}
\end{equation} 
where the effect of the viscosity is implicitly taken into account in the free jet formation assumption.

\paragraph{}To complete the description of the flow through the reed, we now introduce the definition of the useful section.

\subsection{Useful section}
\paragraph{}In this section, we review the different kinds of models for the useful section used in the literature and propose a new one which permits better  temporal simulations.

\subsubsection{Existing models for the useful section}
\begin{figure}[h!tbp]
    \centering
    \includegraphics[height=3cm]{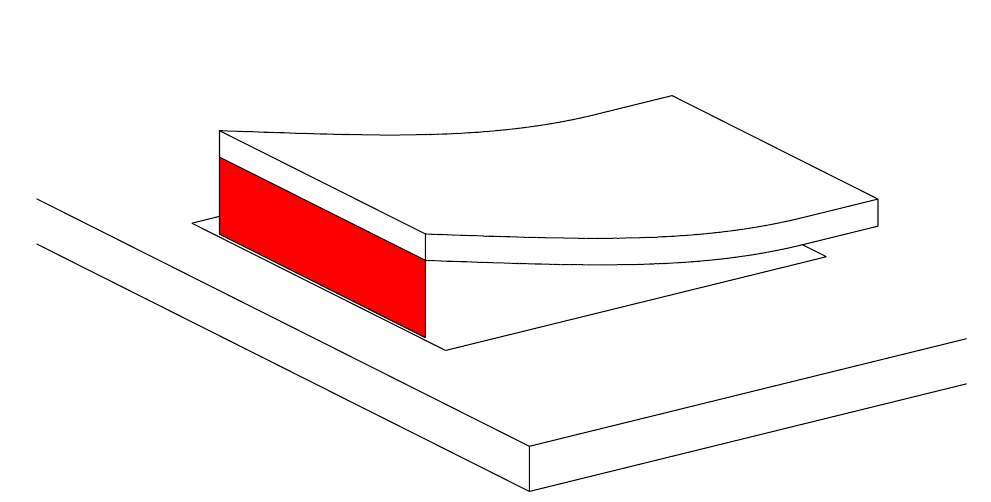}
    \caption{Illustration of the considered escape area in the case of the clarinet-like hypothesis: only a part of the front section is taken into account. This model is still used for the study of clarinet and saxophone, which use beating reeds, where it seems difficult to express the local deformations of the reed as a function of the displacement of the tip of the reed.}
    \label{f2}
\end{figure}

\paragraph{}The first model one may find for the useful section \cite{Backus:63,Wilson:74, Thompson:79, Fletcher:79a, Schumacher:81, Sommerfeldt:88, Hirschberg:90, Fletcher:93, Hirschberg:94, Gazengel:95, Millot:99} considers that the flow issues from the instruments only by the front section of the reed (see Figure \ref{f2}) and, in this case, its generic expression is then:
 \begin{equation}
	S_u=W_r.h_{up},
		\label{eq:SuClassik}
\end{equation}  
where $W_r$ is the width of the reed at its tips and  $h_{up}$ the reed opening at the free-end for the upstream section.  As this model does not consider the sides contribution, Cuesta and Valette \cite{Cuesta:95} replace the width $W$ by the global perimeter $2.L_r+W_r$ in equation (\ref{eq:SuClassik}). To take into account the fact that, during its motion,  the free reed can go inside the thickness of its support or can even protrude through the other face of its support, Millot \cite{Millot:99} assumes that the useful section is zero when the reed is inside the thickness of its support and that, in the other cases, the useful section is given by the product of the absolute value of the tip opening and the global perimeter. Indeed, it is necessary to consider the absolute value of the tip opening and not the algebraic tip opening because it  will give  negative values for $S_u$ each time the reed goes inside the thickness of the support. But the assumption of a closure of the channel, when the reed is inside the thickness of its support, should be rejected because there is always at least a minimal opening due to the clearance gaps between the reed and its support (and, as illustrated by Figure \ref{f2}, the  front section is a not a rectangle of width $W_r$ because of these clearance gaps). But, all these models consider the opening all over the reed constant, which is in contradiction with the nature of the free reed as illustrated on Figure \ref{f1}.

\paragraph{}In the case of the model proposed by Tarnopolsky \textit{et al} \cite{Tarnopolsky:00} and used by Hikichi \textit{ et al} \cite{Hikichi:01}, the useful section is calculated using the following expression:
$$S_u(h_{n})=W_r\sqrt{h^2_{n}+h^2_{min}}+2.L_r\sqrt{a^2(h_{n})+h^2_{min}},$$
where $h_{min}$ is the gap width existing between the  reed and its support when the reed is horizontal and $a(h_{n})$ is the mean aperture under the reed which is calculated as a function of the tip opening $h_{n}$ :
$$ a(h_{n})=h_{000} + \frac{1}{L_r}(h_{n}-h_{n,00})\int_0^{L_r}\psi(s)ds,$$
where $h_{000}$ is the thickness of an optional lay, $L_r$ is the length of the reed, $h_{n}$ and $h_{n,00}$ are respectively the current and the rest openings at the tip of the reed, $\psi(s)$ the reed factor.  If Hikichi uses this expression wherever the reed is (inside or outside the instrument) with an absolute value for $a(h_{n})$, Tarnopolsky \textit{ et al} say that this expression is only valid if the reed does not enter the aperture block but do not give an expression for this case.

\paragraph{}At almost the same time, Debut and Millot \cite{Debut:01a, Debut:01b}  introduced a quasi similar model for $S_u$ considering the case of a reed inside its support and also protruding through the other face of the support. But, in this model,  the useful section is calculated all over the reed sides using a local effective opening noted $h_{eff}(s)$ rather than the mean aperture. The useful section is then given by:
$$ S_u(h_{n}) =W_r\sqrt{h^2_{n}+h^2_{min}} + 2. \int_0^{L_r} \sqrt{h^2_{eff}(s)+h^2_{min}}ds,$$ 
with 
$$h_{eff}=\left\{
\begin{array}{ccl}
h_{n}- e_r & \mathrm{if} & h_{n}>e_r\\
0 & \mathrm{if} & h_{n}\in[-e_s,e_r]\\
-(h_{n}+e_s) & \mathrm{if} & h_{n}<-es
\end{array}
\right.,$$
where $e_s$ is the thickness of the support, $e_r$ is the thickness of the reed.

\begin{figure}[h!tbp]
    \centering
    \includegraphics[height=3cm]{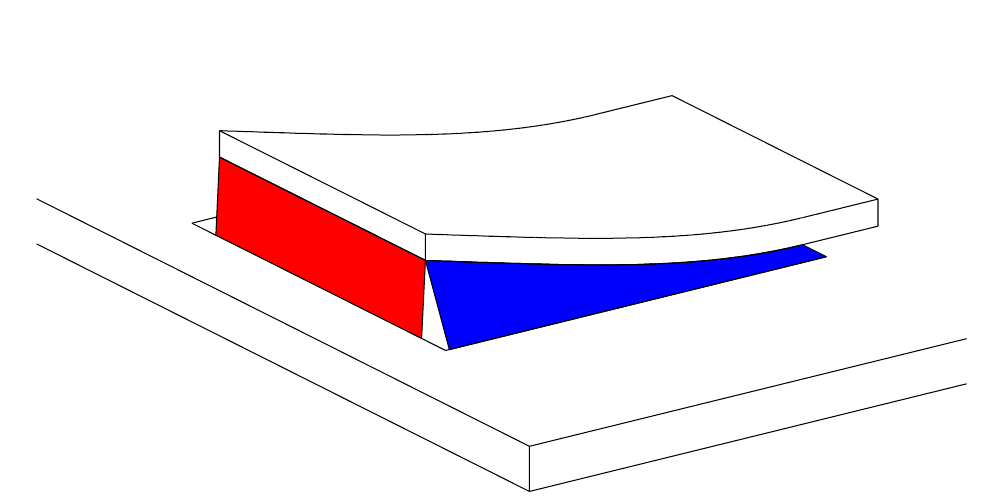}
    \caption{Illustration of the considered escape area in the case of the models used for the study of free reeds \cite{Tarnopolsky:00}, of the sho \cite{Hikichi:01} and for the study of chromatic playing on a diatonic harmonica  \cite{Debut:01a, Debut:01b}. In these models, the flow is assumed to pass by the front of the reed and also by the sides of the reed. But one can see that the considered front and lateral areas are not complete and that the orientation of the local opening changes all over the length of the reeds.}
    \label{f3}
\end{figure}

\paragraph{}But, as illustrated by Figure \ref{f3}, there are still some parts of the section under the reed which are not included in the calculation in these two models: a part of the lateral area and two triangles for the front section. The model proposed by Tarnopolsky \textit{ et al}, using a mean opening $a(h_{n})$, assumes that all the lateral escape areas are oriented in the same direction which is false: the orientation of the normal to the local escape area changes all over the sides, from vertical at the clamped end to almost horizontal at the free end of the reed. This problem is almost solved by the model proposed by Debut and Millot with the use of an integral over the sides of the local openings under the reed. 

\paragraph{}In the model of Debut and Millot, there is still a problem with the definition of the useful section because, even if we do not have a zero $S_u$ when the reed is inside its support,  we change the frontiers of $S_u$ when the reed is at the other side of the support. Indeed, with the definition of $h_{eff}$, we consider the effective height between the downstream face of the support and the downstream face of the reed, when the reed protrudes at the downstream face of the support, while we consider the upstream face of the support and the downstream face of the reed when the reed is inside the instrument. As this change is not natural we propose a new model for the useful section.

\subsubsection{New model for the useful section}
\paragraph{}We have seen that the existing models do not take account of the total section under the reed, as illustrated in Figure \ref{f3}. We also have seen that there is some confusion between the neutral, upstream and downstream  sections of the reed in the case of the definition of the useful section. We now propose another model which may solve this problems.

\paragraph{}We need to distinguish (-,+) and (+,-) reed because the frontiers of the escape area are different and the expressions of the contributions of useful section change. For a (-,+) reed, we consider the escape area between the downstream face of the reed and the upstream face of the support while, for a (+,-) reed, the escape area is chosen between the upstream face of the reed and the downstream face of the reed. But, in both cases, we distinguish three contributions to the useful section: the front section; both sides sections limited by the free end of the reed as proposed by Tarnopolsky or Debut; the missing sides areas.

\paragraph{}The first contribution is the front area, a trapezoid, as the floor frontier in the plane of the support includes the two clearance gaps between the reed and the window in the support (see figure \ref{f3}). The expression for the front contribution is then:
$$S_{end}=(W_r+h_{min}).h_{end},$$
where $h_{end}$ is the height of the trapezoid.

\begin{figure}[h!tbp]
    \centering
   \includegraphics[width=7.5cm]{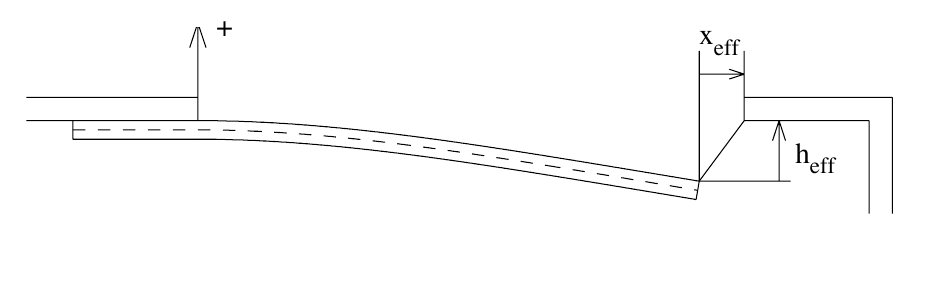}
   \includegraphics[width=7.5cm]{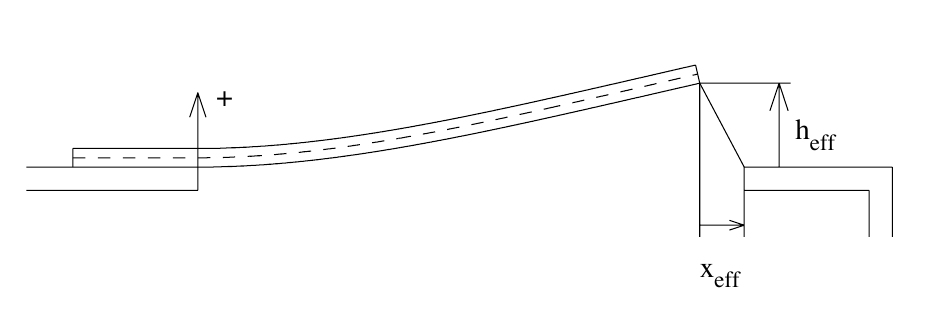}
    \caption{The height $h_{end}$ of the trapezoid is illustrated by the segment linking the end point of the reed and the end point of the aperture in the support. One can also see the definition of both lengths $h_{eff}$ and $x_{eff}$ useful to calculate $h_{end}$. $h_{eff}$ is the opening under the reed limited by the plane of the support. $x_{eff}$ is defined as the length of the segment between the projection of the end point of the reed in the plane of the support and the end point of the aperture. These definitions are illustrated for both (-,+) reed (left) and (+,-) reed (right).}
    \label{f4}
\end{figure}

\paragraph{}We can calculate the height of the trapezoid as the hypotenuse of a triangle whose sides are respectively the effective opening $h_{eff}$ under the reed (illustrated by Figure \ref{f4} for both kinds of reed) and the length $x_{eff}$ of the horizontal segment limited by the projection of the free-end point of the reed we consider and the end of the aperture in the support. The  effective opening under the reed is given, in the (-,+) reed case by:
$$h_{eff}=|h_{do}+e_s|,$$
and, in the (+,-) reed case by:
$$h_{eff}=|h_{up}|.$$
 
 \paragraph{}We must emphasize that we find $e_s$ in the expression of $h_{eff}$ (an after in the expression of $h_{eff}(s)$) in the case of the (-,+) reed,  just because we have chosen the same convention and origin for both reeds. In fact, $e_s$ does not play a role in the whole determination of the useful section because it is also present, with a minus sign, in $h_{do}$ and $h_{do}(s)$. If we had chosen as zero (deflection) point the upwards face of the reed, we would not have $e_s$ in the expressions of $S_u$ and we would be able to point out the fact  that the expressions for $S_u$ for a (-,+) reed are derived by replacing $h_{do}$ by $h_{up}$ in the expressions of $S_u$ for the (+,-) reed:  this change is also equivalent to replacing $e_r$ by $-e_r$ when passing from the (+,-) case to the (-,+) case. So, there is no need to study the influence of the $e_s$ parameter on the behavior of the reed with the proposed model. 
  
\paragraph{}The end of the aperture in the support is situated at a distance $L_r+h_{min}$ from the clamped end and we consider that the projection of the end point of the reed is situated at a distance $L_r+\Delta x$ from the clamped end of the reed which gives for $x_{eff}=h_{min}- \Delta x$. For a (-,+) reed, as we need to consider the projection of the end point of the downstream face of the reed,  $\Delta x$ is then given by $\Delta x= \Delta x_{do}$ (see Appendix for details concerning the expressions of $\Delta x_{do}$ and $\Delta x_{up}$). For a (+,-) reed, we consider the projection of the upstream face of the reed  which gives a $\Delta x= \Delta x_{up}$. In these conditions, the  height of the trapezoid is then:
$$h_{end}=\sqrt{h^2_{eff} + (h_{min}- \Delta x)^2},$$
so that the front section is then given by:
$$S_{end}=(W_r+h_{min}).\sqrt{h^2_{eff} + (h_{min}- \Delta x)^2}.$$

\begin{figure}[h!tbp]
    \centering
   \includegraphics[width=7.5cm]{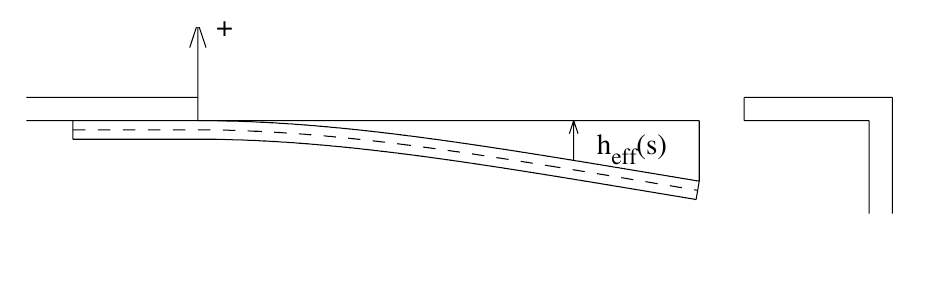}
   \includegraphics[width=7.5cm]{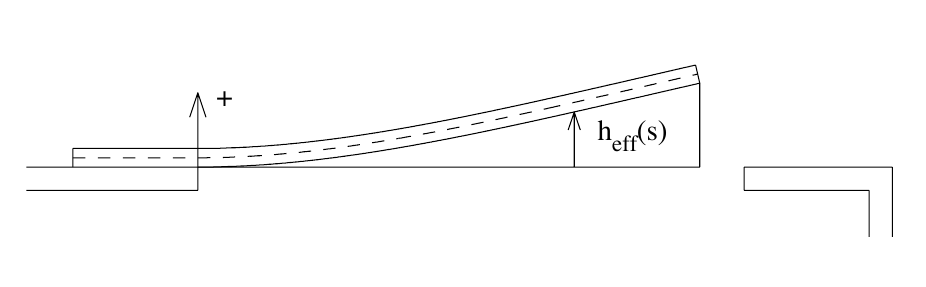}
    \caption{Illustration of the section under the reed and of the definition of the local effective height $h_{eff}(s)$ under the reed for (-,+) reed (left) and (+,-) reed (right). These side areas are limited by the face of the support involved in the calculation, by the associated face of the reed and by a vertical segment linking the plane of the support and the end point of the reed.}
    \label{f5}
\end{figure}

\paragraph{}The second contribution to the useful section is the one associated with the side escape areas delimited by the end of the reed as illustrated by Figure \ref{f5}. To calculate the contribution of each portion of the reed we need to extend the definition of the effective height $h_{eff}$ to $h_{eff}(s)$: $h_{eff}(s)=|h_{do}(s)+e_s|$ for a (-,+) reed and $h_{eff}(s)=|h_{up}(s)|$ for a (+,-) reed. We also need, for each portion of the reed, to consider the gap between the reed and its support to calculate the local effective opening under the reed and, for each side, integrate it over the length of the reed to find the side  area:
$$S_{sides}=2\int_0^{L_r} \sqrt{ h^2_{eff}(s) + h^2_{min} }ds.$$

\paragraph{}There are still two missing areas to complete the useful section: the areas of the triangles 
delimited by the end point of the reed, its projection in the plane of the support and the end of the aperture in the support. With the help of Figure \ref{f4}, one can see that the side lengths of the triangles are respectively $h_{eff}$ and $x_{eff}$ so that the missing area is then derived as:
$$S_{tri}=h_{eff}.(h_{min}-\Delta x).$$

\begin{figure}[h!tbp]
    \centering
   \includegraphics[width=5cm]{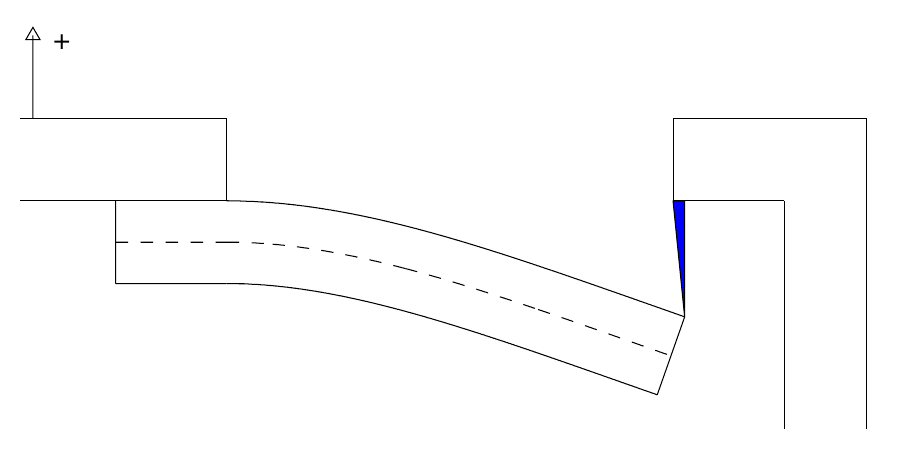}
   \includegraphics[width=5cm]{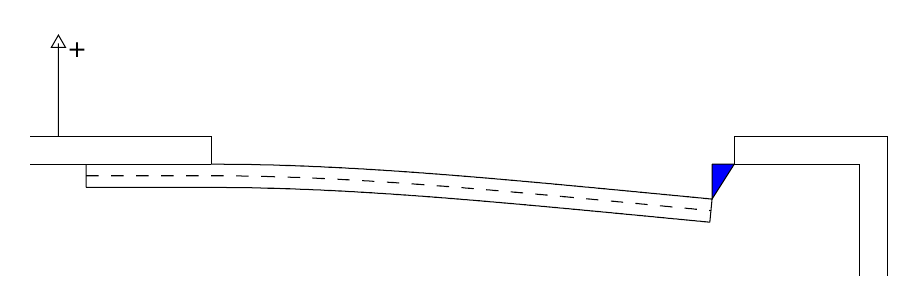}
    \caption{According to the opening of the reed, the triangle area needs to be negative (left) or positive (right) to compensate the defaults of the description of the sides sections under the reed. Indeed, the rotation of the cross section of the reed, for high magnitude openings and thick reed,  can drive the end point further than the limit of the aperture in the support. In such cases, the area under the reed is bigger than the real area and needs to be corrected. This is the reason why $S_{tri}$ is an algebraic area.}
    \label{f6}
\end{figure}

\paragraph{}As shown on Figure \ref{f6}, the area of the triangle needs to be algebraic because for some high magnitude openings of the reed, the end point of the reed is further than the end of the aperture in the support. This possibility is taken into account in the former expression of $S_{tri}$ because $h_{min}-\Delta x$ can be positive or negative according to the position of the useful end point of the reed.  

\paragraph{}Finally we can write the expression of the useful section $S_u$, given by the sum of these three contributions, $S_u=S_{end}+S_{sides}+S_{tri}$, as :
\begin{multline}
S_u=(W_r+h_{min}).\sqrt{h^2_{eff} + (h_{min}- \Delta x)^2}+h_{eff}.(h_{min}-\Delta x)\\
+2\int_0^{L_r} \sqrt{ h^2_{eff}(s) + h^2_{min} }ds.
\end{multline}

\begin{figure}[h!tbp]
    \centering
   \includegraphics[width=10cm]{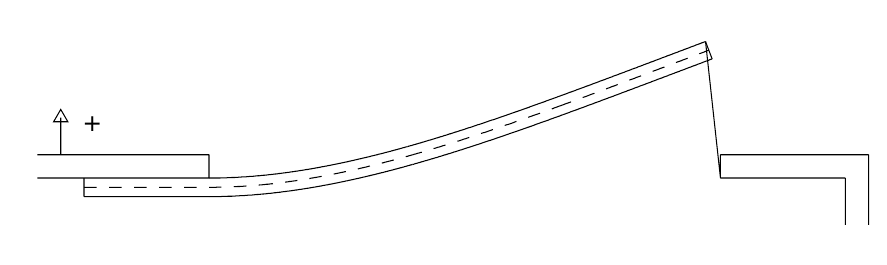}
    \caption{Illustration of a default of the proposed model. For some reed openings, the front area and the reed end intersect so the frontier may be different.}
    \label{f7}
\end{figure}

\paragraph{}One must note that in certain circumstances there is a problem of intersection between the front area we have defined and the corner of the reed, as illustrated by Figure \ref{f7}. To cope with this problem, it may be necessary to modify the definition of the frontiers. But, as we have a corner with sharp edges we can think there may be some flow separation and vortex shedding near the corner but we do not know how it will occur \cite{Gibiat:02}. If we want to take account of these phenomena we may also have to consider the real 3D configuration for the flow through the reed, varying jet velocity along the reed and also variable  over-pressure all over the reed. But flow visualisations performed by Fabre \cite{Fabre:03} clearly indicate that we have some free jets all around the reed which may be a positive argument for our assumptions. Moreover, the pictures of the flows show that the orientation of the free jet depends on the opening of the reed: from a jet normal to the support when the reed is flat, we can see a rotation of the jet down to the support as the reed opening increases for the reed sides views and the reed front views.  The rotation of the free jet for the sides or the front of the reed are explicitly taken into account in our model so we think it may correspond to a fair first approach: one can consider that we unfold both lateral section to consider only a so-called extended front section. As we want to keep a model as simple as possible we do not modify the model for the useful section to adopt a real 3D configuration. 

\paragraph{}With the equations (\ref{eq:Reed}) to (\ref{eq:Bern}) and the definition of the useful section, the local description of the reed is not complete as there are still some missing data: the excitation upstream of the reed and the "load" downstream of the reed. These additional assumptions permit definition of the different reed instruments and identification of easy playing conditions. A minimal configuration is proposed in the following to be able to study both (-,+) and (+,-) reeds.

\section{Discussion of the minimal model}
\paragraph{}In this section, we  illustrate the interest of the modification introduced for the useful section $S_u$  for both (-,+) and (+,-) reeds and examine the assumptions used to derive the whole model. Some first results concerning the nature of the bifurcation for both reed are also given. We also study the influence of the volume length $L_1$, of the excitation velocity $v_0$ and of the reed thickness $e_r$.

\subsection{Minimal configuration}
\begin{figure}[h!tbp]
    \centering
    \includegraphics[height=6cm]{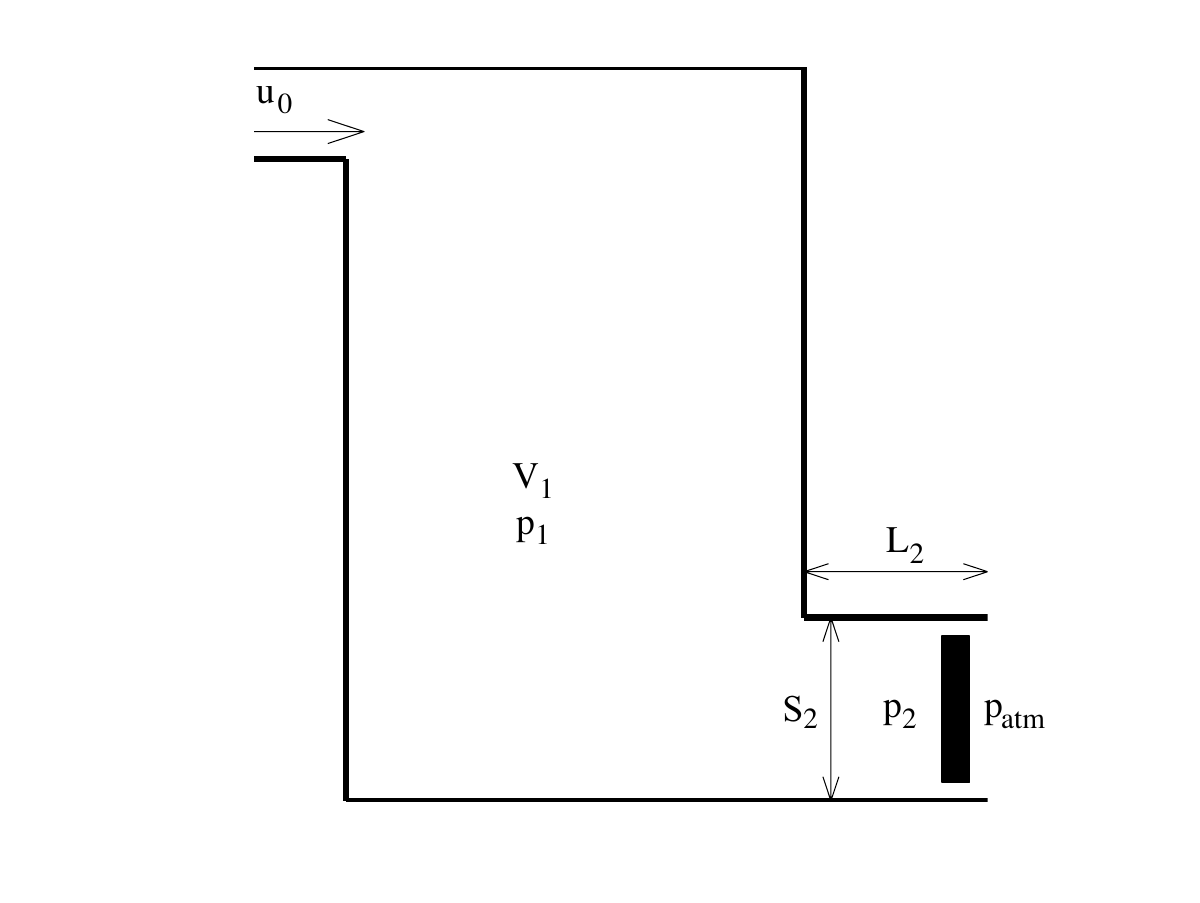}
    \caption{Configuration introduced to study both kinds of free  reeds. The excitation is the constant volume flow $u_0$ supplying the volume $V_1$ where the pressure $p_1$ is assumed to be uniform. After the volume, one encounters a pipe of length $L_2$ and cross area $S_2$ upstream of the reed, in which the airflow is considered incompressible. Downstream of the reed the pressure is atmospheric.}
    \label{f8}
\end{figure}

\paragraph{}To be able to study both (-,+) and (+,-) reeds with the same conditions we introduce the configuration of Figure \ref{f8}. In this configuration, we first encounter a volume $V_1$ in which the pressure $p_1$ is assumed to be uniform. This volume is supplied by a fixed volume flow $u_0$ and we adopt the adiabatic hypothesis for the volume: we use in fact the low Mach number approximation for the adiabatic hypothesis to derive the mass conservation equation. After the volume, we find a pipe of length $L_2$ and cross section $S_2$. In this pipe, we consider an irrotational and  incompressible airflow ($p_2$ is uniform in the entire pipe) and also assume  that the kinetic energies are much less important than the unsteadiness. We then find the reed, either a (-,+) or a (+,-) one, and, finally, the outside of the configuration where we assume to find the atmospheric pressure. 

\paragraph{}In such conditions, the missing equations to get a complete reed problem are then:
\begin{description}
	\item[$\bullet$] $\di \frac{V_1}{c^2_0}\frac{d (p_1-p_{atm})}{dt}= \rho_0 (u_0 -u)$, mass conservation for the volume $V_1$;
	
	\item[$\bullet$] $\di p_1 =p_2 + \rho_0\frac{L_2}{S_2}\frac{du}{dt}$, Bernoulli equation for the pipe;

	\item[$\bullet$] $\di p_2 = p_{atm} + \frac{1}{2} \rho_0 v^2_j$, Bernoulli equation between the upstream and the downstream of the reed;

	\item[$\bullet$] $\di \frac{d^2 \zeta}{dt^2}+Q^{-1}\omega_0 \frac{d \zeta}{dt} + \omega^2_0 \zeta= \mu (p_2 - p_{atm})$, reed motion equation;

	\item[$\bullet$] $\di u=S_r\frac{d \zeta}{dt}+\alpha S_uv_j$, total output volume flow equation. 
\end{description}

\paragraph{}To determine the conditions of reed instabilities, we calculate the expression of the upstream pressure $p_2$ as a function of $\zeta$ and its derivatives, using the Fourier transform of the linearized versions of the equations (see Appendix for details). The part of  the over-pressure which is proportional to the derivative  $\di \frac{d\zeta}{dt}$ may permit instability if its value is greater than $\di \frac{Q^{-1} \omega_0}{\mu} \frac{d \zeta}{dt}$. 

\paragraph{}Given the symmetry of the  laws for the useful section (see Figure \ref{f9}), we assume that $S_u=S_{u00}+a_1(h_{n}-h_{n,00})^2$ is a  fair approximation of the useful section near the mean opening. In the following, to simplify the expressions we introduce $\zeta_0=h_{n,0}-h_{n,000}$, $\zeta_{00}=h_{n,00}-h_{n,000}$ where $h_{n,0}$ is the mean opening, $S_{u0}=S_{u00}+a1(\zeta_0-\zeta_{00})^2$. We also use the prime suffix for the variable or acoustical part of quantities  used in the calculation: $\zeta'$ is for instance the acoustical part of $\zeta$ while $\zeta_0$ is the mean value of $\zeta$. 

\paragraph{}With linearization of the equations and some further calculation, the Fourier transform of $p'_2$ can be written:
\begin{multline}
P'_2(\omega)=\frac{\di -\rho_0 \big( 1- \frac{V_1L_2}{c^2_0S_2} \omega^2\big)}
{\di \big(\frac{\alpha S_{u0} }{ v_{j0} }\big)^2 \big( 1 - \frac{V_1L_2}{c^2_0S_2} \omega^2 \big)^2+ \big(\frac{V_1}{c^2_0}\omega  \big)^2}.\bigg[ \alpha A \frac{\alpha S_{u0}}{ v_{j0}} \big( 1- \frac{V_1L_2}{c^2_0S_2} \omega^2 \big) + \omega^2 S_r \frac{V_1}{ c^2_0}\\
+ \mathrm{j}\omega \bigg(S_r \frac{\alpha S_{u0}}{ v_{j0}} \big( 1- \frac{V_1L_2}{c^2_0S_2} \omega^2 \big)
- \alpha A \frac{V_1}{ c^2_0} \bigg) \bigg].\zeta'(\omega),
\end{multline}
 where $A=2a_1(\zeta_0-\zeta_{00})v_{j0}$.

\paragraph{}One can see that the condition for instability implies that:
$$\big( 1- \frac{V_1L_2}{c^2_0S_2} \omega^2\big)\big[ 
S_r \frac{ \alpha S_{u0} }{v_{j0} } \big( 1- \frac{V_1L_2}{c^2_0S} \omega^2 \big) - \alpha A \frac{V_2}{ c^2_0}
\big]<0,
$$
which may be verified for (-,+) and (+,-) reeds because we can  play, at least, on the signs of two factors: $\di 1- \frac{V_1L_2}{c^2_0S_2} \omega^2$, depending of the volume and pipe dimensions and $A$ depending of the nature of the reed. In fact, as $A>0$ for a (+,-) reed we must have $\di 1- \frac{V_1L_2}{c^2_0S_2} \omega^2>0$ while, as $A<0$ for a (-,+) reed, we must have $\di 1- \frac{V_1L_2}{c^2_0S_2} \omega^2<0$.  

\paragraph{}The keypoints for the numerical simulations are given in the Appendix.

\paragraph{}As the proposed configuration may permit the instabilities of both kinds of reed, we can now  give some numerical results with this minimal model.

\subsection{Illustration of the minimal model}
\paragraph{}In this subsection we apply the model for two reeds, with the same geometrical and mechanical properties but with a different settlement: one (-,+) reed and a (+,-) another one. The geometrical and mechanical properties are given by Table \ref{t2}. 

\begin{table}
\begin{tabular}[h!tbp]{lrr}
description & notation & value\\
\hline
length of the reed & $L_r$ & 12.95 mm\\
width of the reed & $W$ & 2.1 mm\\
thickness of the reed & $e_r$ & 110 $\mu$m\\
thickness of the support & $e_s$ &900 $\mu$m\\
rest departure from flat reed position & $\Delta h_{n,00}$ & 528 $\mu$m\\
rest reed opening & $h_{n,00}$ &\\
& (-,+) reed & $ -\Delta h_{n,00}-\frac{e_r}{2}-e_s$\\
& (+,-) reed & $\Delta h_{n,00}+\frac{e_r}{2}$\\
clearance gaps width & $h_{min}$ & 50 $\mu$m\\
eigenfrequency & $f_r$ & 444 Hz\\
equivalent reed stiffness & $K$ & 47.9 N.m$^{-1}$\\
quality factor & $Q$ & 95\\ 
\end{tabular}
\caption{Mechanical and geometrical characteristics common for both tested reed. These characteristics  were found for a (+,-) reed of a G diatonic harmonica on channel 4 (Lee Oscar$^{\mathrm{TM}}$ model) \cite{Millot:01}.}
\label{t2}
\end{table} 

\paragraph{}For this comparison, we compute, in time domain, the system of equations for the minimal model. We assume constant the parameters given by Table \ref{t3} but we consider variable values for the feeding velocity $v_0$ ($u_0=S_0.v_0$), the length of the volume 1  according to the considered reed.

\begin{table}
\begin{tabular}[h!tbp]{lrr}
description & notation & value\\
\hline
cross section area of the feeding pipe &  $S_0$ & 30 mm$^2$\\
cross section area of the volume 1&  $S_1$ & 800 mm$^2$\\
length of the pipe 2 & $L_2$ & 20 mm\\
cross section area of the pipe 2 & $S_2$ & 25 mm$^2$\\
\end{tabular}
\caption{Constant geometrical dimensions  for the configuration used to study both (-,+) and (+,-) reeds.}
\label{t3}
\end{table}

\begin{figure}[h!tbp]
    \centering
   \includegraphics[width=7cm]{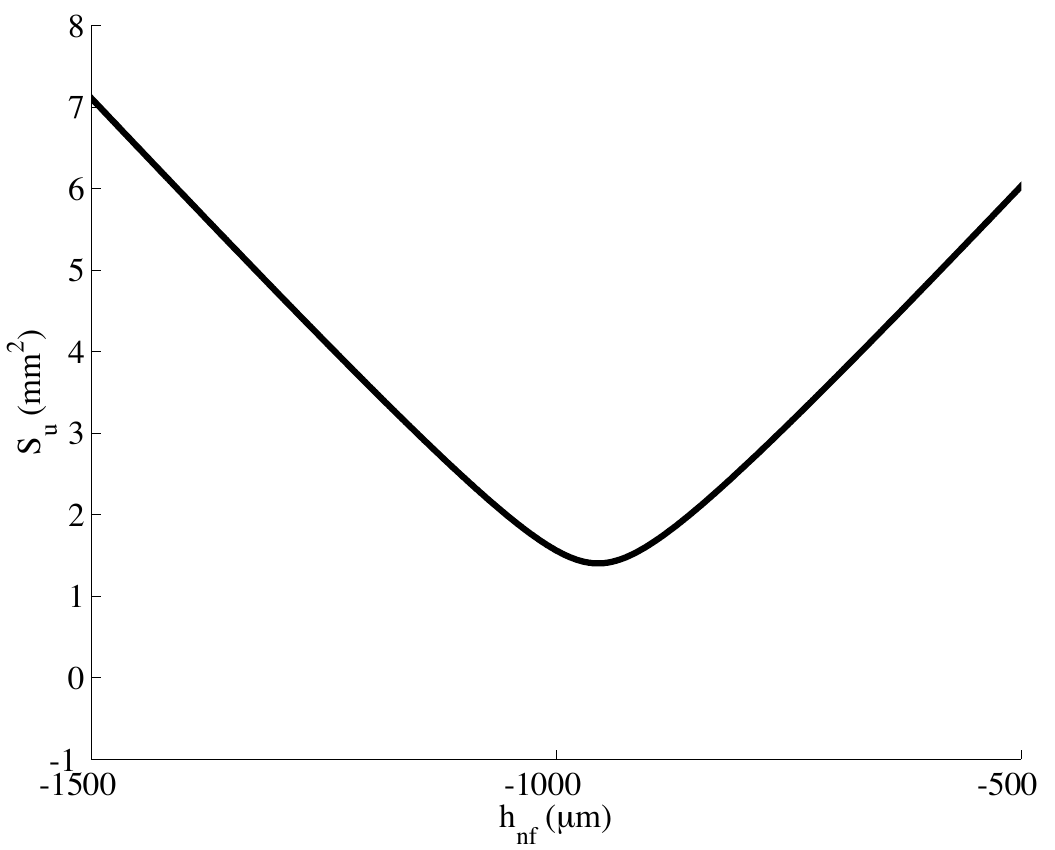} 
  \includegraphics[width=7cm]{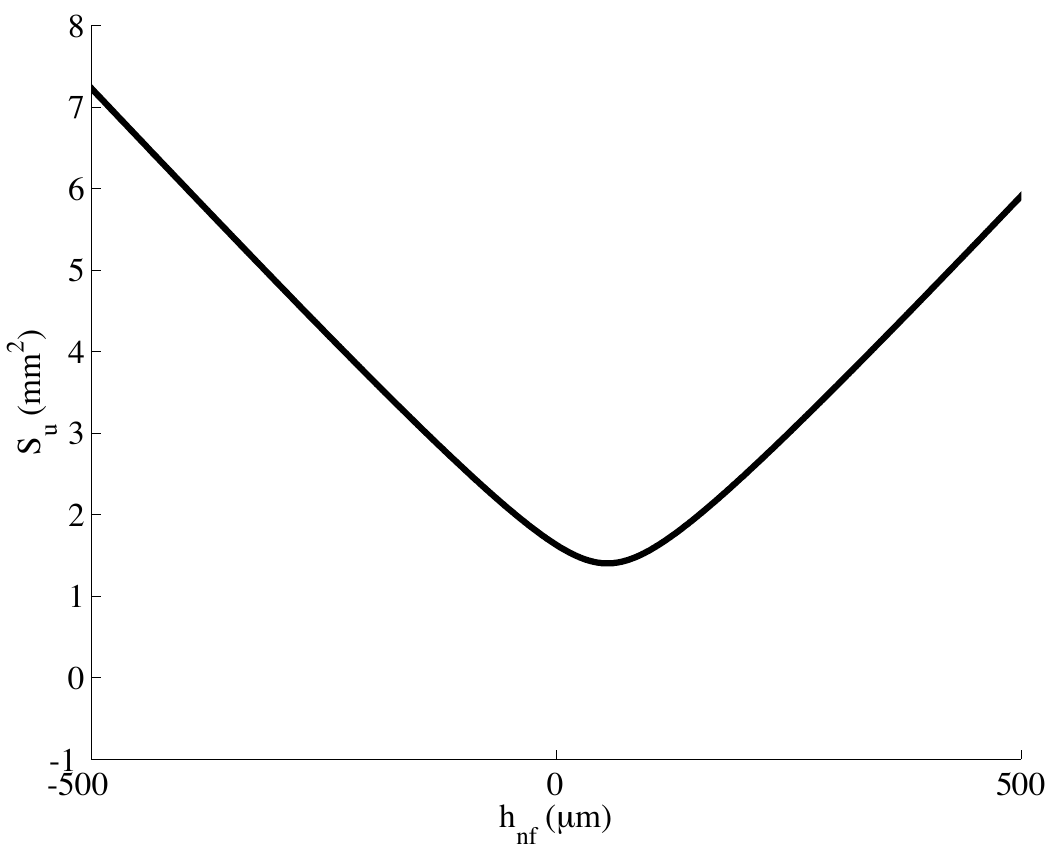}
   \caption{Illustration of the laws for the useful section  $S_u$ as a function of the neutral section displacement $h_{n}$ in the case of  a (-,+) reed (left) and in the case of a (+,-) reed (right).}
    \label{f9}
\end{figure}

\paragraph{}The laws of the useful section for both reeds are given in Figure \ref{f9}. One can see that, as the differences between the reeds is only the settlement, we have the same laws but translated: for the (-,+) reed the law is centered around $\di h_{n}=-e_s-\frac{e_r}{2}$ while, for the (+,-) reed, the center is at  $\di h_{n}=\frac{e_r}{2}$. In fact, in both cases, we find that the centre corresponds to the flat reed position and that the law seems symmetrical which is normal because the neutral section is at the centre of the thickness which gives symmetrical displacements of points of both upstream and downstream sections around the points of the neutral section.

\begin{figure}[h!]
    \centering
   \includegraphics[width=4.5cm]{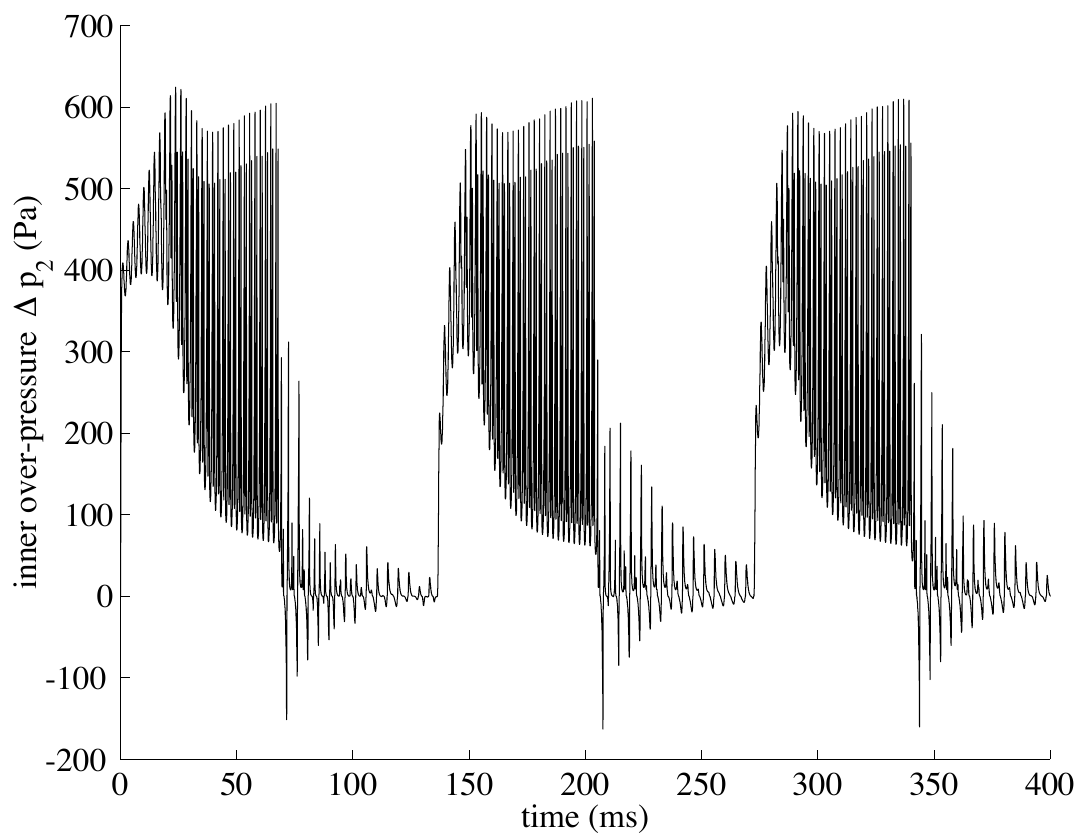} 
  \includegraphics[width=4.5cm]{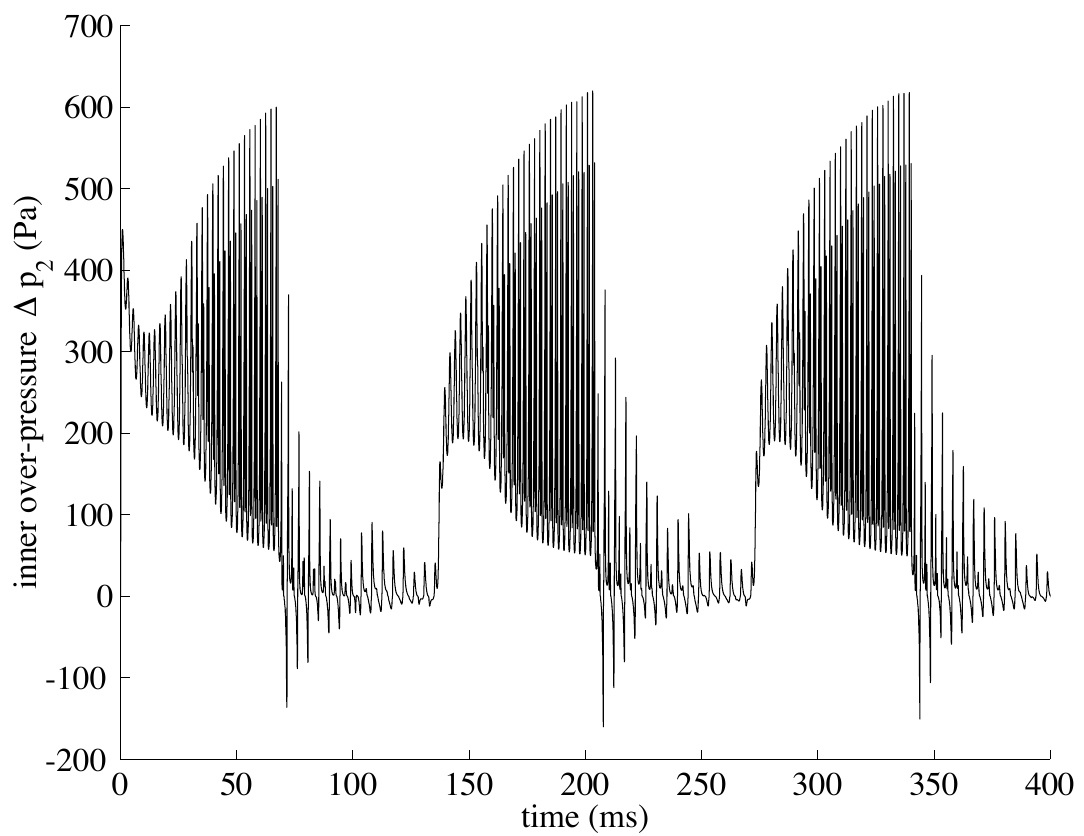}
  \includegraphics[width=4.5cm]{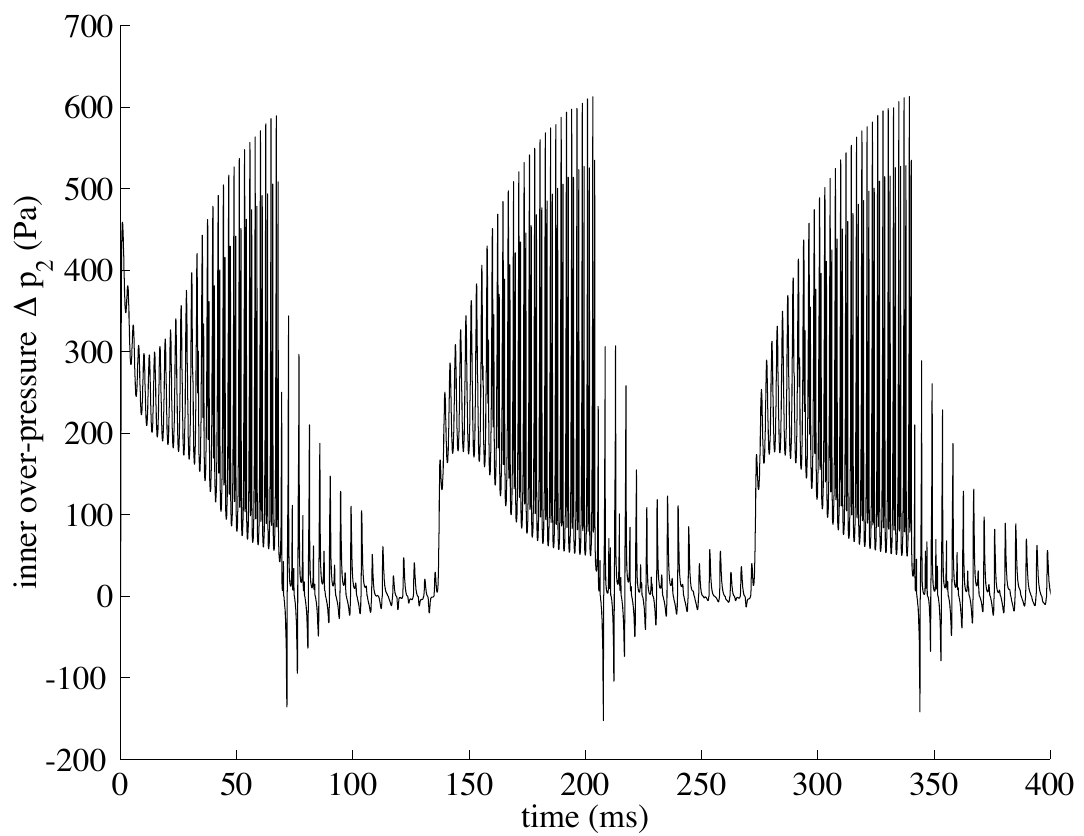}
  \includegraphics[width=4.5cm]{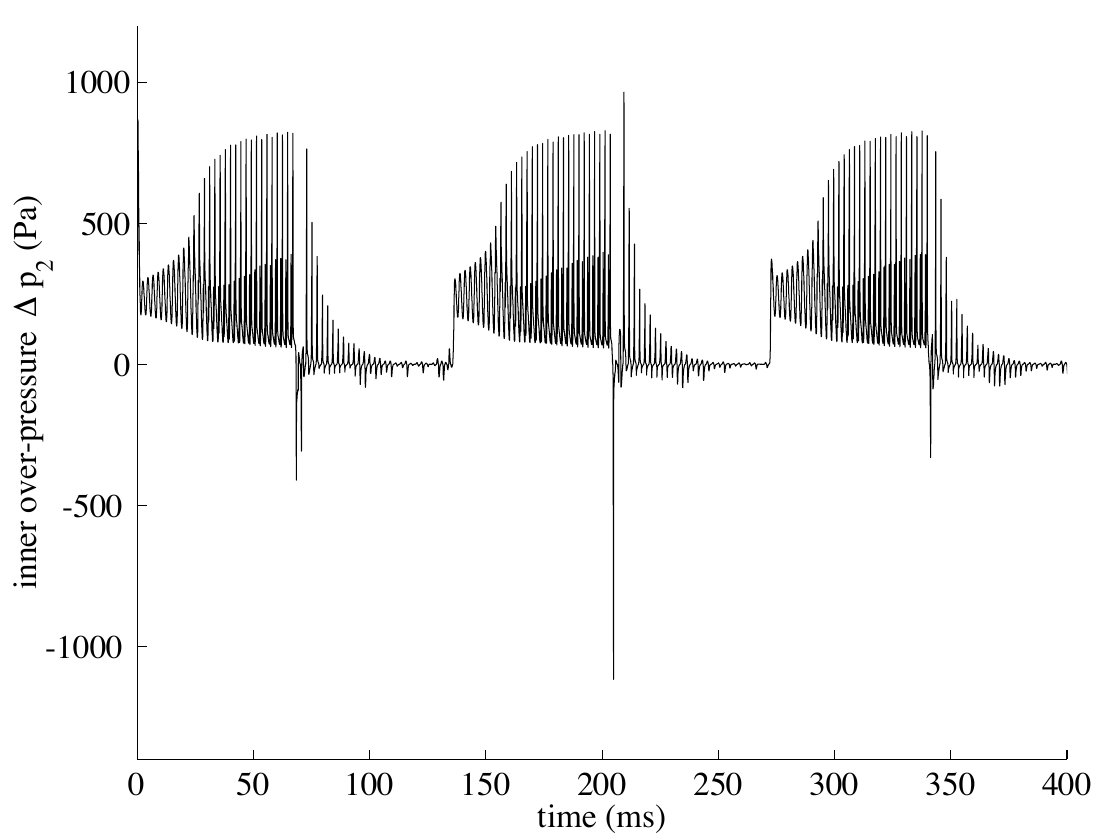}
  \includegraphics[width=4.5cm]{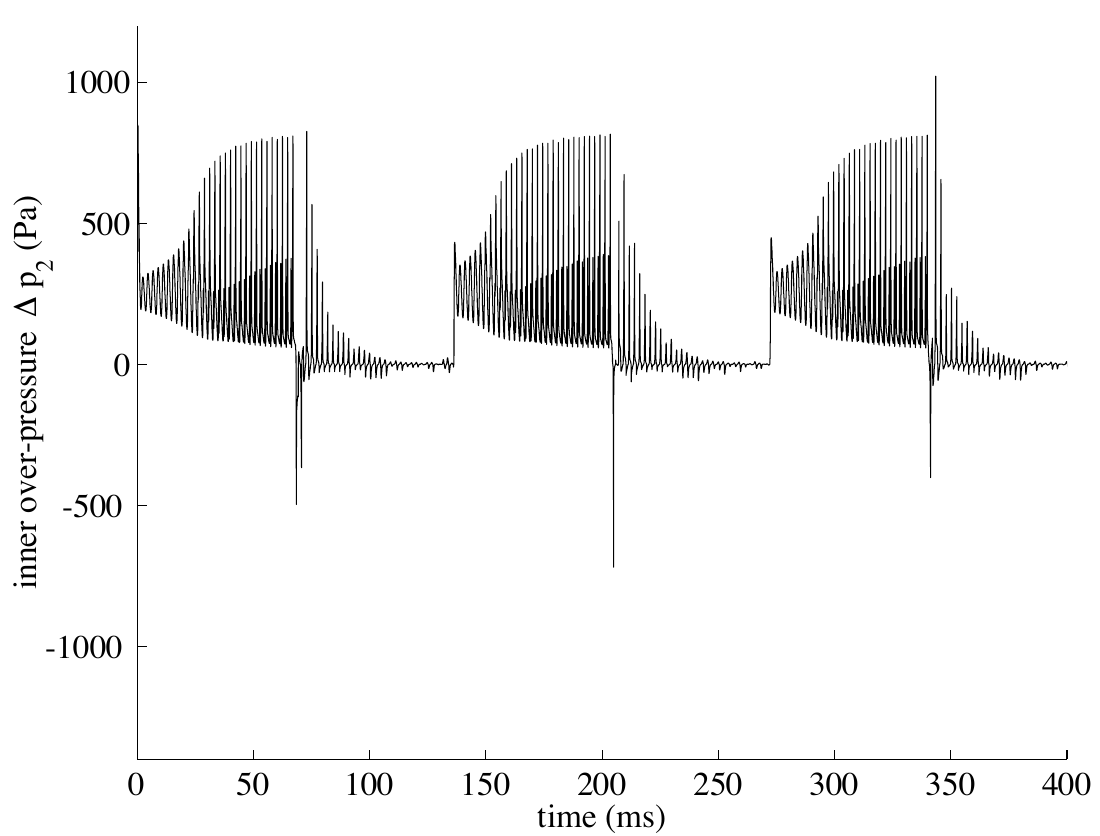}
  \includegraphics[width=4.5cm]{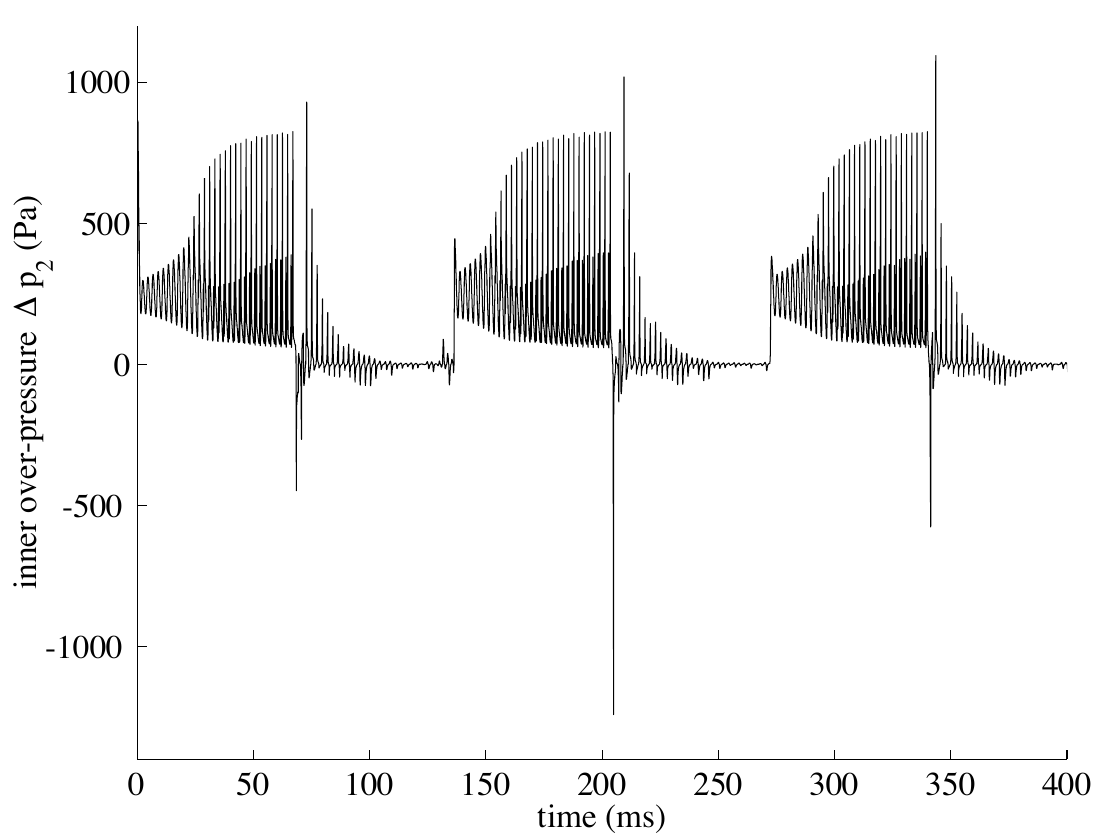}
   \caption{Illustration for blown notes of the different dynamical behaviors of  a reed while using Millot (left), Hikichi (centre) or Debut (right) model for the useful sections. (top) waveforms for the (-,+) reed: $L_1=1.5$ cm, $v_0=3$ m.s$^{-1}$. (bottom) waveforms for the (+,-) reed: $L_1=8$ cm, $v_0=2.5$ m.s$^{-1}$. Numerical simulations are performed using three repetitions of 3000 samples with constant  excitation signal $v_0$ followed by 3000 zeros in order to simulate an artificial but dynamic excitation close to a musical triplet.}
    \label{f10}
\end{figure}

\paragraph{}Plots of figure \ref{f10} illustrate the behavior of the three models of useful sections (Millot, Hikichi and Debut) for an excitation composed of the three steady portions separated by zero portions: we tried to be as close as possible to a triplet using steady excitations. These simulations are performed for both (-,+) and the (+,-) reeds and one can notice that while the three models seem to give quite similar simulations for a (+,-) reed, the results are rather different for a (-,+) reed. The Hikichi and Debut models give rather similar results compared to the Millot model which reacts more quickly. 

\begin{figure}[h!]
    \centering
   \includegraphics[width=4.5cm]{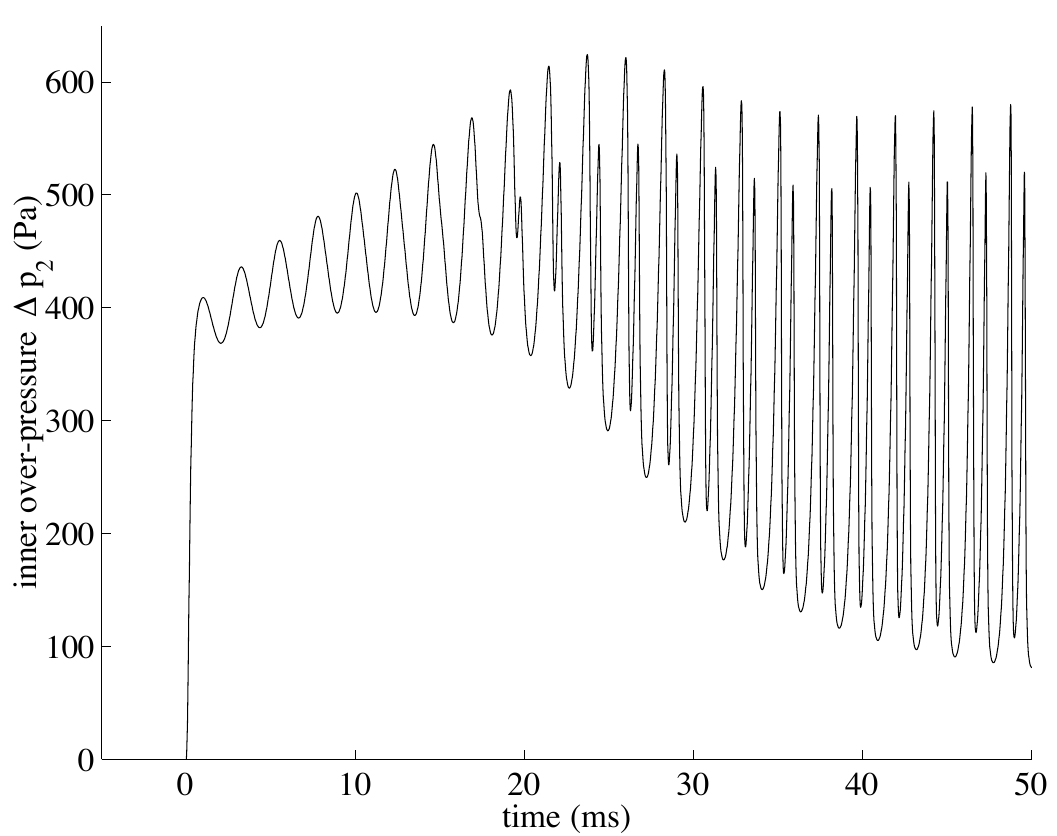} 
  \includegraphics[width=4.5cm]{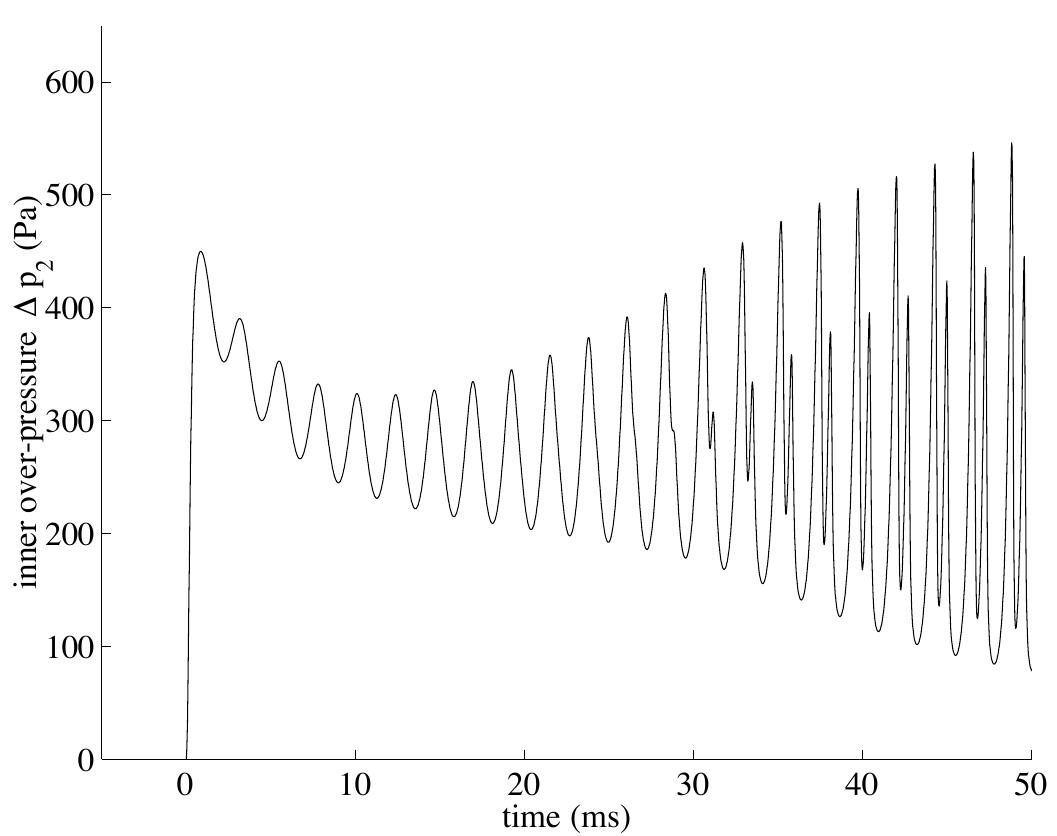}
  \includegraphics[width=4.5cm]{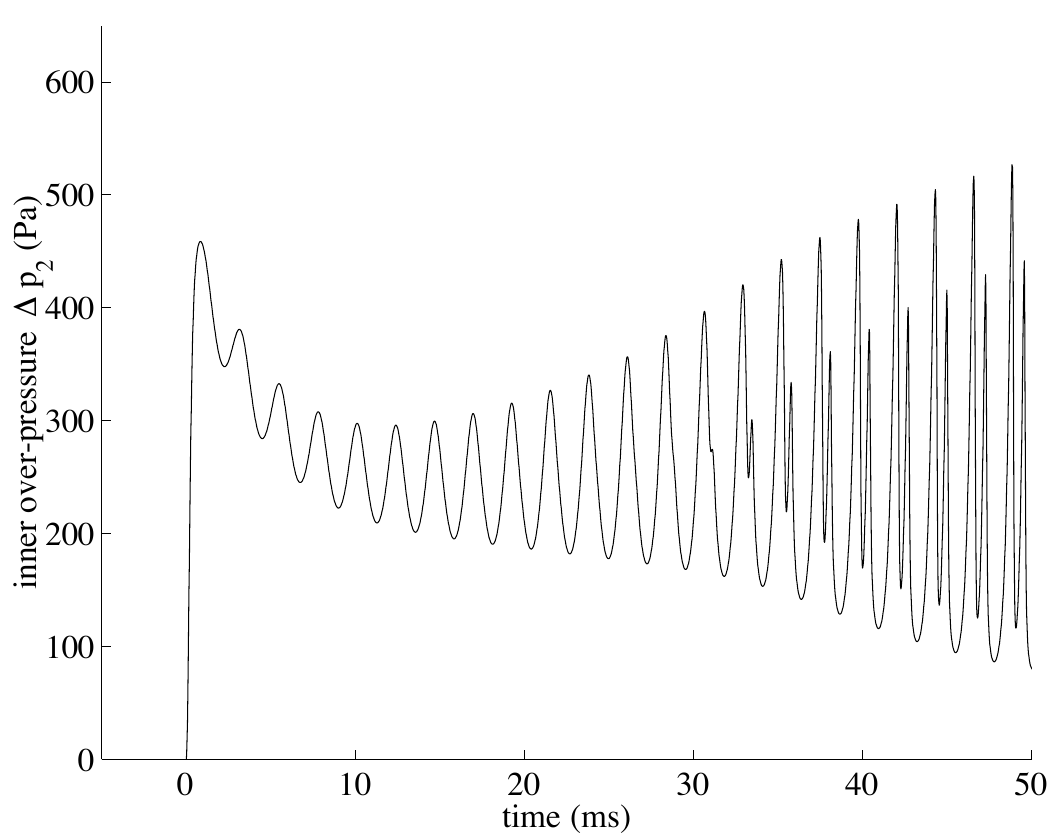}
  \includegraphics[width=4.5cm]{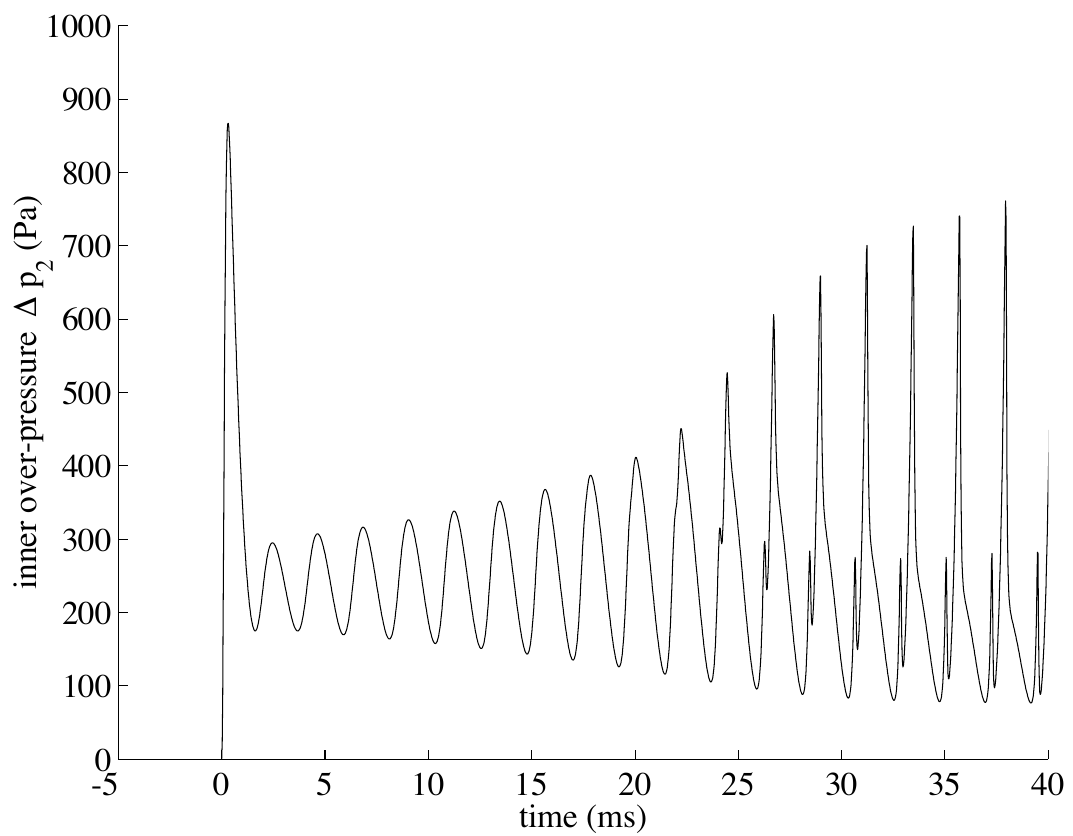}
  \includegraphics[width=4.5cm]{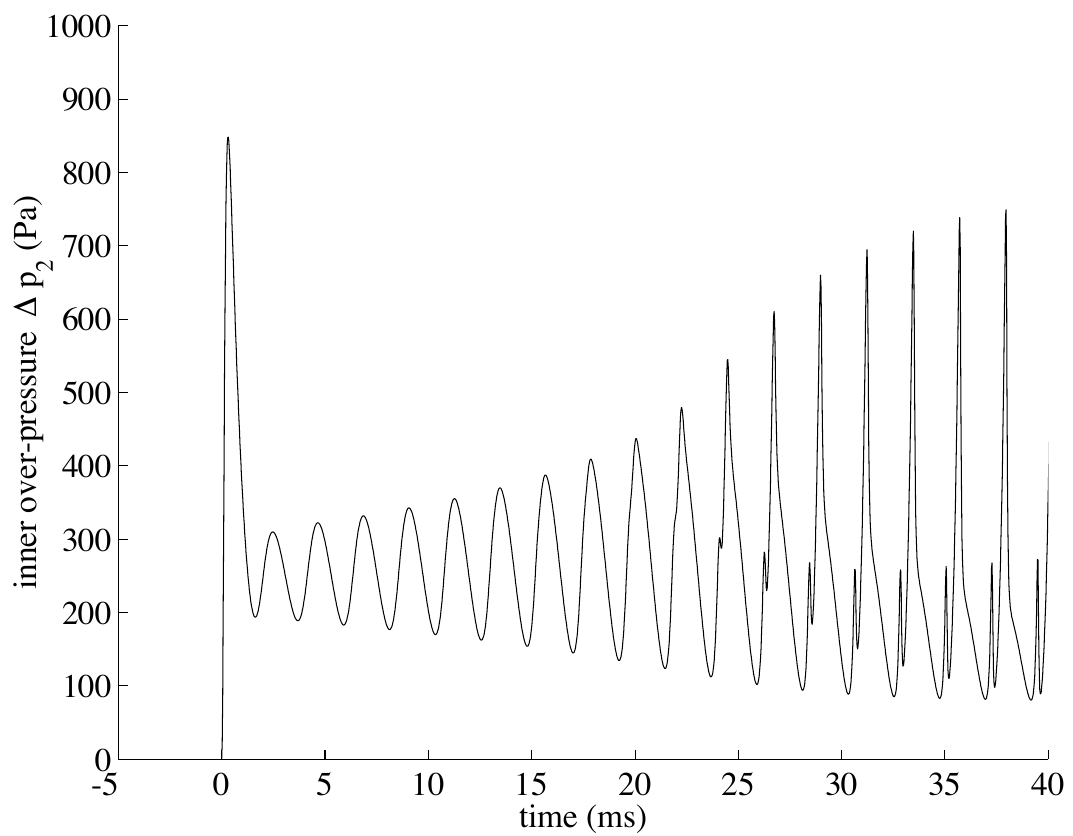}
  \includegraphics[width=4.5cm]{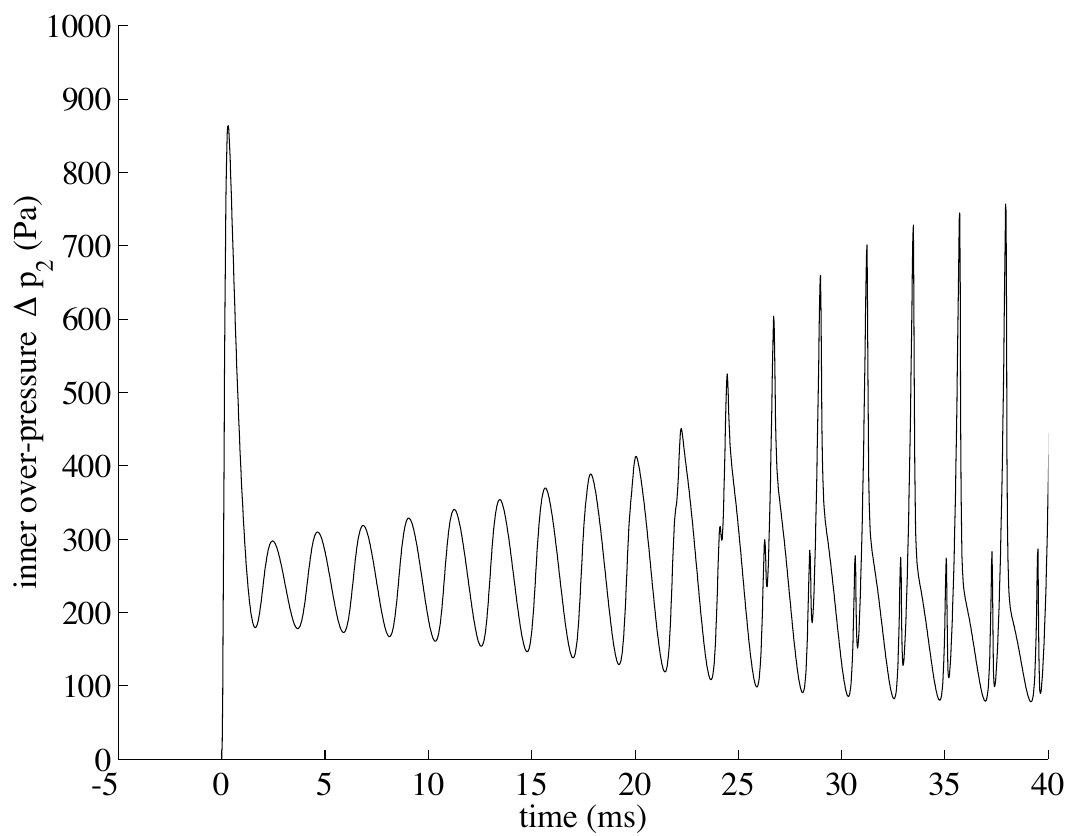}
   \caption{Zooms on the beginnings of the plots of figure 10 which illustrate the different dynamical behaviors of  a reed while using Millot (left), Hikichi (centre) or Debut (right) model for the useful sections. (top) waveforms for the (-,+) reed ; (bottom) waveforms for the (+,-) reed.}
    \label{f11}
\end{figure}

\begin{figure}[h!]
    \centering
   \includegraphics[width=4.5cm]{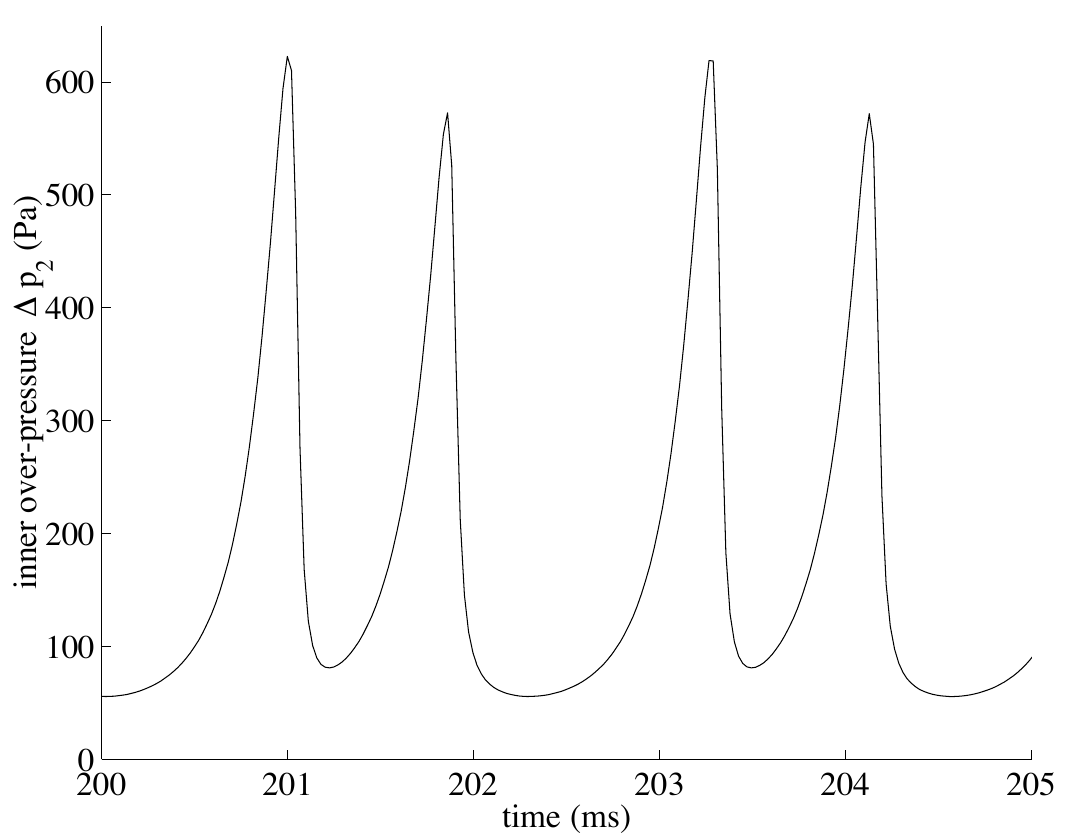} 
  \includegraphics[width=4.5cm]{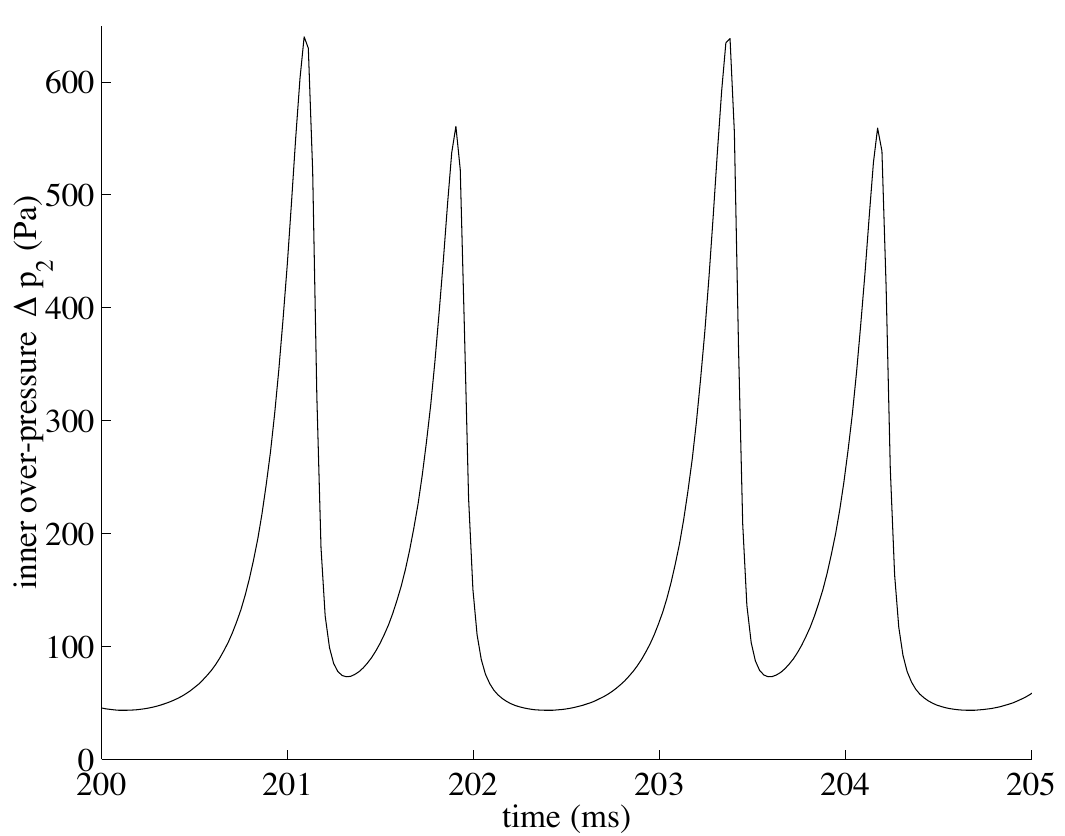}
  \includegraphics[width=4.5cm]{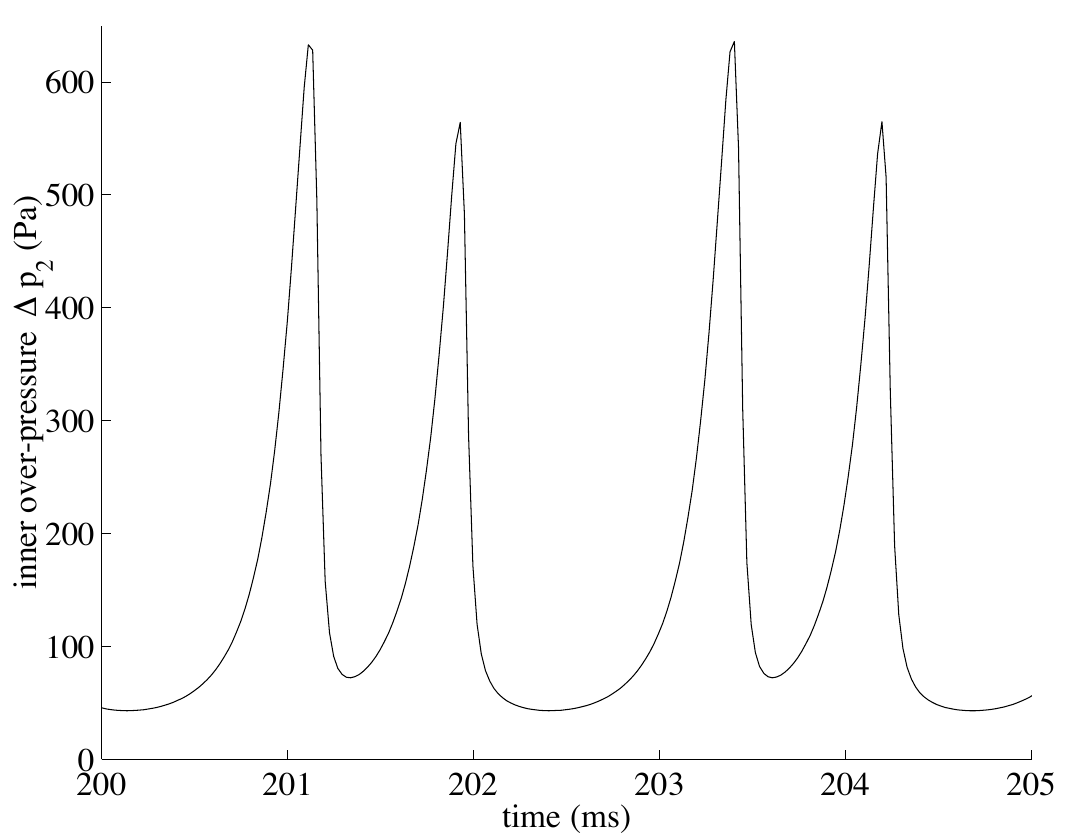}
  \includegraphics[width=4.5cm]{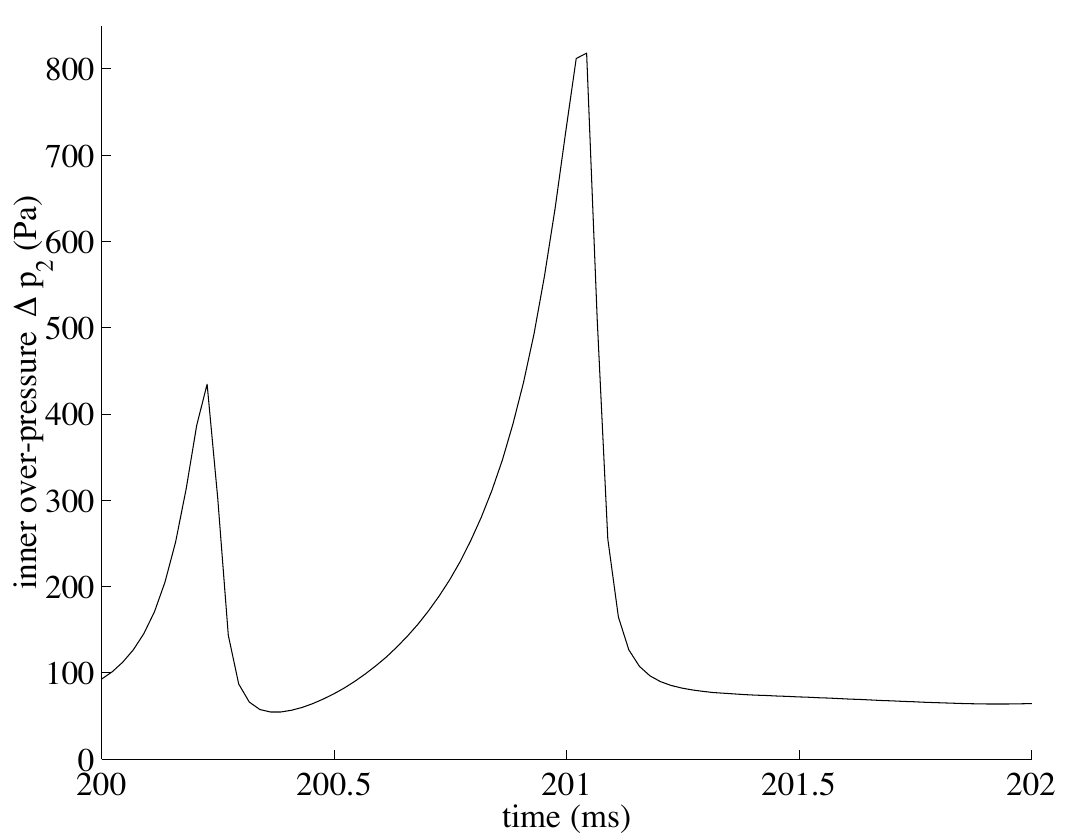}
  \includegraphics[width=4.5cm]{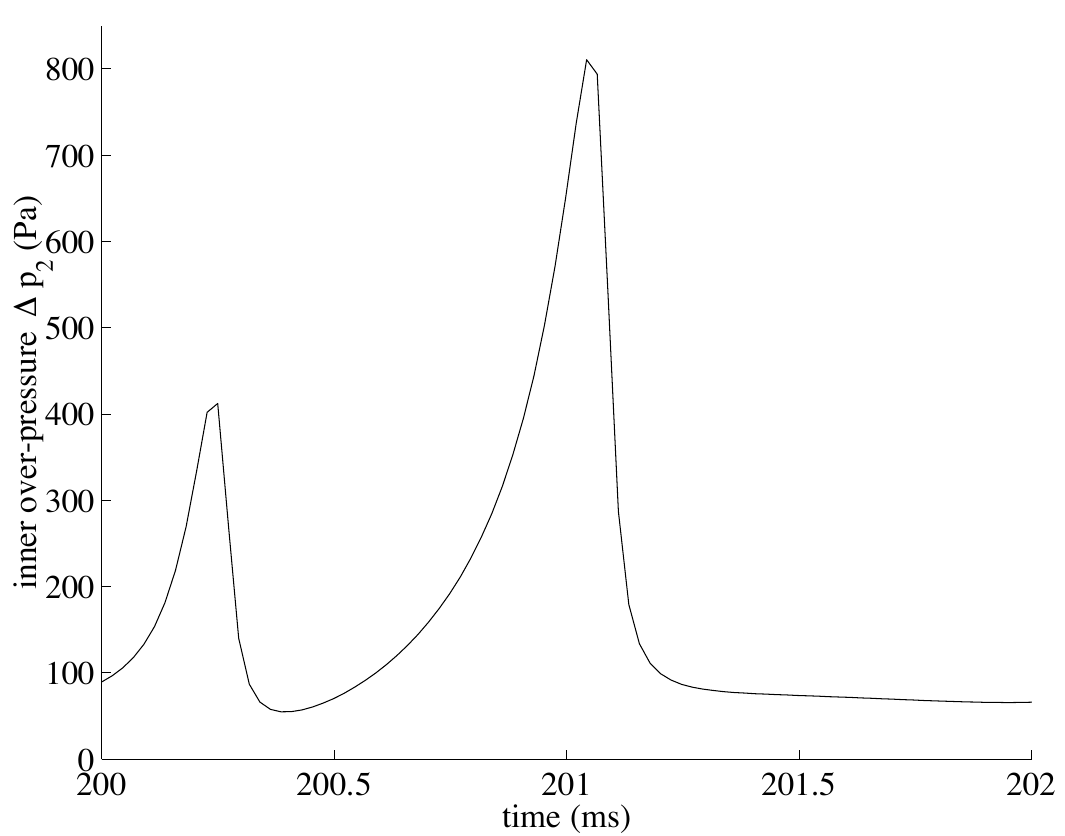}
  \includegraphics[width=4.5cm]{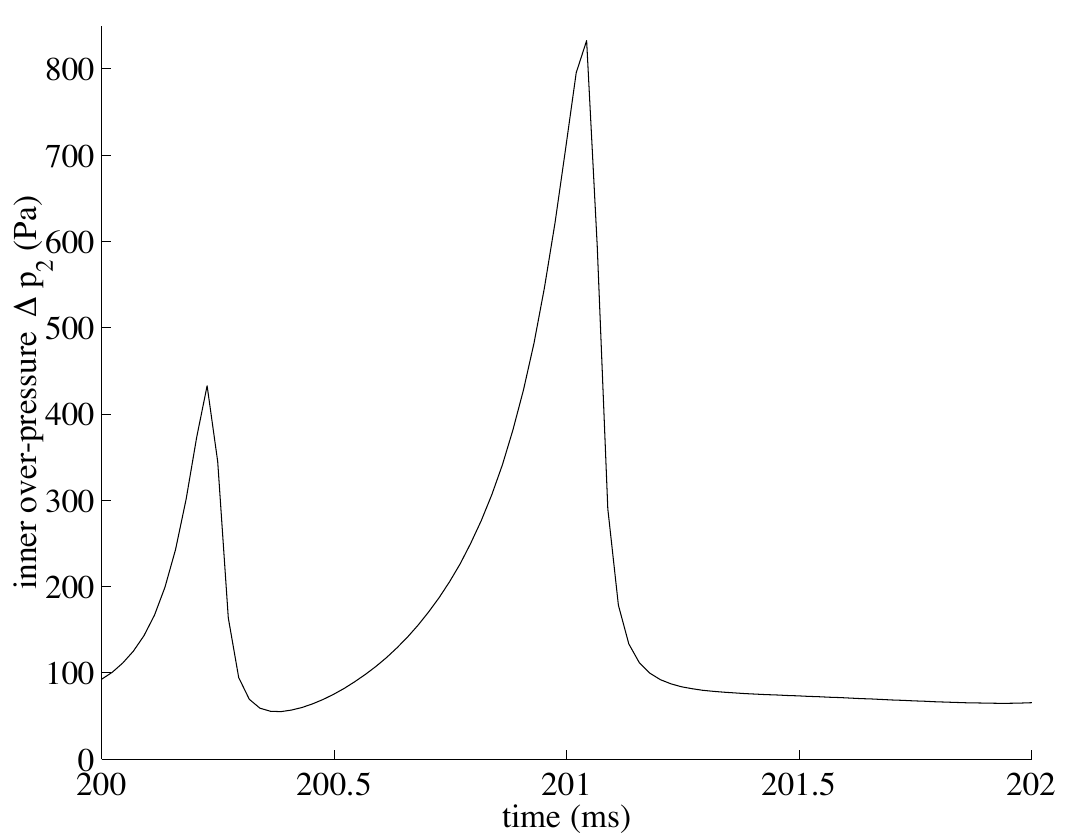}
   \caption{Zooms on the steady parts of the plots of figure 10 which illustrate the different dynamical behaviors of  a reed while using Millot (left), Hikichi (centre) or Debut (right) model for the useful sections. (top) waveforms for the (-,+) reed ; (bottom) waveforms for the (+,-) reed.}
    \label{f12}
\end{figure}

\paragraph{}Figure \ref{f11} present zooms on the attack phases for the three models and both reeds. For the (-,+) reed the difference of dynamic behavior for the Millot model is quite obvious and shows that the reed answer is stronger and faster. For the (+,-) case, one can notice that there are some small temporal shifts in the answer of the reed and some tiny differences for the magnitude of the response. These small temporal shifts can also be seen with the plots of zooms on the steady part of the signals (see figure \ref{f12}) for both reeds and the three useful section models. With these three figures, we can suggest that our new model present some significant differences at least for the transient part of the reed response in pressure. 

\begin{figure}[h!]
    \centering
   \includegraphics[width=8cm]{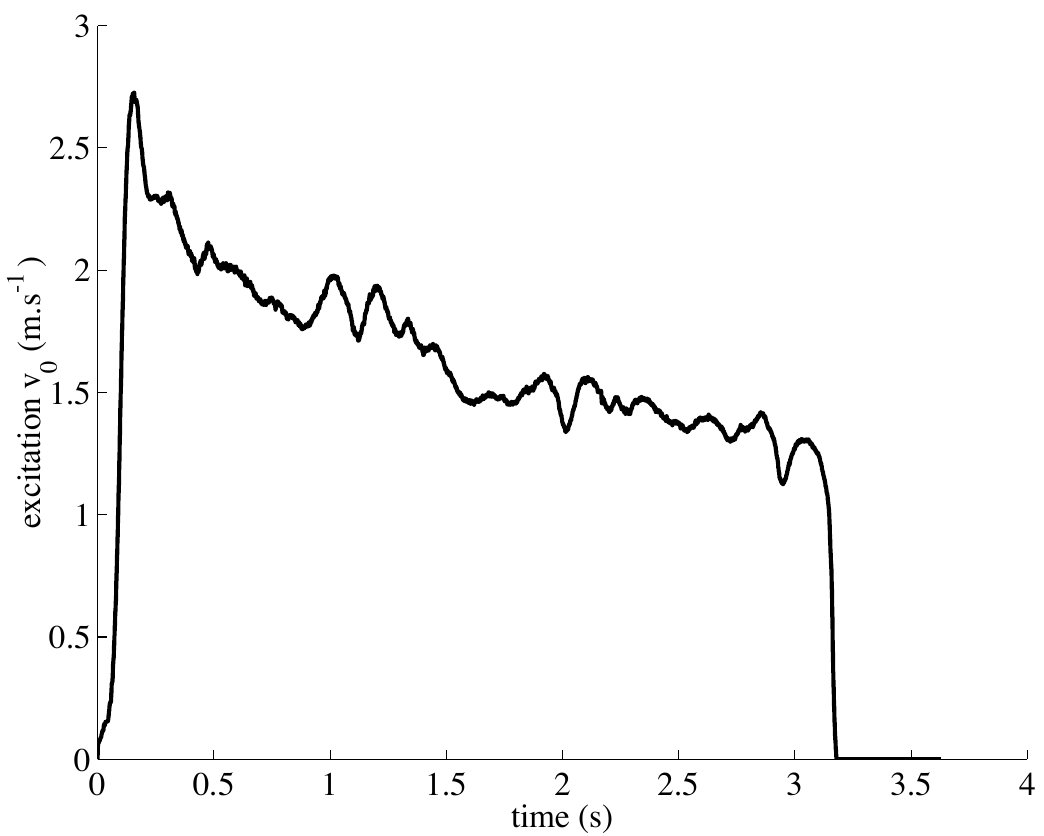} 
   \caption{Plot of the dynamic excitation used for the following numerical simulations.}
    \label{f13}
\end{figure}

\paragraph{}To demonstrate this difference for the dynamic behavior, we also present the numerical simulations performed using a more realistic excitation, the one presented in figure \ref{f13} and derived from the very lowpass filtering of a measured over-pressure signal inside a diatonic harmonica.

\begin{figure}[h!t]
    \centering
   \includegraphics[width=5cm]{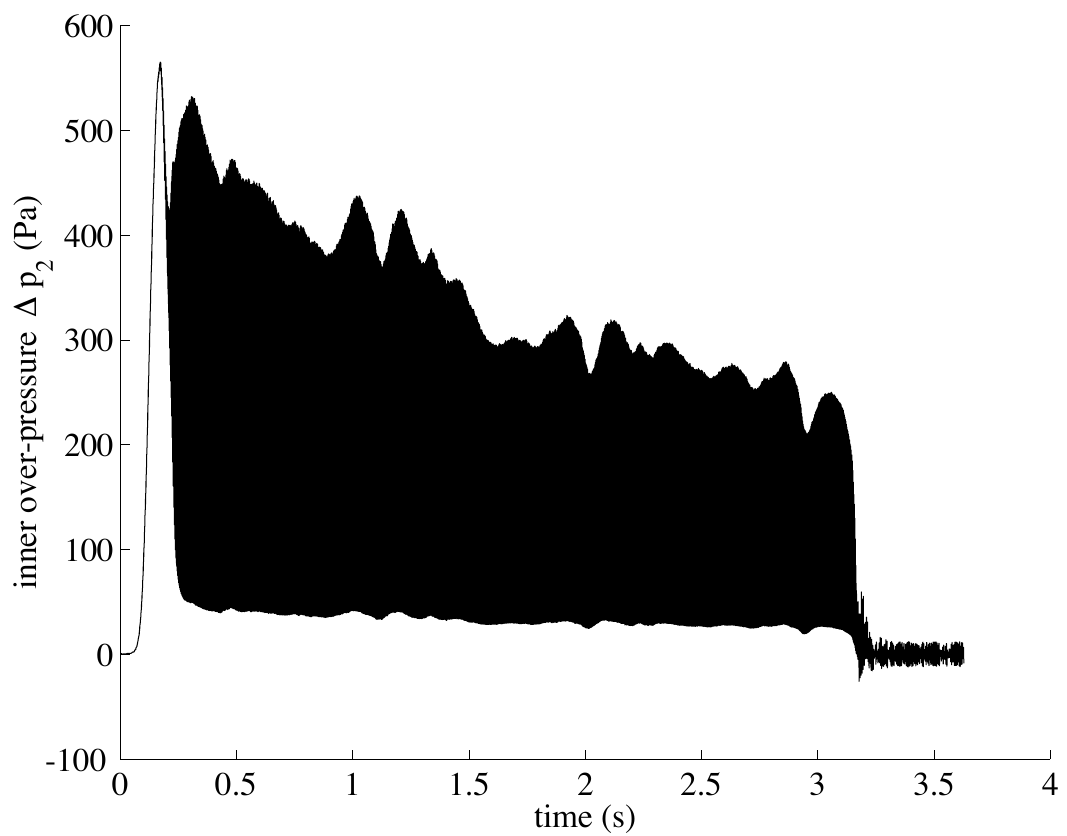} 
  \includegraphics[width=5cm]{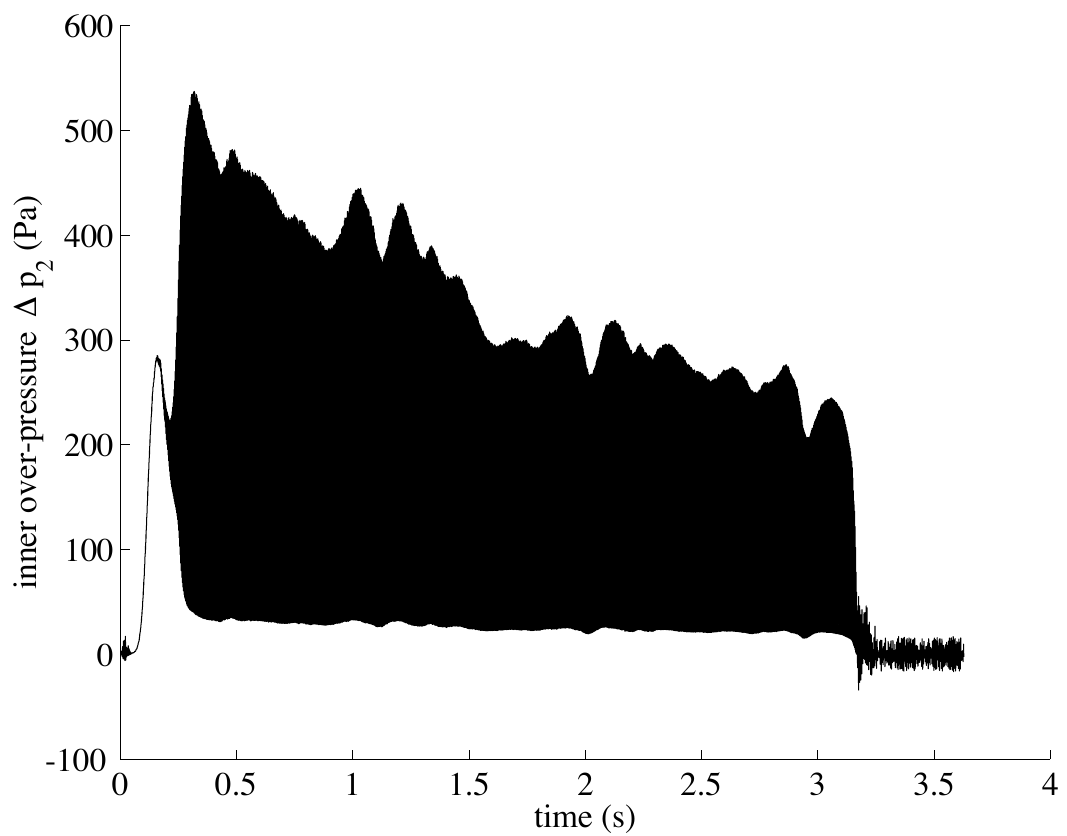}
  \includegraphics[width=5cm]{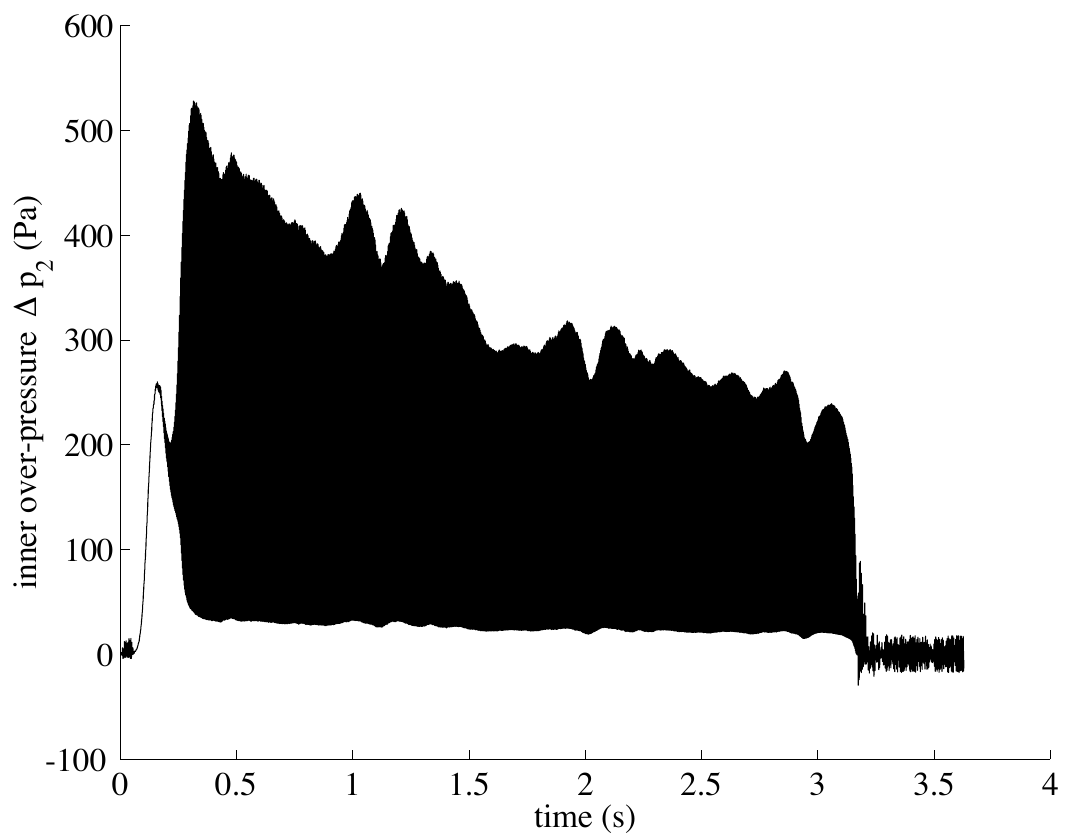}
   \caption{Another illustration, using the dynamic excitation (see figure \ref{f13})  of the different dynamical behaviors of  a (-,+) reed while using Millot (left), Hikichi (centre) or Debut (right) model for the useful sections. The parameters are the following: $L_1=8$ cm, mean $v_0=1.35$ m.s$^{-1}$. One can notice that the reed does not have the same temporal response during the attack, according to the useful section model : fastest reaction is found for Millot model but the rest of the waveform seem quite similar.}
    \label{f14}
\end{figure}

\paragraph{}On figure \ref{f14}, the whole signals for the over-pressure are only presented in the case of a (-,+) reed. Indeed, the differences are quite tiny in the (+,-) case as noticed for the artificial "triplet" excitation so we have chosen to concentrate on the (-,+) case.  While the attacks are obviously different, the rest of the waveforms appear rather similar for the three models. It is also interesting to notice that the over-pressure envelope seems to follow the excitation curve in three cases after the transient phases. With the whole waveform plots, we can only note that the Millot model gives a more reactive reed while the Hikichi model gives a later departure. 

\begin{figure}[h!t]
    \centering
   \includegraphics[width=5cm]{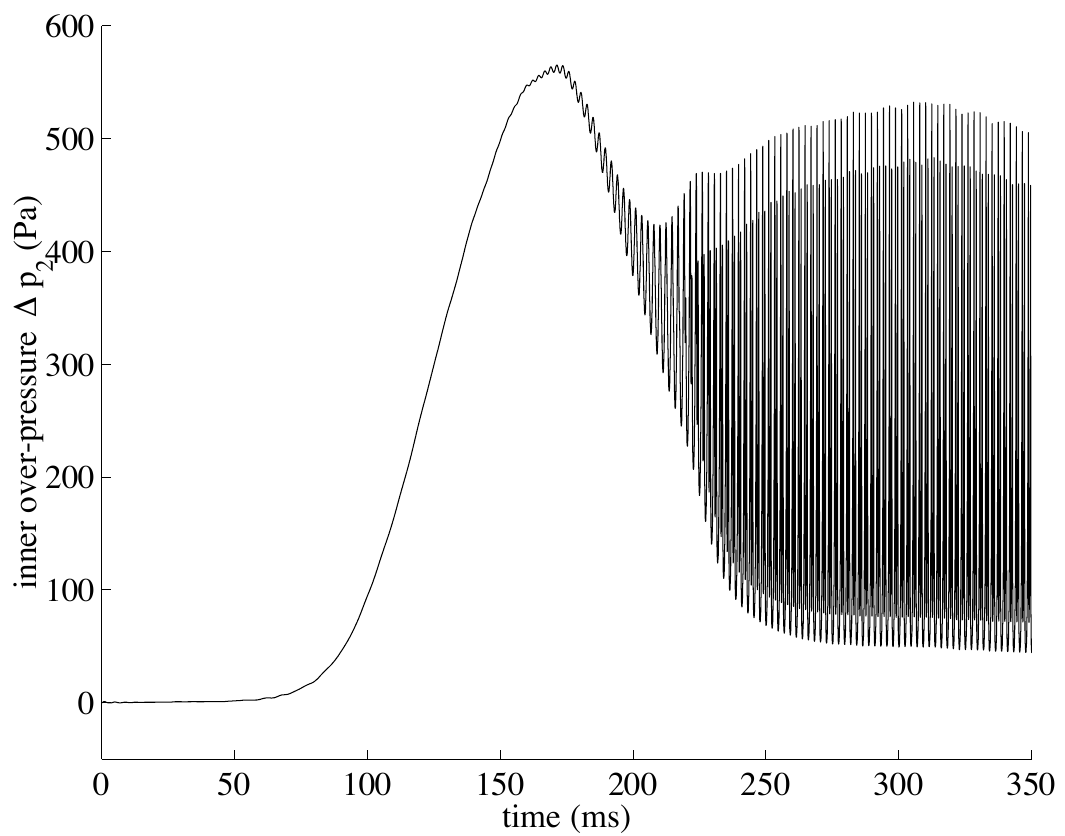} 
  \includegraphics[width=5cm]{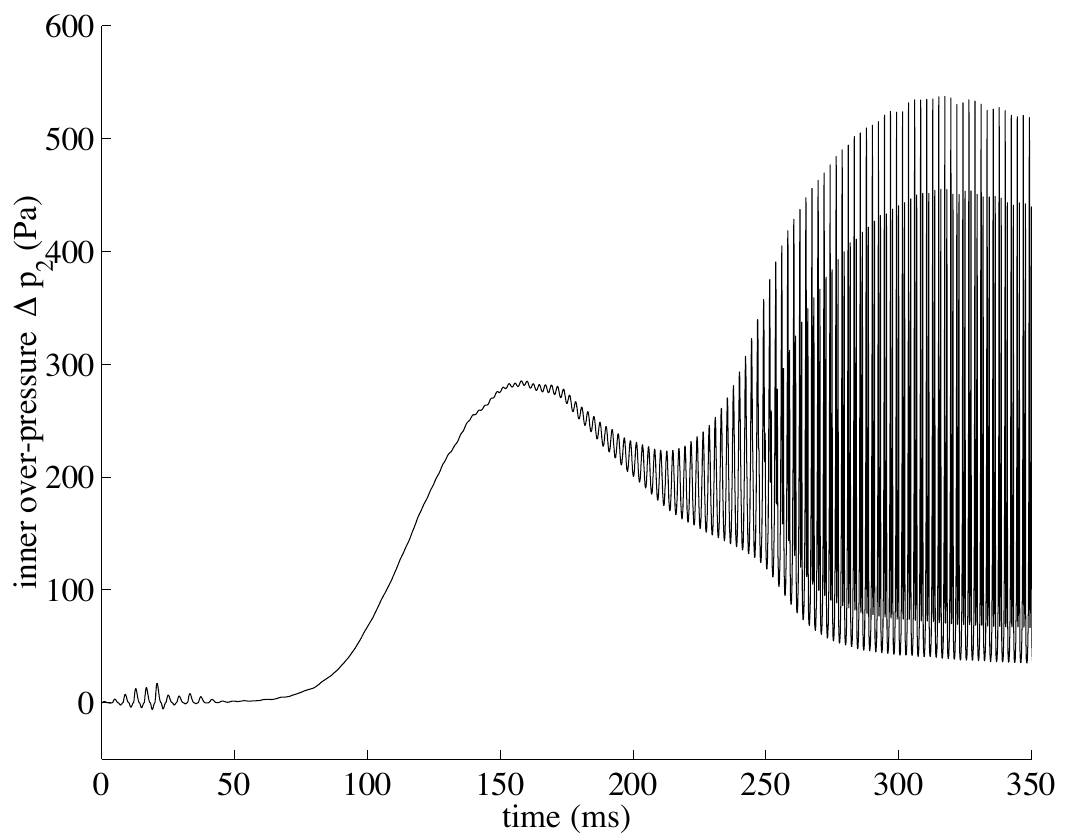}
  \includegraphics[width=5cm]{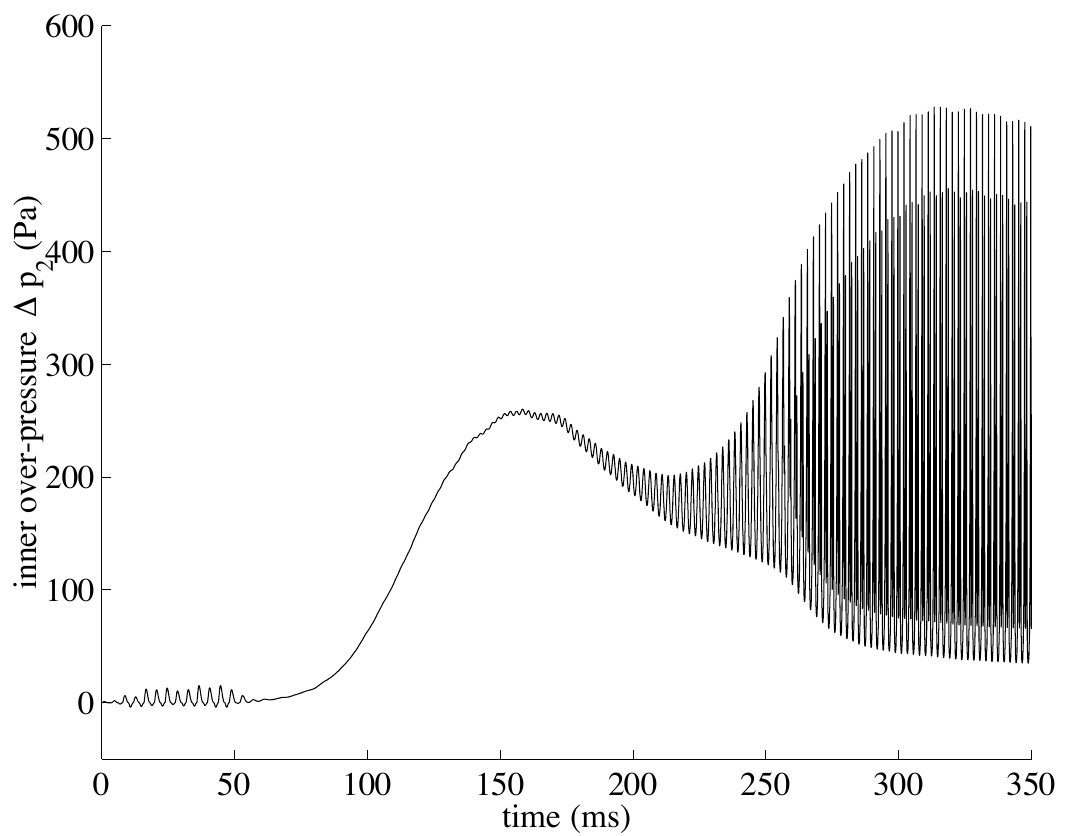}
   \caption{Zooms on the beginnings of the plots of figure \ref{f13} which illustrate the different dynamical behaviors of  a reed while using Millot (left), Hikichi (centre) or Debut (right) model for the useful sections. One can notice that the Millot model gives a fast and more reactive reed with a rather different behavior, Hikichi model gives a reed which reacts the most slowly.}
    \label{f15}
\end{figure}

\paragraph{}With figure \ref{f15}, the attention is focused on the transients which really point out the great differences of the dynamical behavior of the reed according to the useful section model. The Millot model reacts earlier and more strongly than the two other which may correspond to a more reactive reed configuration. The Debut model follows, and the third reed to react is the one using the Hikichi model. These differences for the reed transients constitute a strong argument to search a useful section as precisely as possible even in the case of a simplified model. So, even if our model presents a more complex formulation we do think it must be carefully considered. Moreover, to perform numerical simulations, it is always possible to calculate tables for the useful sections before beginning the temporal simulations which permits access to shorter times for the calculation.

\subsection{Discussion of the assumptions}
\paragraph{}In this section, we review the assumptions we have made to derive the proposed model and discuss their validity.

\paragraph{}For all the results presented in this section, the parameters for the (-,+) reed were $L_1=8$~cm and $v_0=2.5$ m.s$^{-1}$ while for the (+,-) reed we used $L_1=1.5$~cm and $v_0=3$ m.s$^{-1}$. 

\begin{figure}[h!]
    \centering
    \includegraphics[width=6cm]{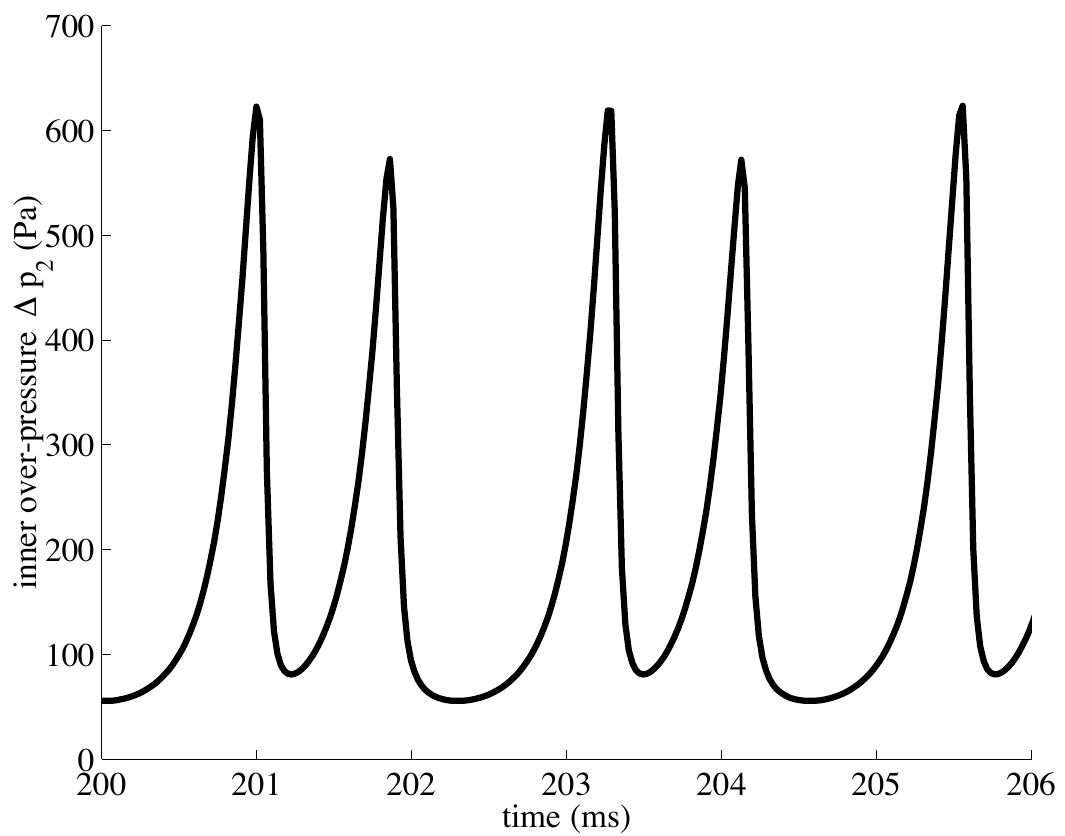}
    \includegraphics[width=6cm]{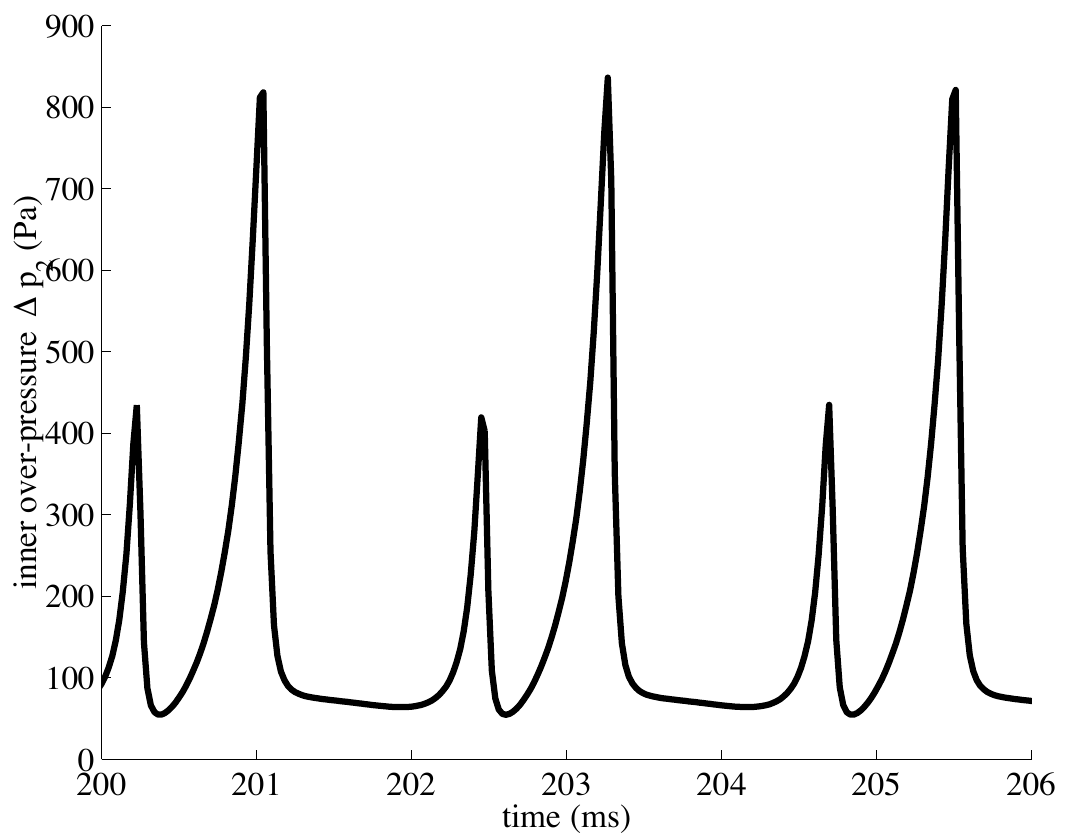}
    \includegraphics[width=6cm]{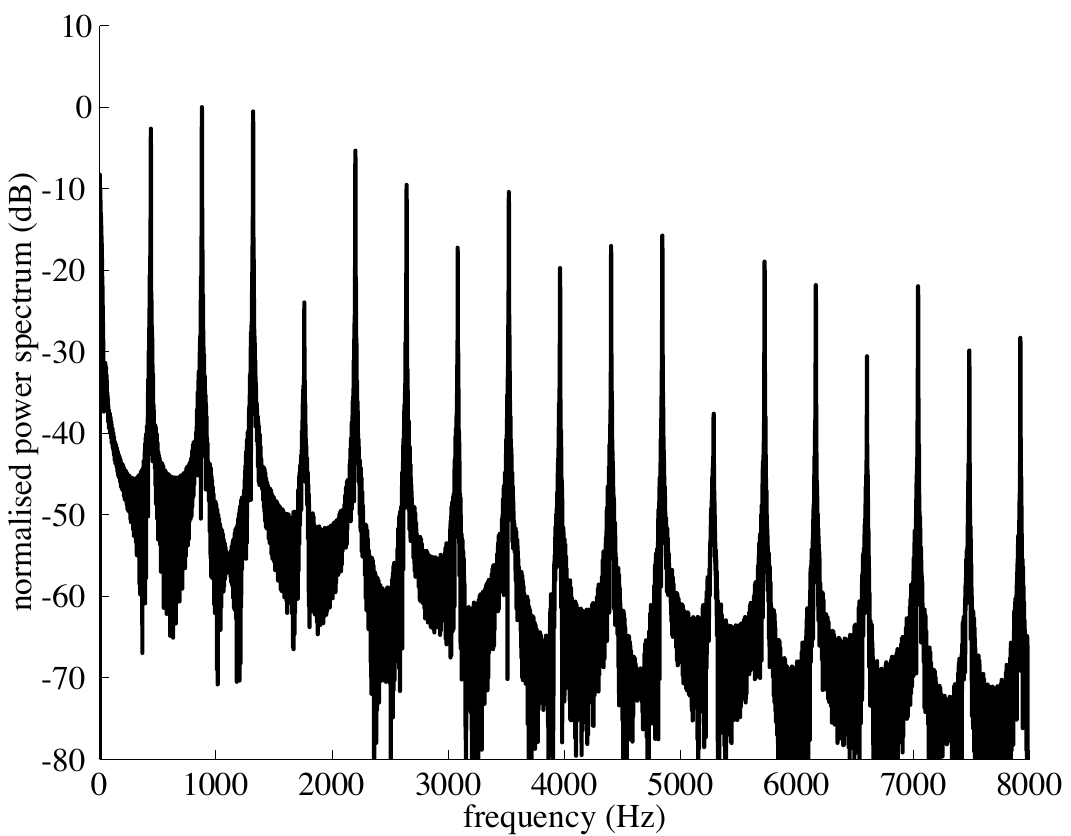}
    \includegraphics[width=6cm]{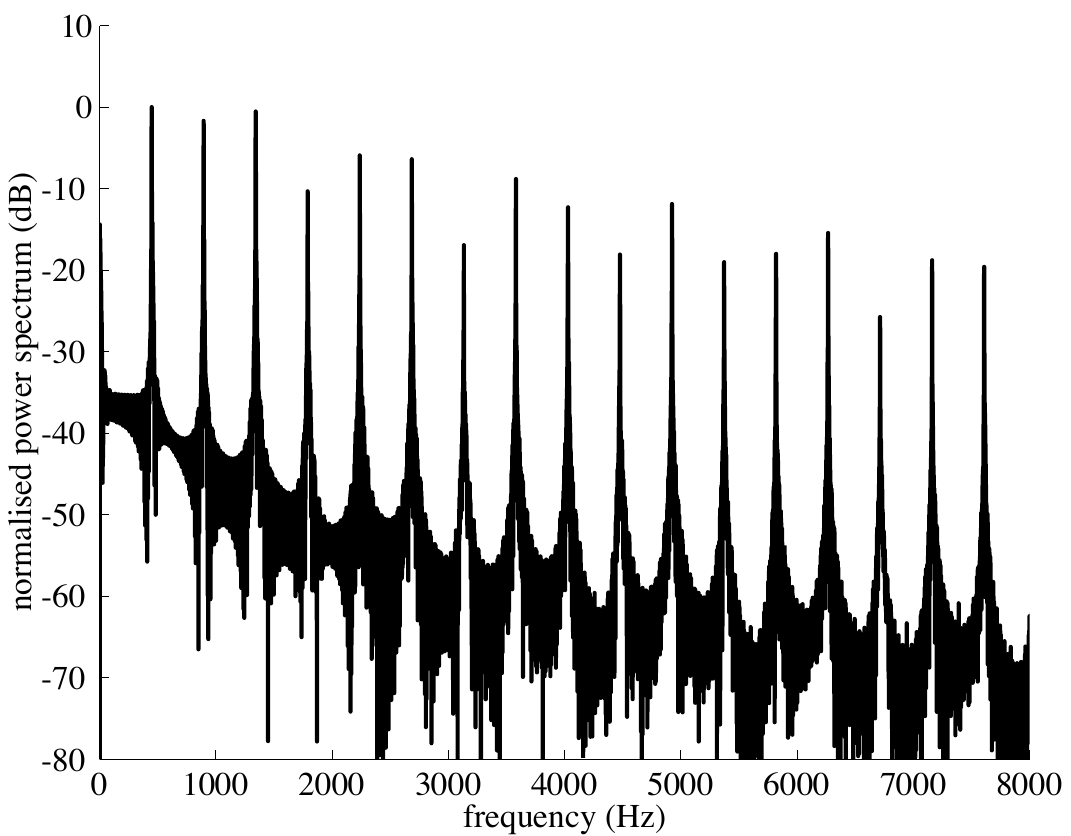}
    \caption{(top) Waveforms for the over-pressure $\Delta p_2$ for the (-,+) reed (left) and the (+,-) reed (right). (bottom) Related normalized spectra.}
    \label{f16}
\end{figure}

\paragraph{}Figures \ref{f16}  respectively illustrate the waveform and the power spectrum for the upstream/downstream over-pressure $\Delta p_2=p_2 -p_{atm}$ in the case of a (-,+) or a (+,-) reed. We can not compare these results with experimental results because  we adopt the reed properties for the (+,-) reed  of  the fourth channel of a diatonic harmonica while both (+,-) and (-,+) reeds have different properties on a diatonic harmonica. Moreover, we do not have the experimental setup associated with the loading configuration we propose. Yet the waveforms and the spectrums found for both (-,+) and (+,-) are realistic compared to real blows and overblows produced on a diatonic harmonica. And, when one uses a time-varying excitation, the sound produced is quite realistic for the listener. But an experimental validation should be done for the one reed case; an experimental validation for a two reeds case will be presented elsewhere, in the case of chromatic playing on diatonic harmonica.

\begin{figure}[h!]
    \centering
    \includegraphics[width=6cm]{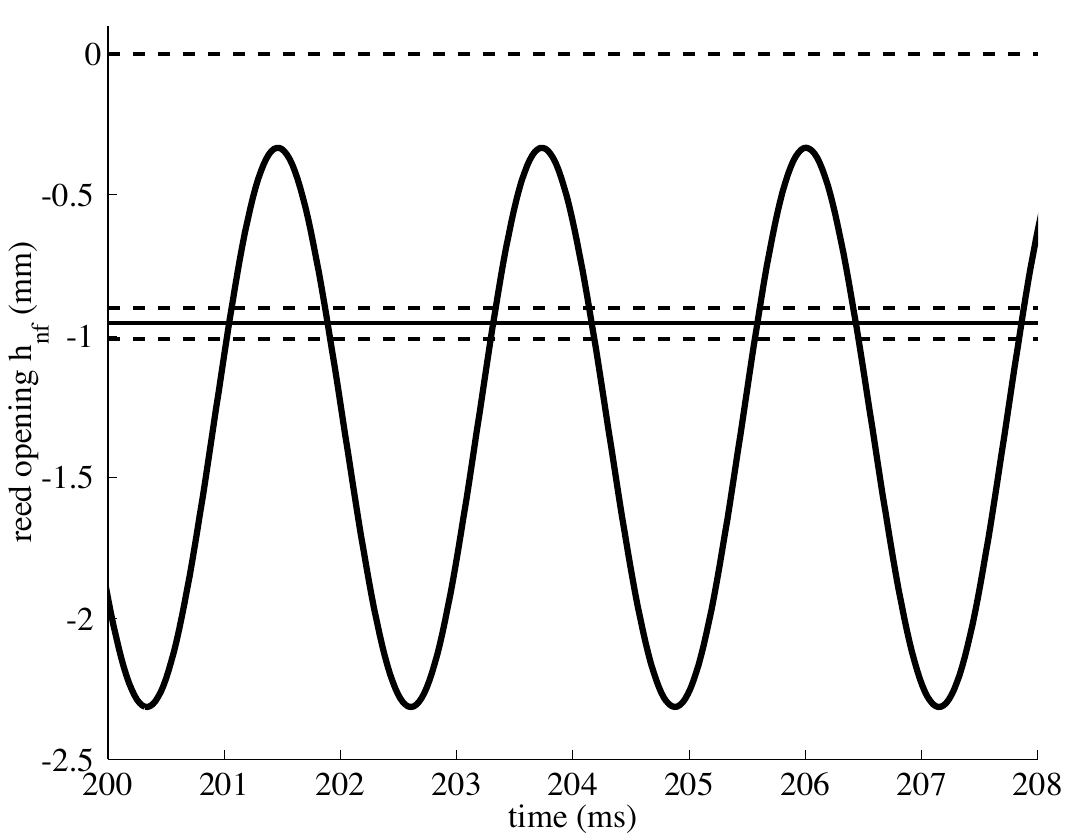}
    \includegraphics[width=6cm]{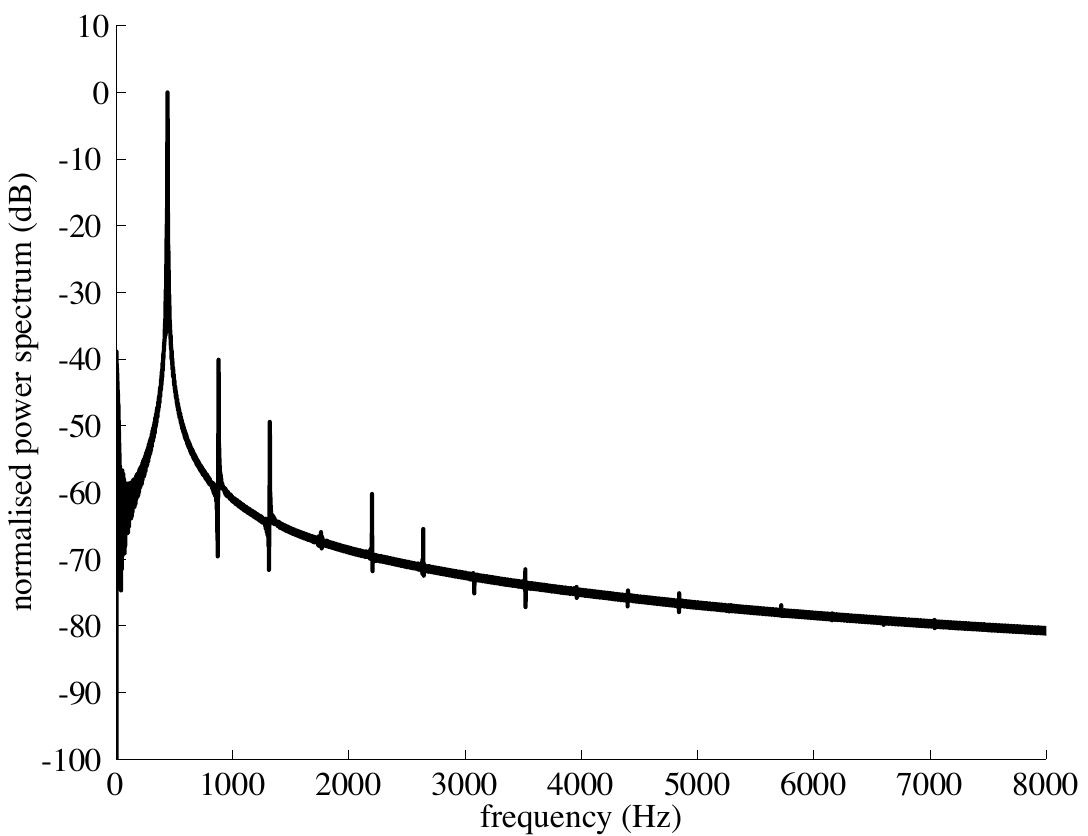}
    \includegraphics[width=6cm]{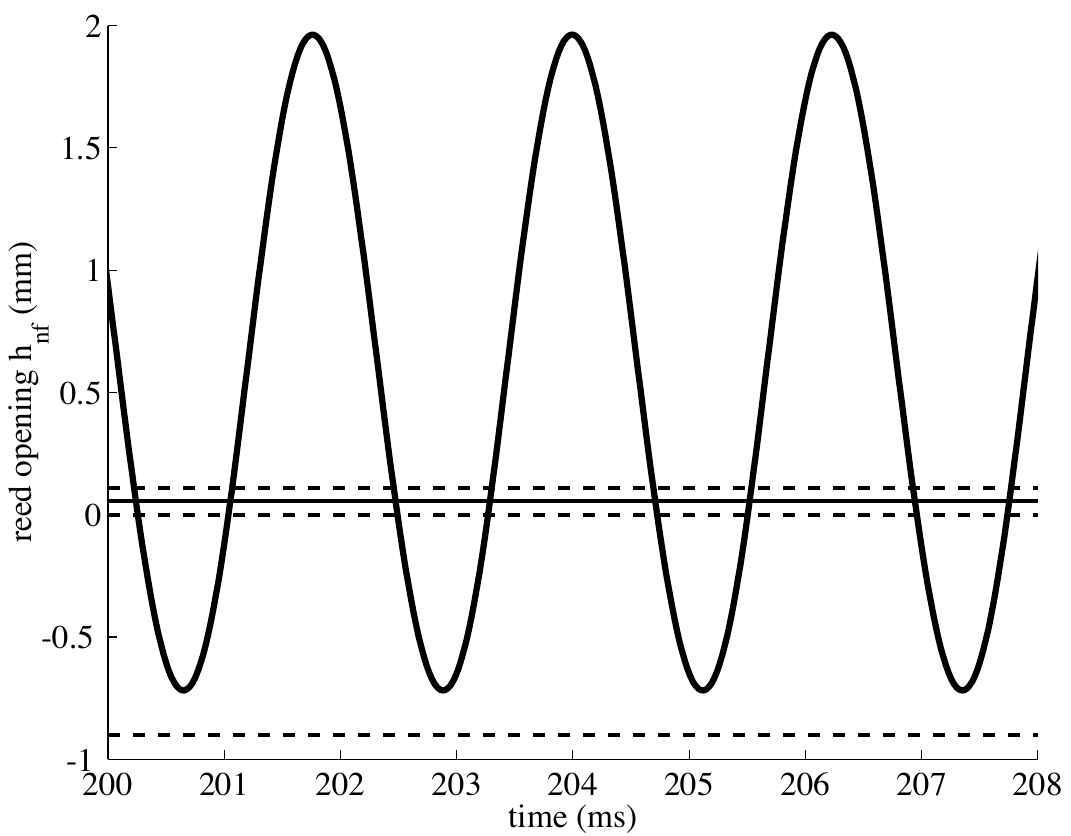}
    \includegraphics[width=6cm]{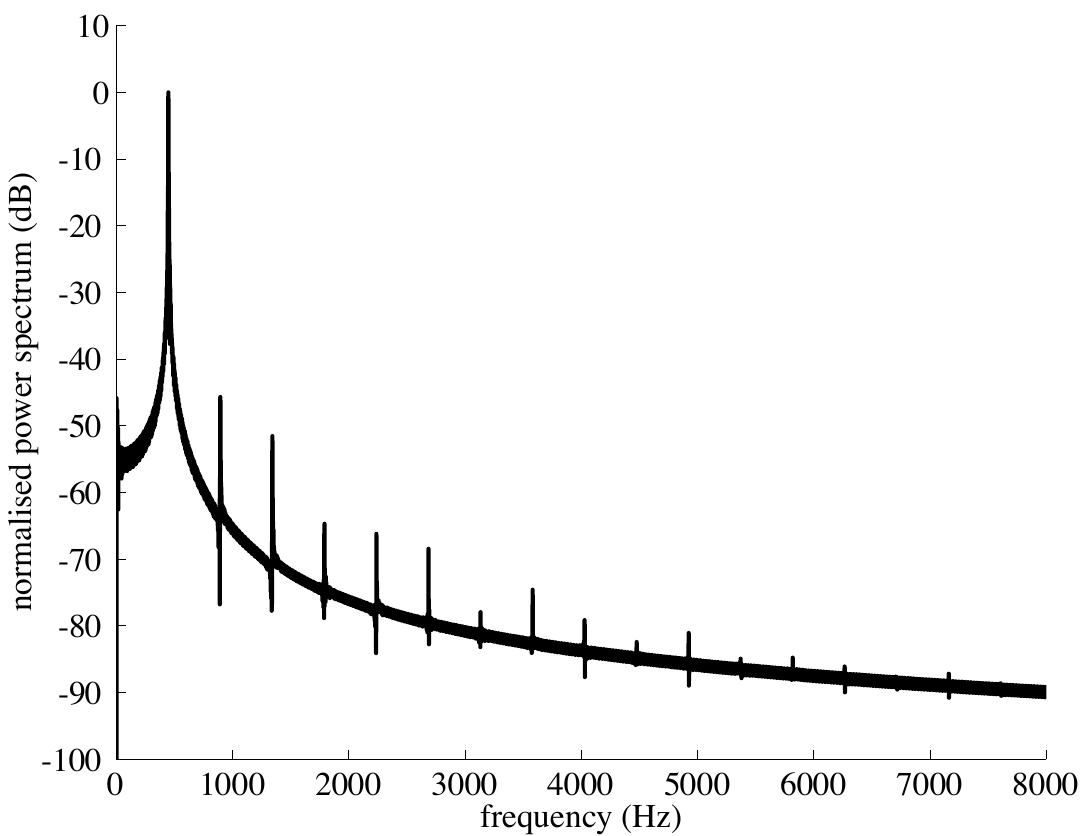}
    \caption{Waveforms for the neutral section opening and related normalized power spectrum in a case of a (-,+) reed (top) and of a (+,-) reed (bottom). For the waveforms, the solid plot indicates the flat position for the reed (no deformation case) while the dashed plots correspond respectively to the zero reference line, and the $\pm\frac{e_r}{2}$ departure from the flat position.}
    \label{f17}
\end{figure}

\paragraph{}With the Figure \ref{f17}, the sinusoidal nature of the neutral section displacement is confirmed as proposed in \cite{Millot:01} for harmonica reeds with a playing frequency under the reed eigenreed frequency $f_r$ in the case of the (-,+) reed and over $f_r$ in the case of the (+,-) reed as it is shown in the following. In both cases, the reed displacement is important and the reed goes inside the thickness of the support during a noticeable part of the period. The thickness of the reed is illustrated by both dashed lines around the solid line, associated with the flat position for the reed. This thickness does not seem so tiny compared to the whole magnitude of the reed displacement or to the support thickness so we think we should not neglect it in the reed model. 

\paragraph{}The space between the zero line and the first other dashed line corresponds to the inside of the support so both reeds vibrate inside the thickness of the support and, for very long reeds (low frequency ones on a diatonic harmonica) the reeds can even go through the whole thickness of the support.

\paragraph{}One can note that the signal is calculated with a little length $L_1$ (1.5 cm) in the case of the (+,-) reed while we have a  long length $L_1$ (8 cm) in the case of the (-,+) reed. This is in accordance with the instability conditions we have found in the precedent subsection: little $L_1$ makes easier the verification of the instability condition in the case of a (+,-) reed while it makes harder the verification of the instability condition in the case of a (-,+) reed. This discussion is more developed in next subsection with the study of the influence of the variation of $L_1$ for both reeds.

\begin{figure}[h!t]
    \centering
    \includegraphics[width=7cm]{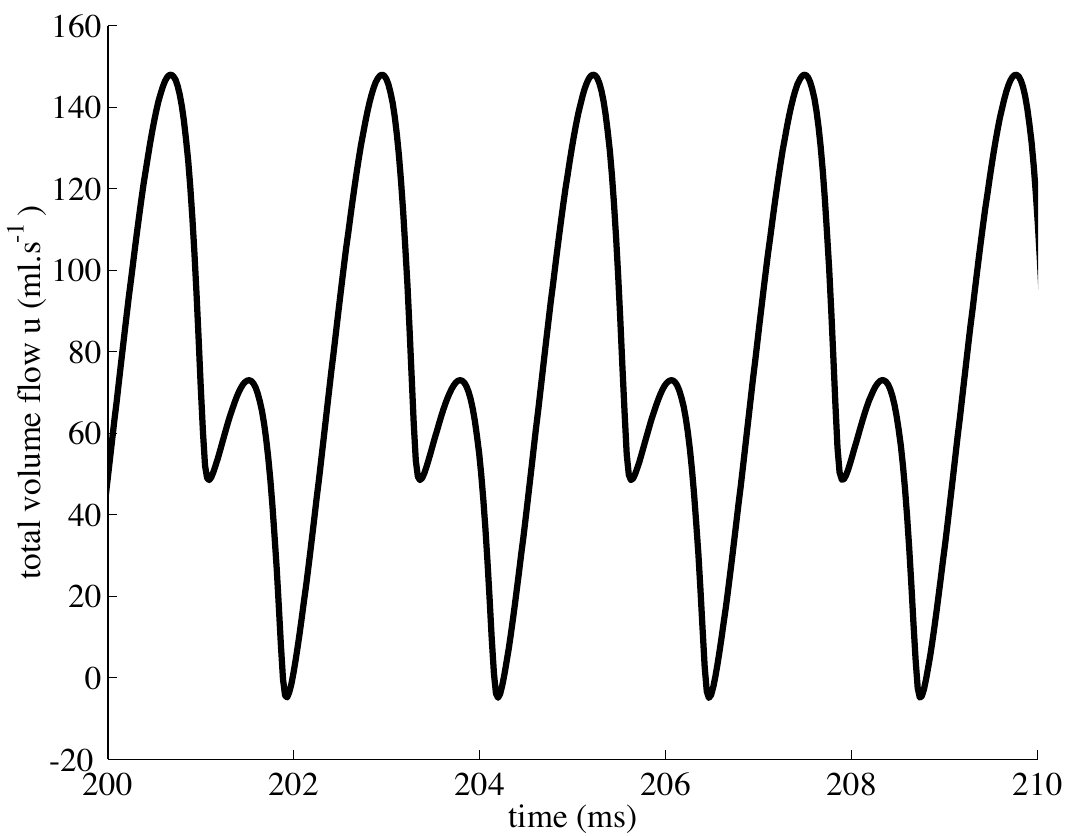}
    \includegraphics[width=7cm]{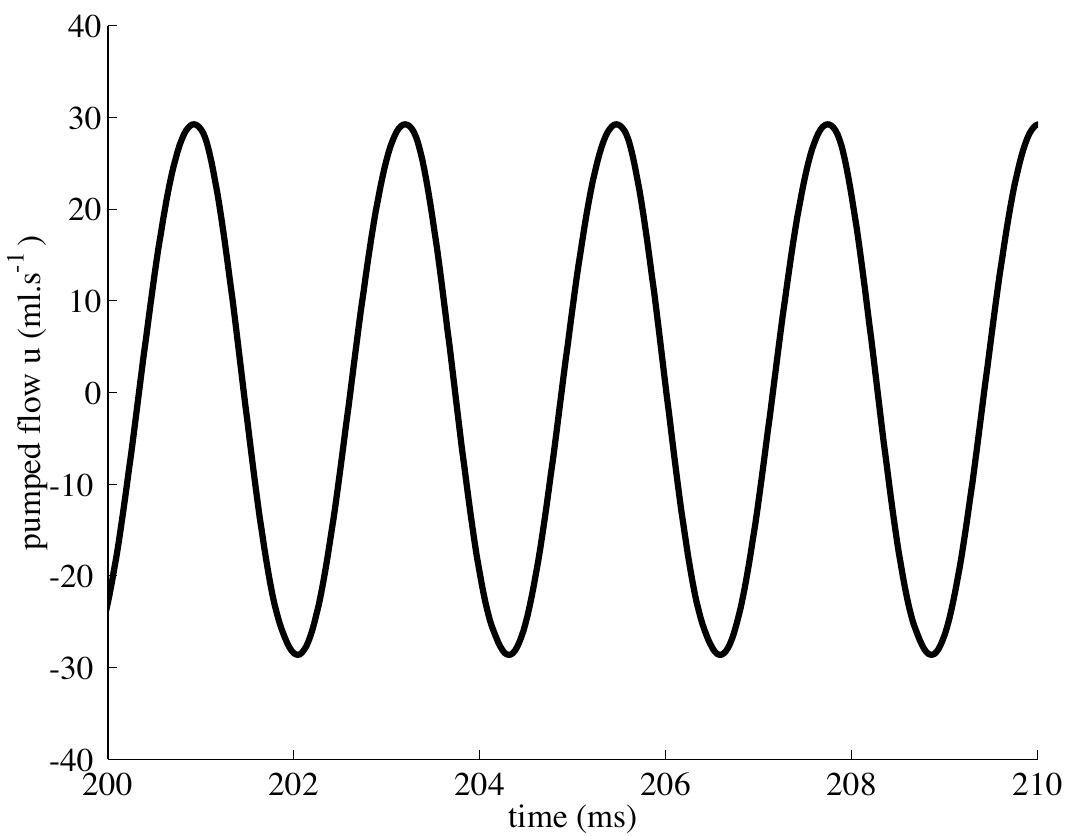}
     \includegraphics[width=7cm]{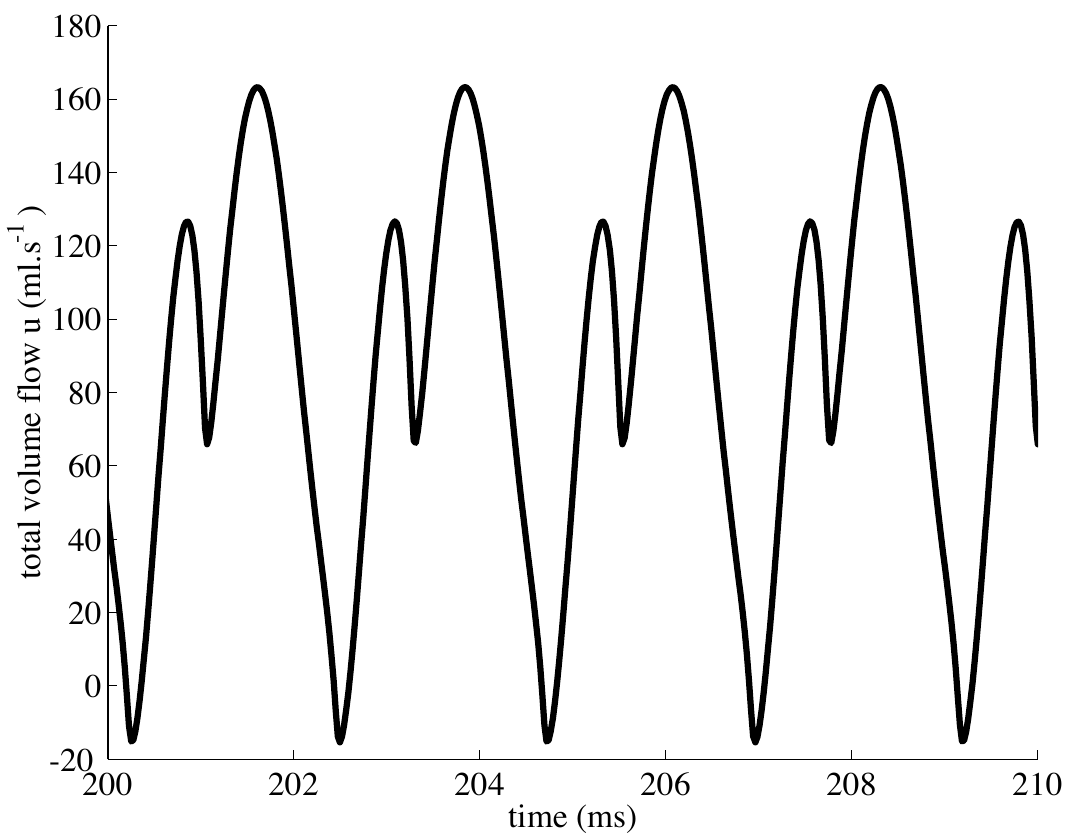}
    \includegraphics[width=7cm]{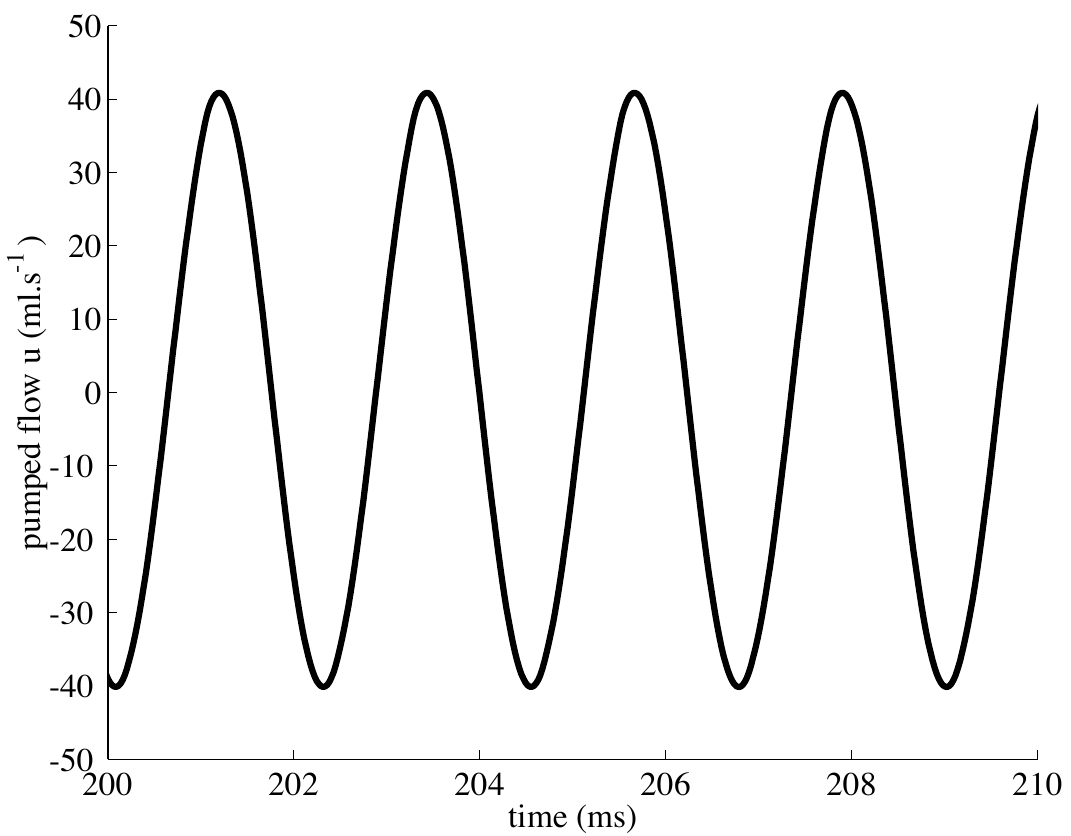}
   \caption{Waveforms for the total volume flow $u$ (left) and for the pumped flow $u_p$  (right) in a case of a (-,+) reed  (top) and of a (+,-) reed (bottom). One can note that the pumped flow plays a significant role in the temporal evolutions of the total volume flow.}
    \label{f18}
\end{figure}

\paragraph{}In the model, we propose to take account of the flow displaced by the reed, the pumped flow. This assumption is justified because the pumped flow represents a significant contribution to the output volume flow as illustrated in Figure \ref{f18} so this phenomenon must be taken into account in a minimal  free reed model. It may be interesting to quantify the role played by the pumped flow in the radiated sound, because we have two contributions per reed: the pumped flow and the flow passing through the reed, whose natures are quite different. We suspect that the pumped flow contributes mainly to the low frequency part of the sound (because of its sinusoidal character) while the flow passing through the reed may mostly feed the high frequency part of the sound because it is related to the high non linearity of the flow through the reed (non linearity associated with the useful section and the jet velocity).  

\begin{figure}[h!t]
    \centering
    \includegraphics[width=5cm]{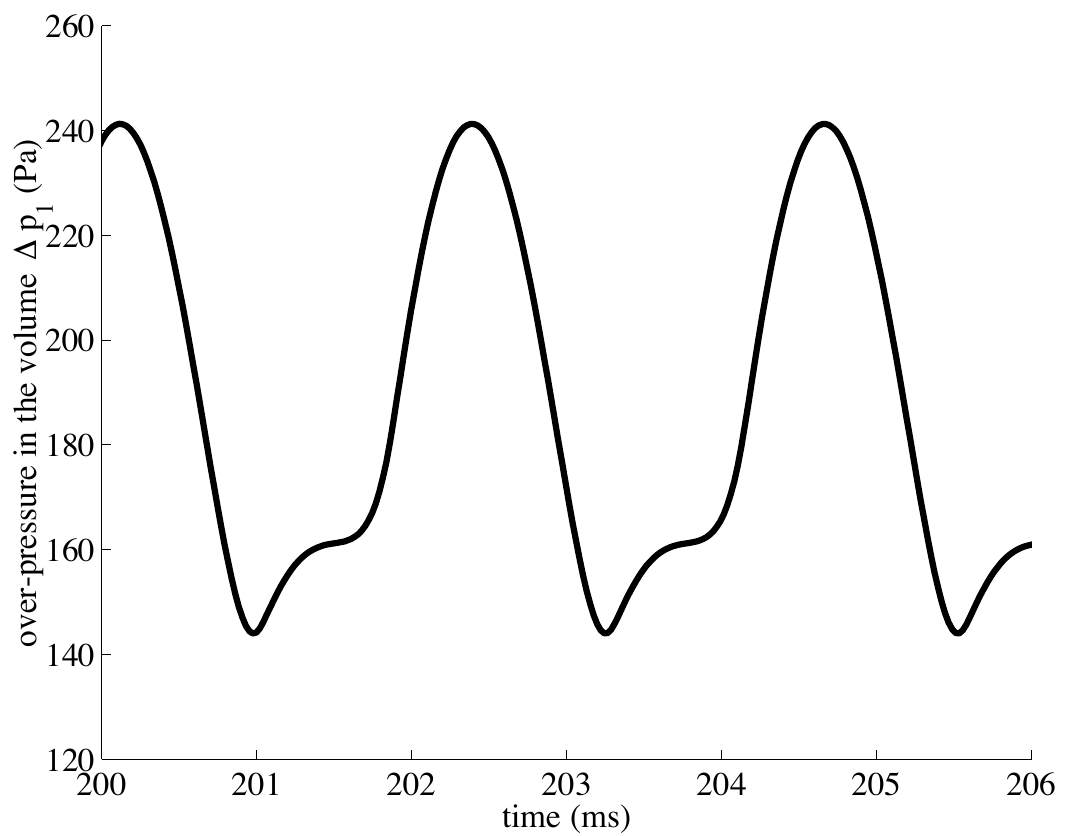}
    \includegraphics[width=5cm]{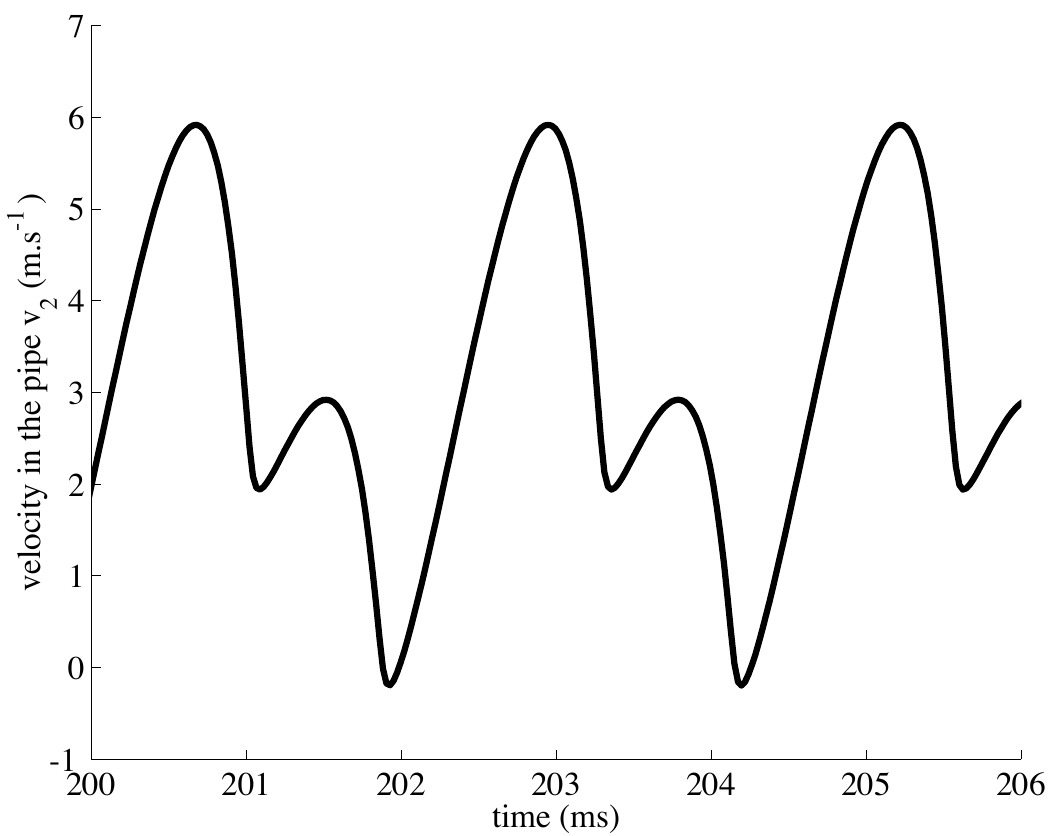}
    \includegraphics[width=5cm]{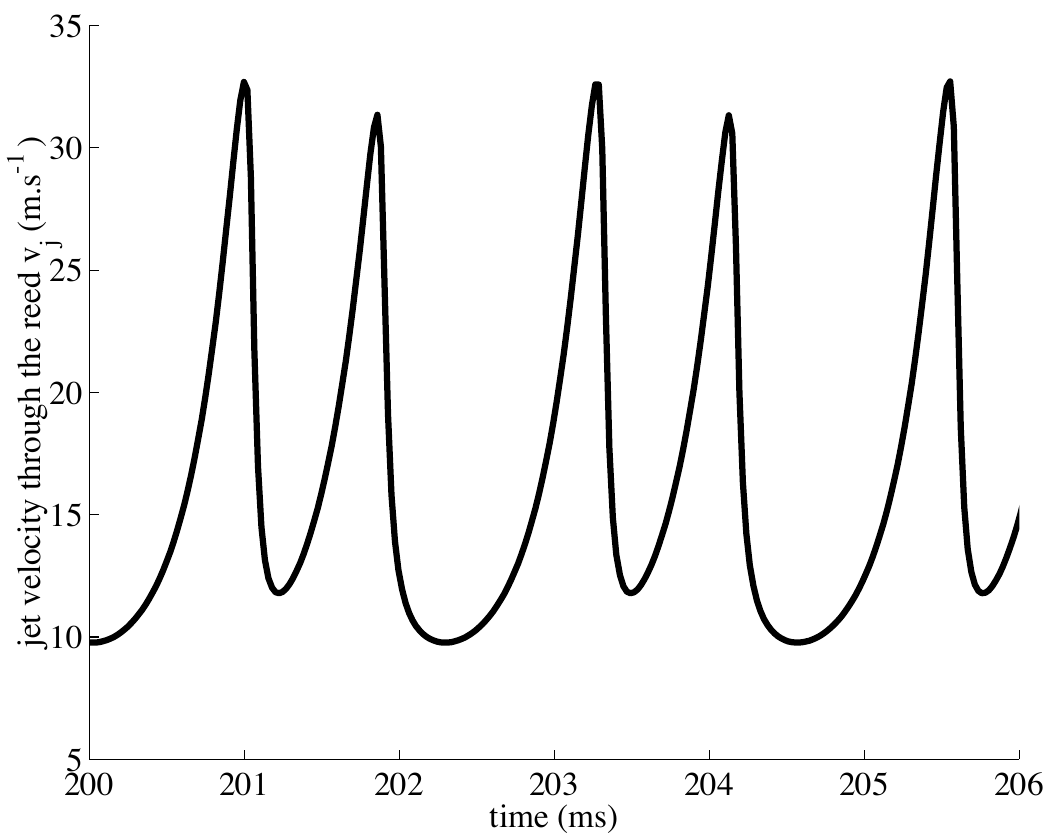}
    \includegraphics[width=5cm]{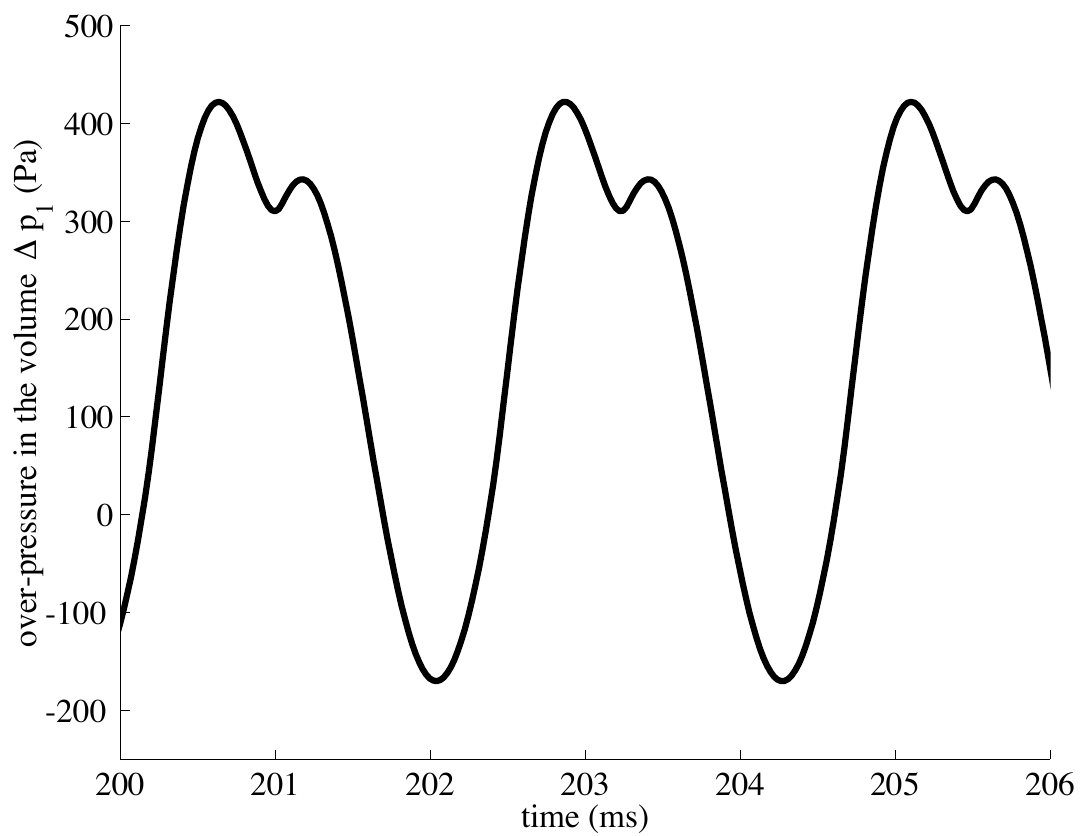}
    \includegraphics[width=5cm]{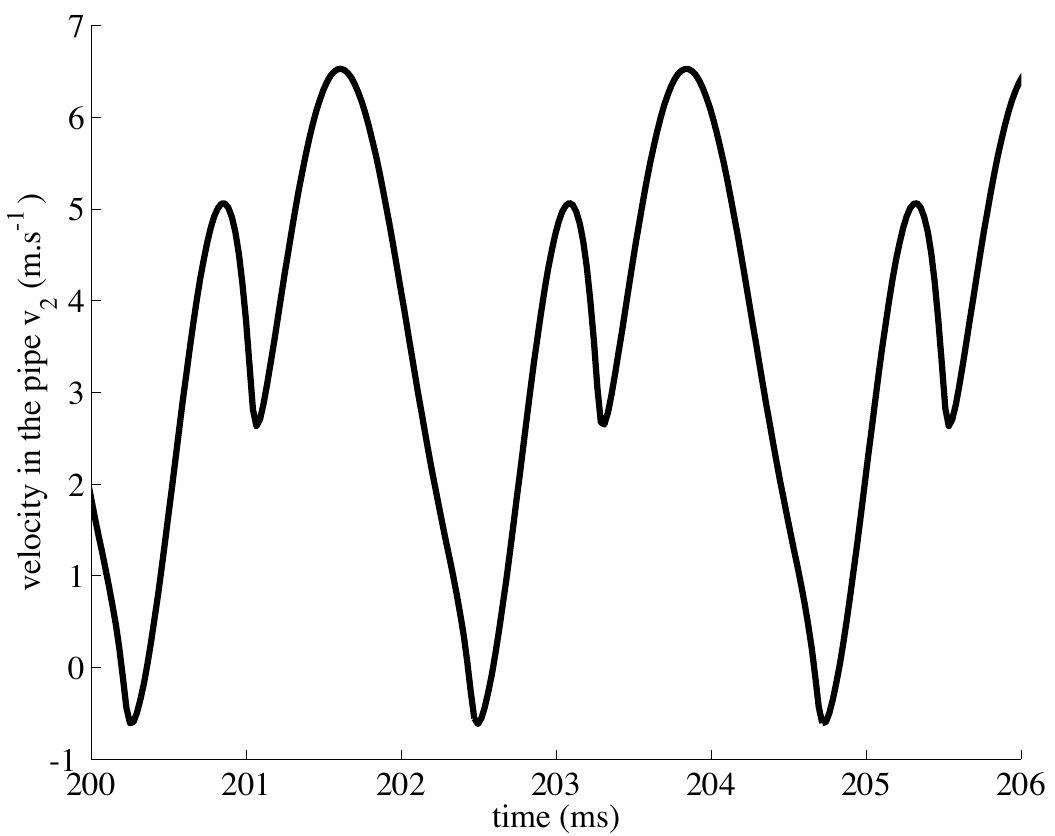}
    \includegraphics[width=5cm]{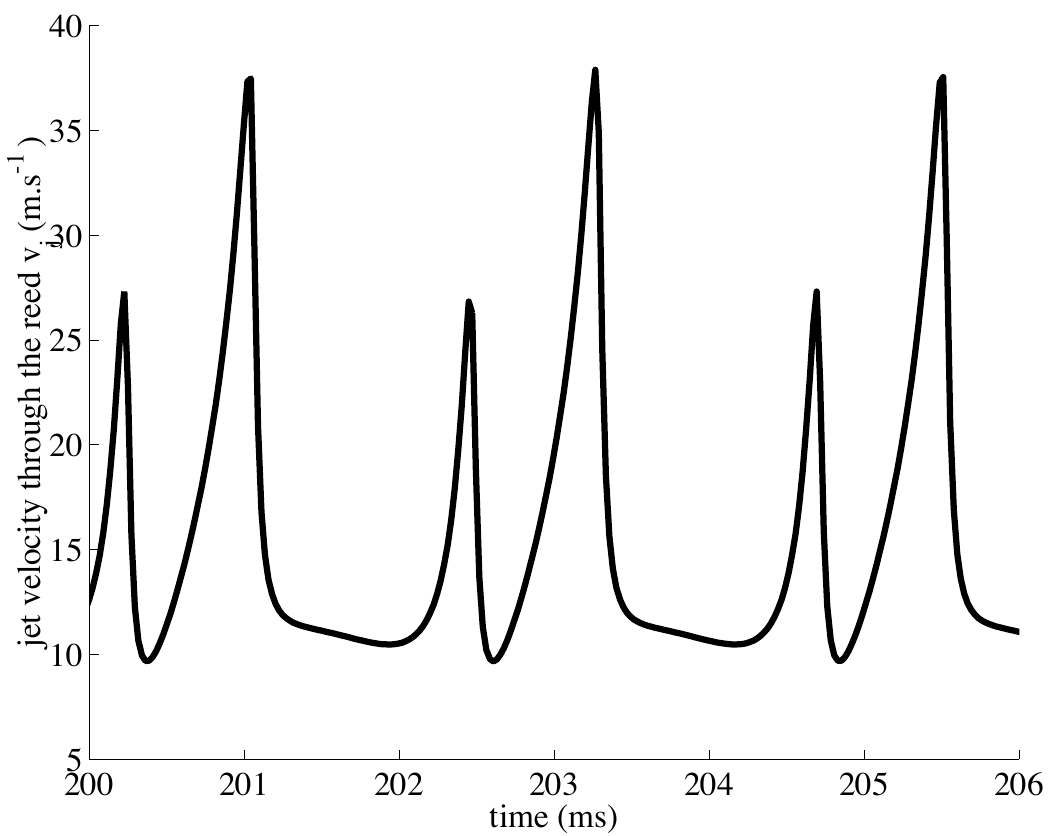}
    \caption{Waveforms for the neutral over-pressure $\Delta p_1$ (left) in the volume $V_1$, for the velocity $v_2$ in the pipe 2 (middle) and for the jet velocity $v_j$ (right)  in a case of a (-,+) reed (top) and of a (+,-) reed (bottom). One can note that the variations of the over-pressure are significant and that the velocities are important, above all in the free jet at the downstream of the reed which give high Reynolds numbers.}
    \label{f19}
\end{figure}

\paragraph{}With the Figure \ref{f19} we illustrate the behavior of the over-pressure $\Delta p_1$ for the volume 1, the behavior of the airflow velocity in the pipe and of the velocity in the free jet forming downstream of the reed. For the derivation of the Bernoulli equations for the pipe and through the reed we used the incompressible assumption. This assumption seems realistic considering the order of magnitude of the over-pressure $\Delta p_2$ (respectively 570 Pa and 783 Pa) and the fact that the Mach numbers are lower than 0.1: the compressibility of the air may be lower than 1 \%.  It seems also a fair first approach to consider the low Mach number approximation of the adiabatic relation to derive the mass conservation for volume $V_1$. We neglect the kinetic terms for the Bernoulli equation  for the pipe 2 and this seems a valid assumption considering the differences of order of magnitude between the velocity $v_2$ and the over-pressure $\Delta p_2$: the contribution of the unsteadiness is dominant in $\Delta p_2$ and neglecting the role of $\di \frac{1}{2}\rho_0 v^2_2$ seems reasonable as the Strouhal number $\di S_{t2}=\frac{L_2.f_{play} }{v_{20}}$ is greater than one (respectively 2.93 and 2.49). For the Bernoulli equation between upstream and downstream of the reed, we also do not consider the upstream kinetic energy and this is a fair assumption when considering the respective orders of magnitude for $v_2$ and $v_j$. We also consider the airflow quasi-stationary through the reed which seems realistic with a Strouhal number $\di S_{tr}=\frac{e_s.f_{play}}{v_{j0}}$ much lower than unity (respectively 0.024 and 0.027). The Reynolds numbers in the pipe  and through the reed (considering the square root of the minimal useful section as equivalent diameter) are greater than 1235 which may mean that the viscosity effect are much lower than the other phenomena in the case of the free reed, which justifies that we do not consider explicitly the viscosity in this model.

\subsection{Influence of the length $L_1$}
\begin{figure}[h!tbp]
    \centering
    \includegraphics[width=7cm]{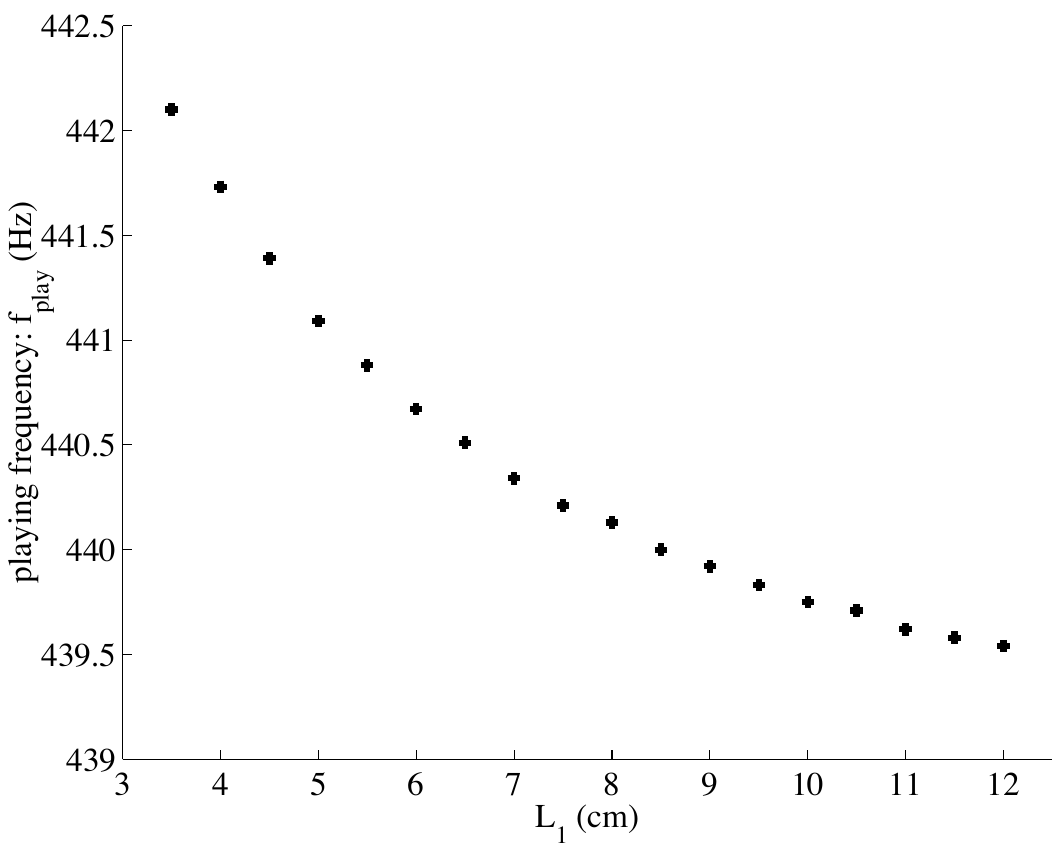}
    \includegraphics[width=7cm]{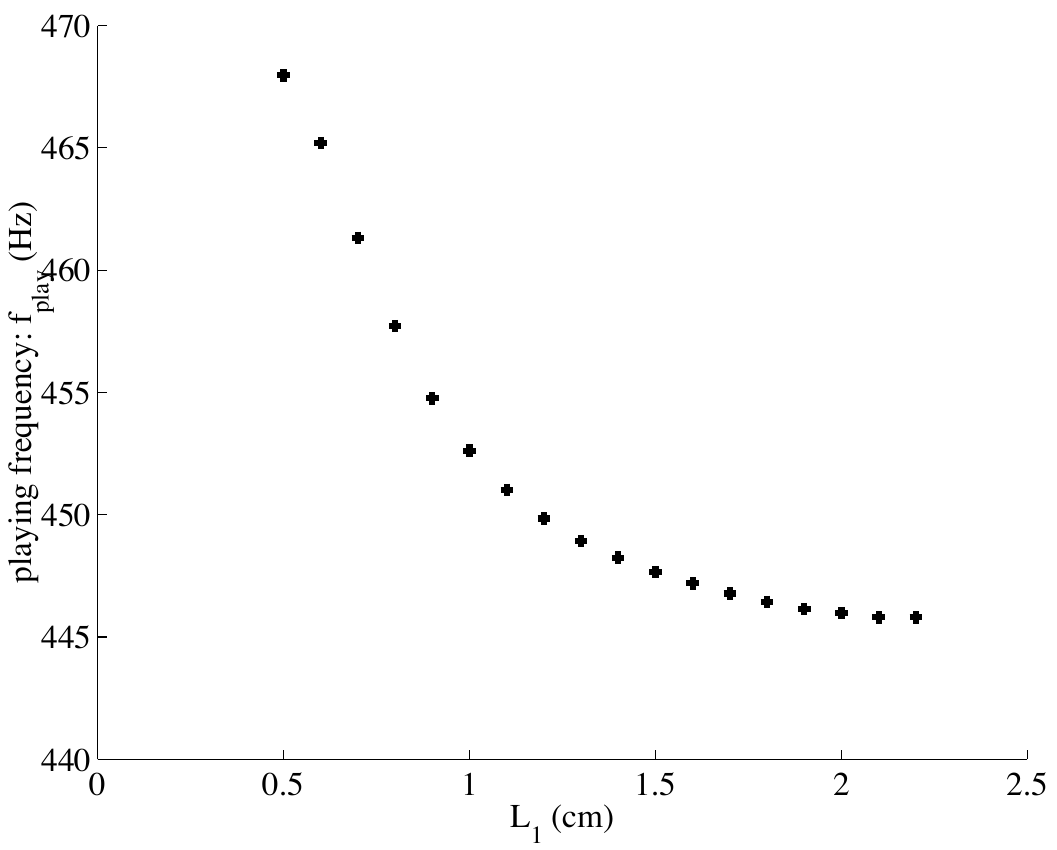}
     \caption{Influence of $L_1$ length on the playing frequency for a (-,+) reed (left) and a (+,-) reed (bottom) using respective excitation velocities $v_0$ of 2.5 and 3 m.s$^{-1}$. The playing frequency weakly varies for the (-,+) reed but presents great variations for a (+,-) reed.}
    \label{f20}
\end{figure}

\paragraph{}In Figure \ref{f20}, we study the influence of the pipe length $L_1$ over the playing frequency for both (-,+) and (+,-) reeds. We still used an excitation velocity of 2.5 m.s$^{-1}$ for the (-,+) reed and of 3 m.s$^{-1}$ for the (+,-) one and the other parameters were kept constant.

\paragraph{}The playing frequency is almost independent of the volume $V_1$ in the case of the (-,+) reed (variation of  2.6 Hz) but it decreases when $V_1$ increases in the case of the (+,-) reed with a great variation of 22.2 Hz. It is also interesting to note that the playing frequency is lower than the eigenfrequency for the (-,+) reed and higher for the (+,-) reed as predicted by Helmholtz \cite{Helmholtz:54} and experimentally verified by Jonhston \cite{Johnston:87}, Millot \cite{Millot:99, Millot:01} or Bahnson \cite{Bahnson:98}.

\begin{figure}[h!tbp]
    \centering
    \includegraphics[width=7cm]{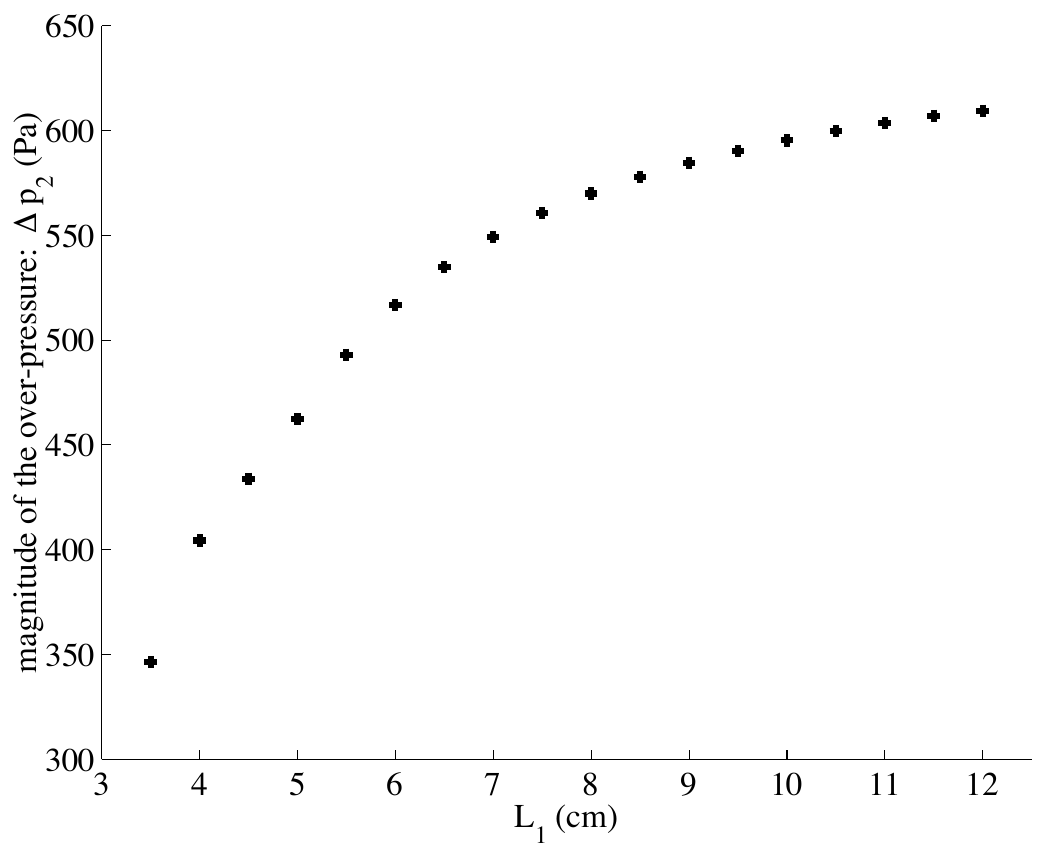}
    \includegraphics[width=7cm]{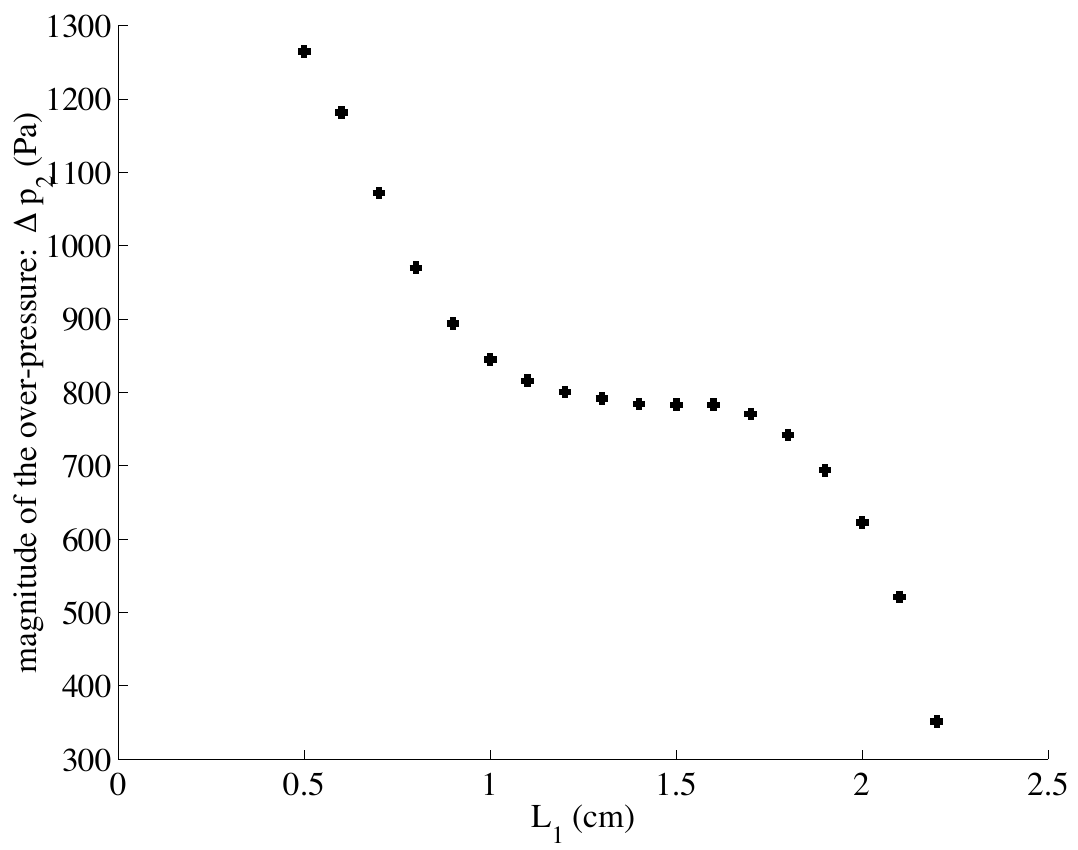}
     \caption{Influence of $L_1$ length over the magnitude of the over-pressure $\Delta p_2$ for (-,+) reed (left) and a (+,-) reed (right) using respective excitation velocities $v_0$ of 2.5 and 3 m.s$^{-1}$. Both reeds present great variations of the magnitude of $\Delta p_2$ but there is a flat-like part in case of the (+,-) reed.}
    \label{f21}
\end{figure}

\paragraph{}On Figure \ref{f21}, we have plotted the evolution of the magnitude of the over-pressure $\Delta p_2$ as a function of $L_1$. We can note that the influence is different for both reeds because the magnitude increases with $L_1$ for the (-,+) reed while it decreases for the (+,-) reed. This result seems coherent with the instability conditions derived for both reeds: having a little $V_1$ is a good thing  for the instability of a (+,-) reed  because we must have $\di 1- \frac{V_1L_2}{c^2_0S_2} \omega^2>0$ while, for a (-,+) reed,  a little $V_1$ is a worse thing because we must have $\di 1- \frac{V_1L_2}{c^2_0S_2} \omega^2<0$.  We can say that there are some convergence between the instability conditions and the results of the simulations.

\paragraph{}We can also note that the variation of $\Delta p_2$ is greater in the case of the (+,-) reed (914~Pa) compared to the (-,+) reed (263 Pa) which may be related to the bigger variation of the playing frequency in the (+,-) case and with the fact that the (+,-) reed vibrates above its eigenfrequency when blown.

\subsection{Influence of the excitation velocity $v_0$}
\begin{figure}[h!tbp]
    \centering
    \includegraphics[width=7.5cm]{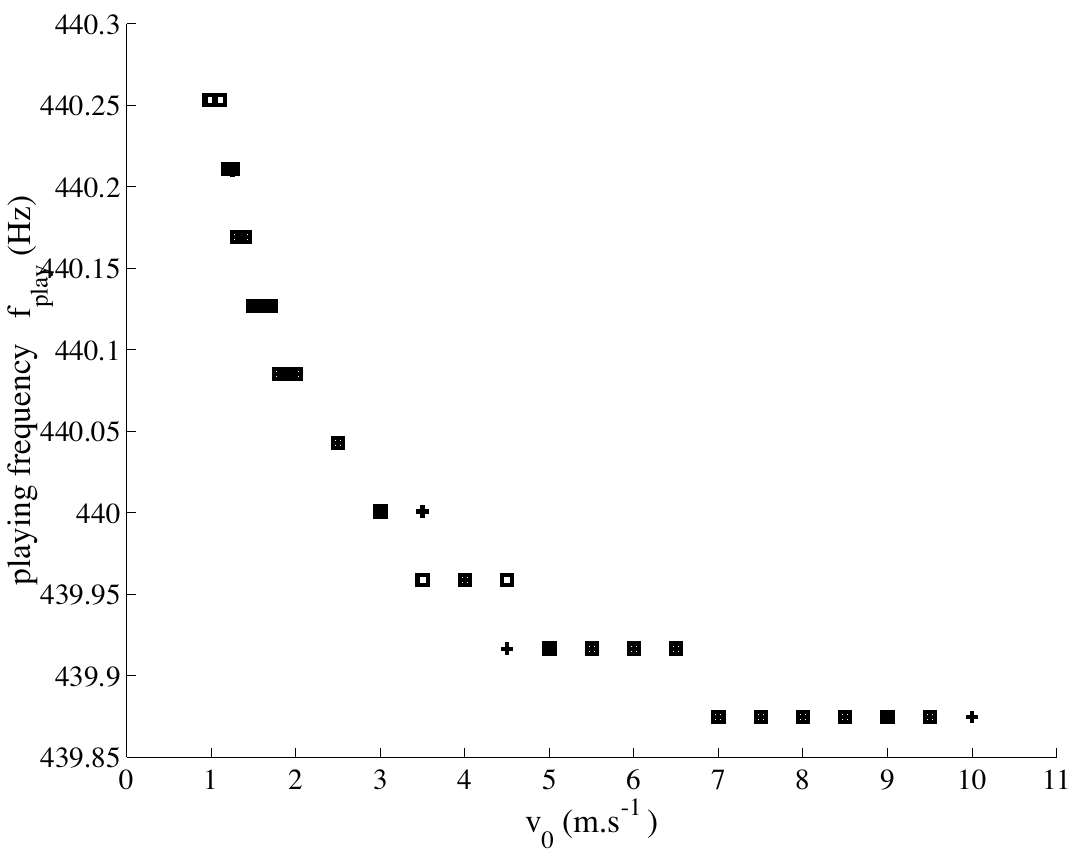}
    \includegraphics[width=7.5cm]{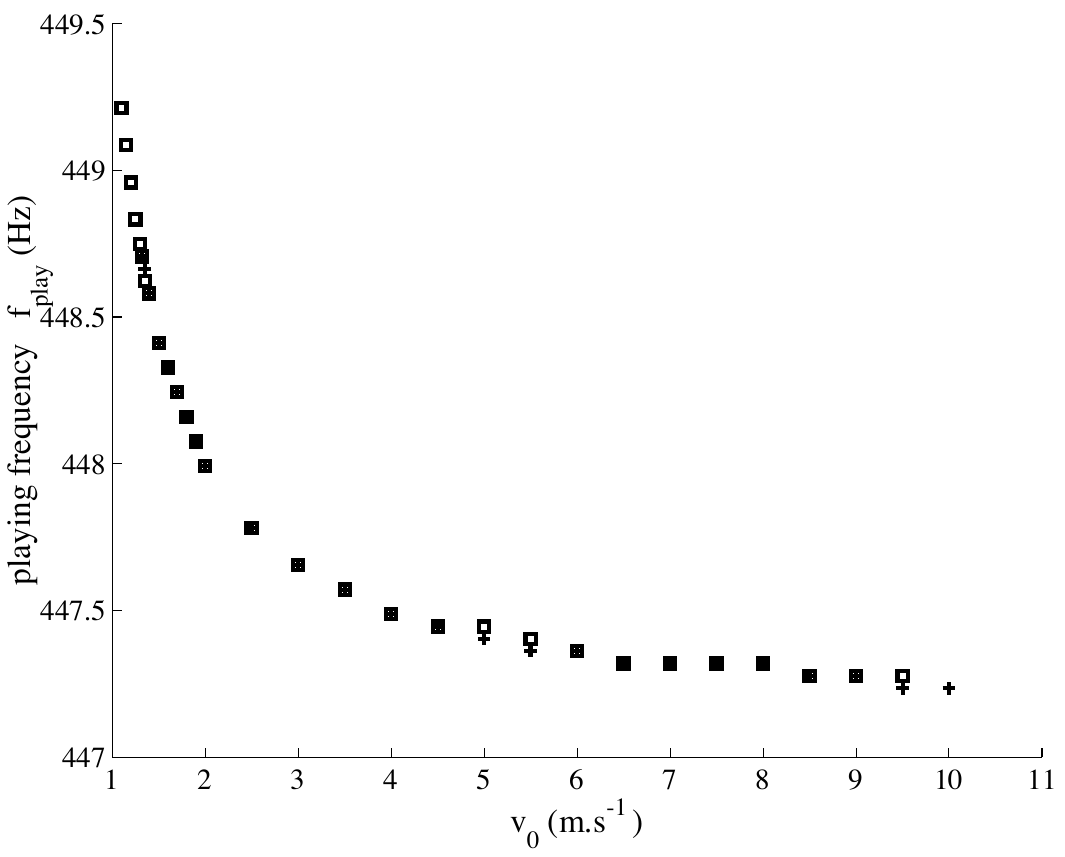}
     \includegraphics[width=7.5cm]{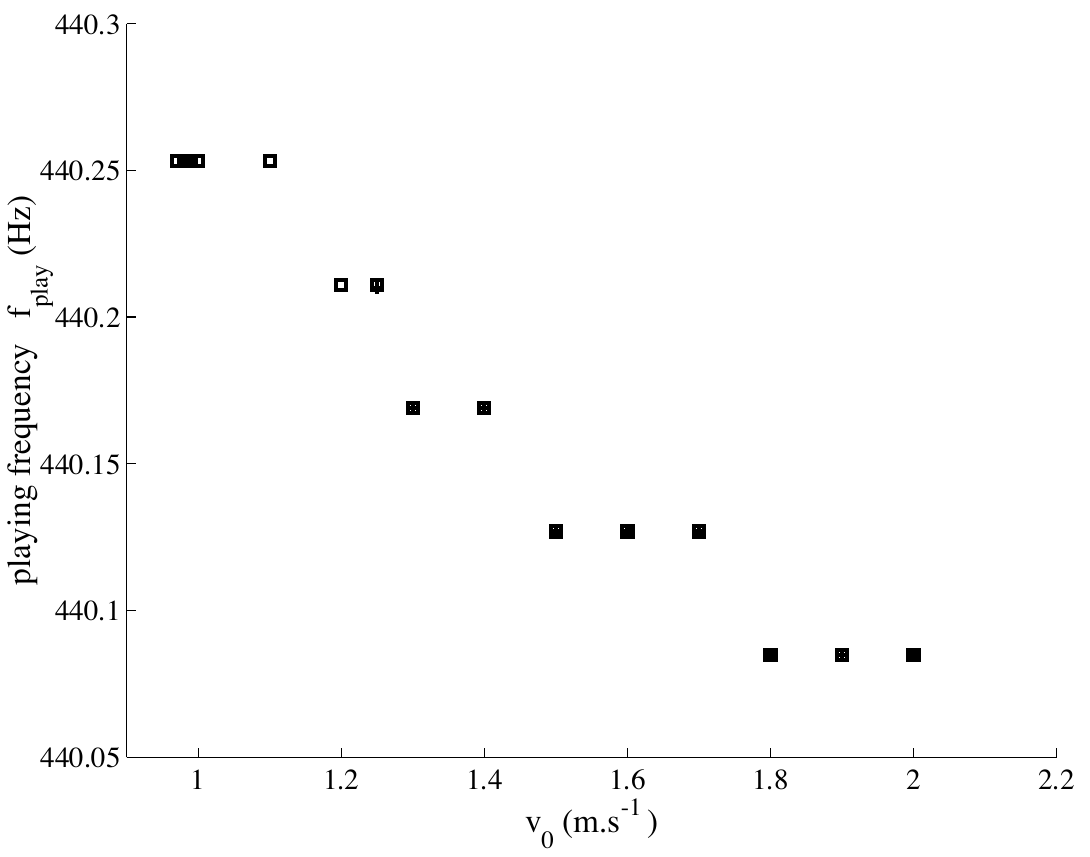}
   \includegraphics[width=7.5cm]{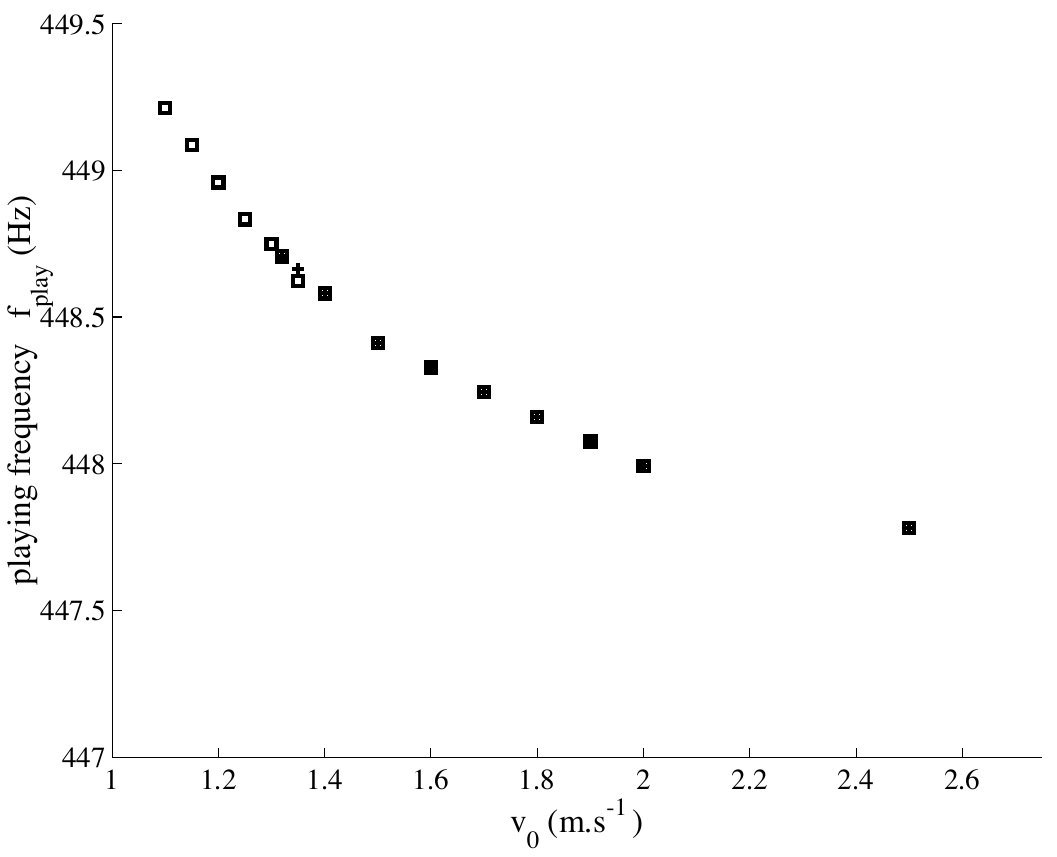}
    \caption{Plots of the variation of the playing frequency when the velocity excitation $v_0$ varies for a (-,+) reed (left plots) and a (+,-) reed (right plots). Top curves illustrate the case for the whole range of excitation while bottom curves present a zoom on the first values of $v_0$. We use crosses when $v_0$ increases and boxes when $v_0$ decreases.}
    \label{f22}
\end{figure}

\begin{figure}[h!tbp]
    \centering
    \includegraphics[width=7.5cm]{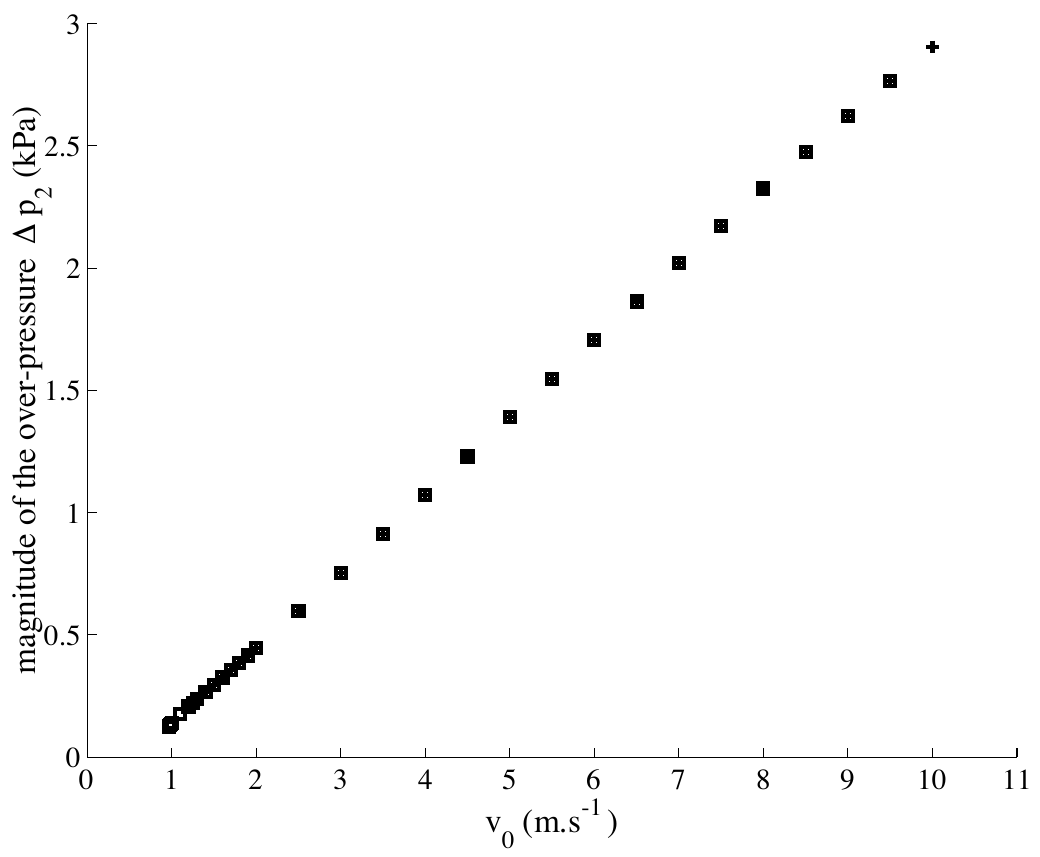}
   \includegraphics[width=7.5cm]{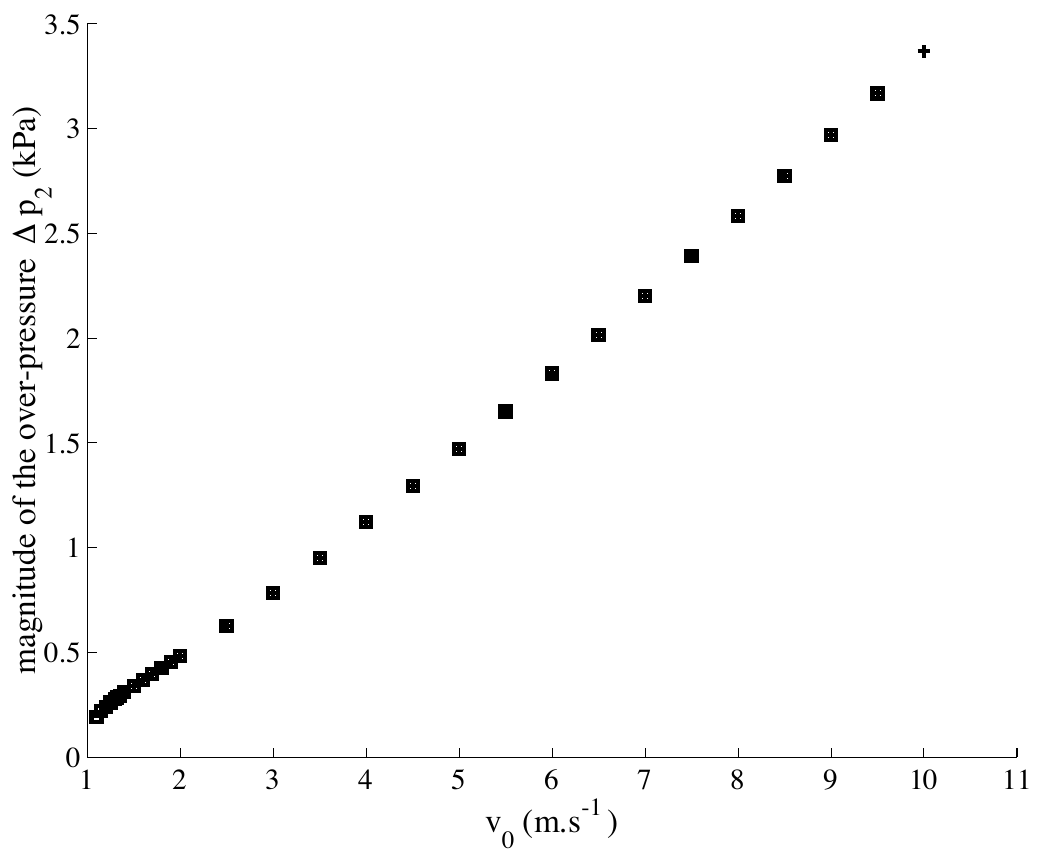}
    \includegraphics[width=7.5cm]{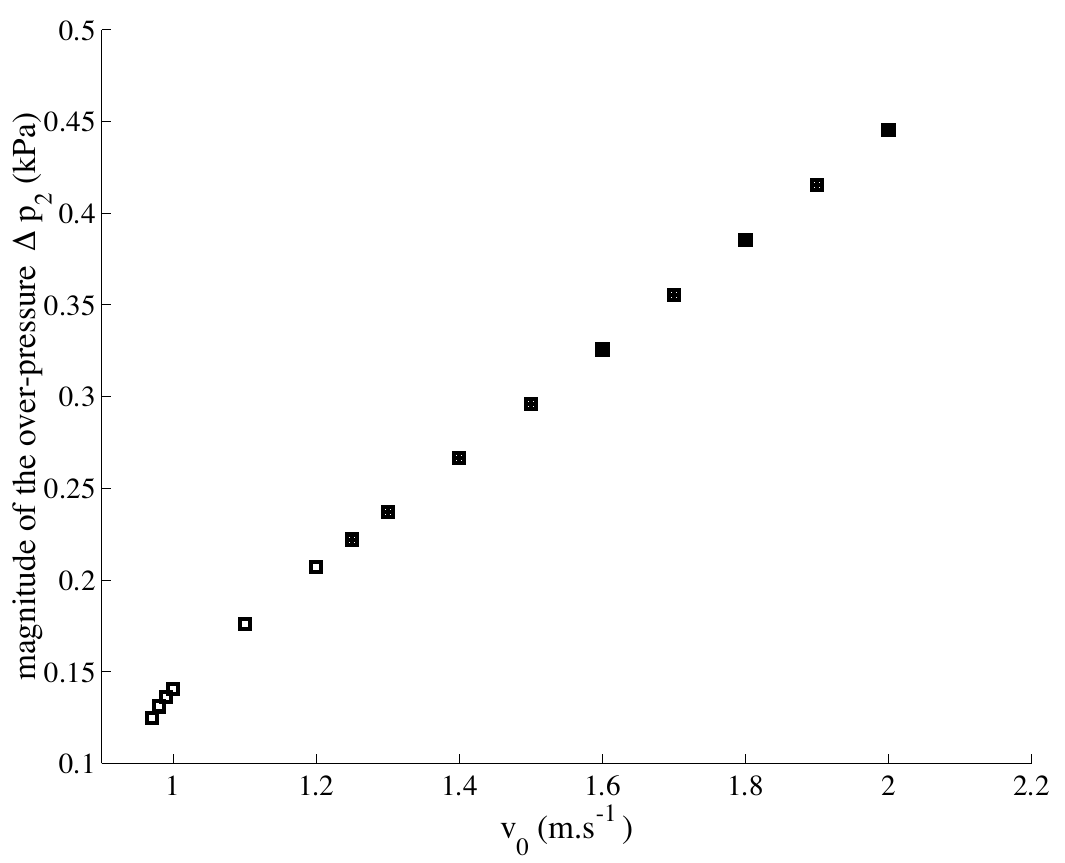}
    \includegraphics[width=7.5cm]{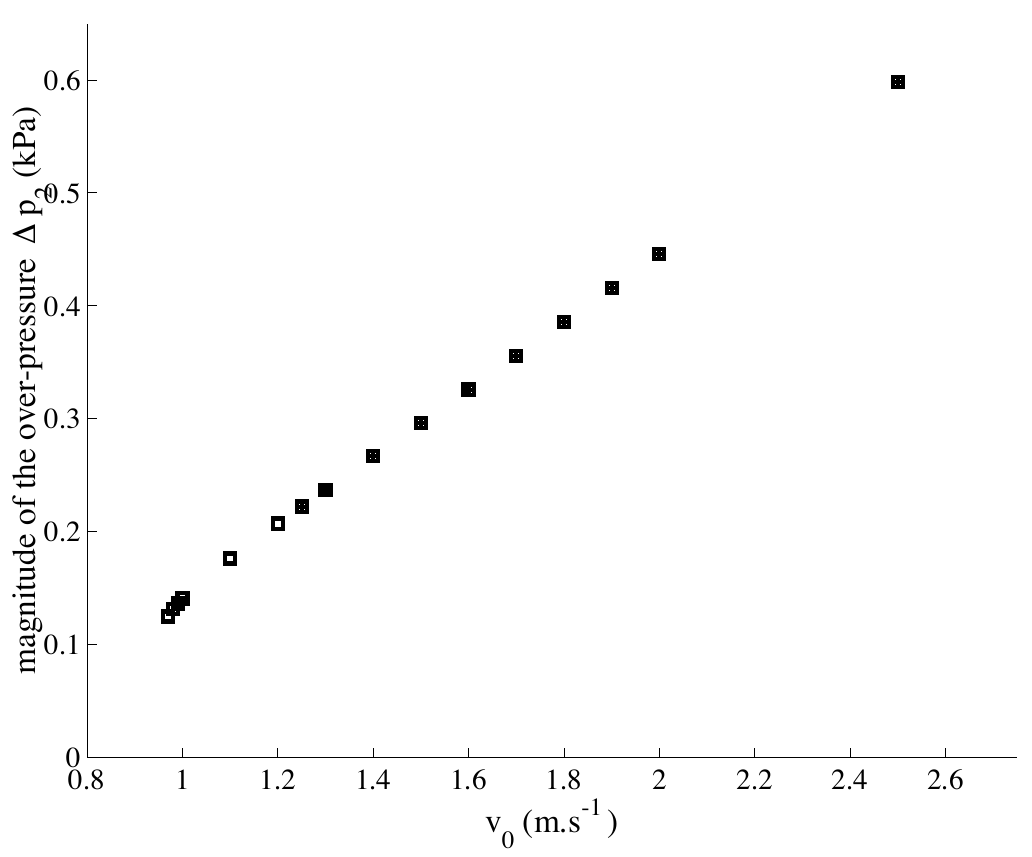}
    \caption{Plots of the variation of the magnitude of the over-pressure $\Delta p_2$ when the velocity excitation $v_0$ varies for a (-,+) reed (left plots) and a (+,-) reed (right plots). Top curves illustrate the whole range of excitation while bottom curves present a zoom on the first values of $v_0$. We use crosses when $v_0$ increases and boxes when $v_0$ decreases.}
    \label{f23}
\end{figure}

\paragraph{}We also have studied the influence of the excitation parameter $v_0$ for both reeds and the results for the playing frequency and the magnitude of $\Delta p_2$. Results for the playing frequency for both reeds are shown on Figure \ref{f22} while the ones for the magnitude of $\Delta p_2$ are given on Figure \ref{f23}.

\paragraph{}For each reeds, we started from rest and first searched the $v_0$ for which the reed  produces sound. Then, we increased the excitation velocity up to 10 m.s$^{-1}$ and plotted all these points with crosses. After, we decreased the excitation velocity $v_0$ until the velocity for which it was impossible to maintain sound production and plotted all these plots with boxes. 

\paragraph{}The variation of the playing frequency is quite weak when the excitation velocity varies whatever is the type of the reed: 0.4 Hz for the (-,+) reed and 2 Hz for the (+,-) reed. But, we can note that, for both reeds, we can decrease the excitation velocity and still produce sound once the sound production mechanism is launched.

\begin{figure}[h!tbp]
    \centering
    \includegraphics[width=10cm]{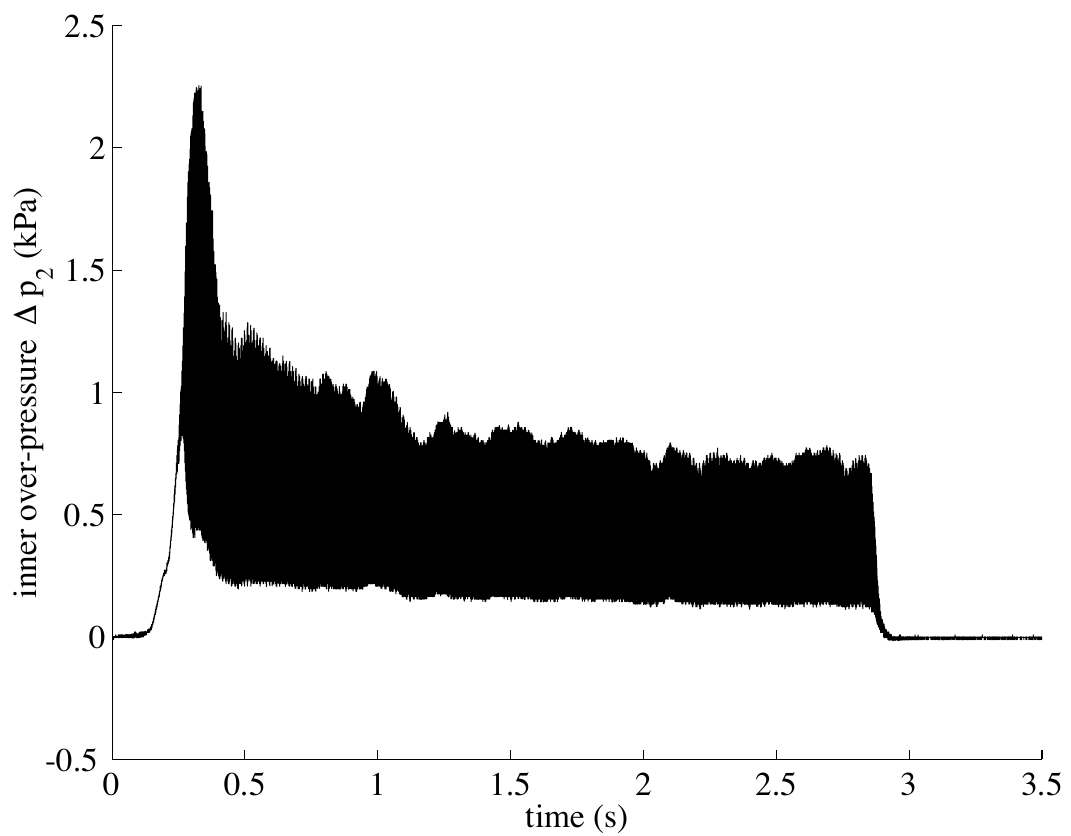}
    \caption{Example of a waveform for the over-pressure $\Delta p_2$ measured on a diatonic harmonica as played by a musician. This waveform shows a strong attack level for the over-pressure and then a lower "sustain" part of the signal. We found such configuration for all the sounds produced by the musician which may indicate that it is important to take a violent attack to get the production of the sound.}
    \label{f24}
\end{figure}

\paragraph{}One can notice that we have a similar phenomenon for the magnitude of the over-pressure $\Delta p_2$ but with great variations according to excitation velocity:  2780 Pa for the (-,+) reed and 3177 Pa for the (+,-) one. We still observe that we can decrease the excitation once the sound production is established for both reeds. And we observed, probably, the same kind of phenomenon during the recordings with a real musician on a diatonic harmonica for all kinds of note (normal blown and drawn notes, blown and drawn bends, overblows and overdraws). For each recorded note, we had a waveform quite similar to the one given in Figure \ref{f24}.

\paragraph{}In Mechanics, there are the notions of direct and inverse bifurcations. A direct bifurcation corresponds to the situation where the oscillation is obtained from a threshold value of the excitation but with a magnitude of the response growing continuously from zero to the maximal possible value. And, for a direct bifurcation, while decreasing of the excitation we follow exactly the same curve up to the threshold point. For an inverse bifurcation, it is impossible to observe a continuous departure from a zero magnitude response: there is a step in the response. And, moreover, it is possible to maintain the response for lower excitation levels once the response is established: one can decrease the excitation once the sound is established (for a musical instrument for instance) and still observe sound production  for a little range of excitations lower than the threshold one. 

\paragraph{}From the study of the influence of the excitation parameter, we can see that both reeds seem to present an inverse bifurcation because the sound production begins with a violent step in over-pressure and also because sound production can be maintained while decreasing the excitation level over a small range. 

\subsection{Influence of the reed thickness}
\paragraph{}To keep constant the reed mass while varying the thickness we must also change the value of either the reed length or the reed width. So, in the following, we study the influence of the variation of the thickness with, firstly, a related variation of the reed length $L_r$ and, secondly, a related variation of the  reed width $W_r$.

\paragraph{}We do not study the influence of the variation of the support thickness $e_s$ because this thickness is only present in the model to consider the same reference for both kinds of free reeds. Indeed, if we carefully consider the expression of the effective height, at the reed tip or for a local point on the reed side, we can verify that the support thickness  has absolutely no effect on the useful section within the proposed model. 

\subsubsection{Case of varying reed length $L_r$}
\begin{figure}[h!tbp]
    \centering
    \includegraphics[width=7.5cm]{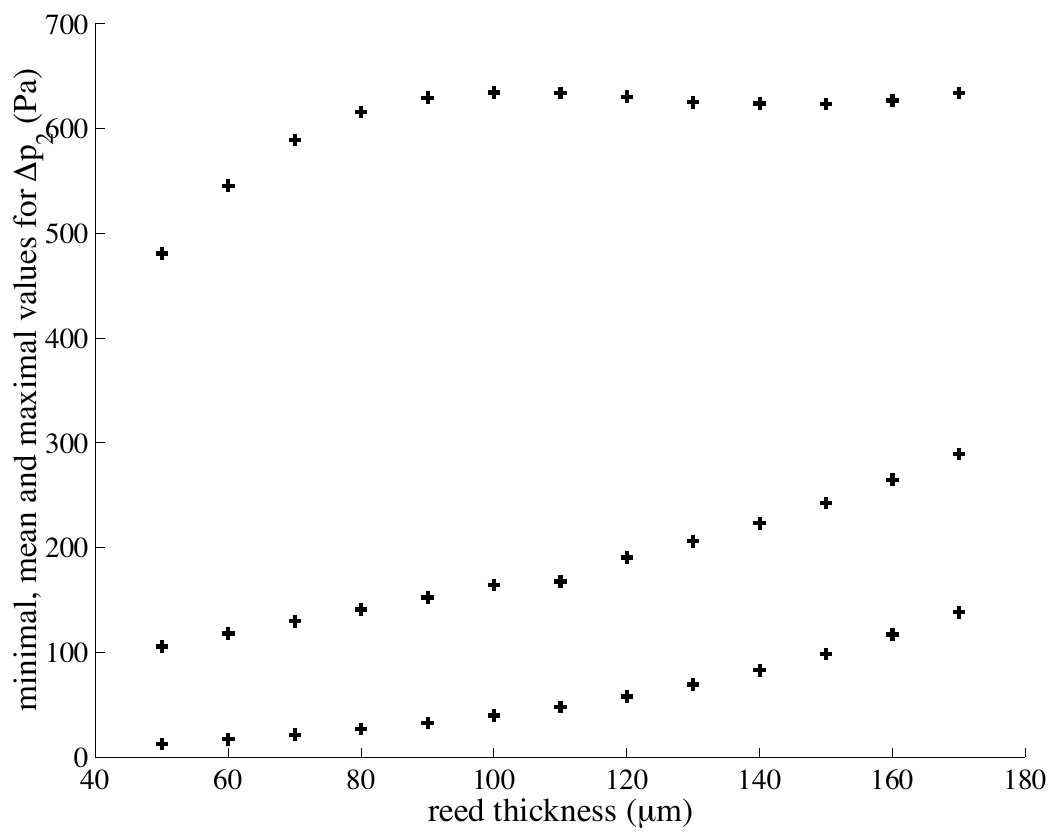}
   \includegraphics[width=7.5cm]{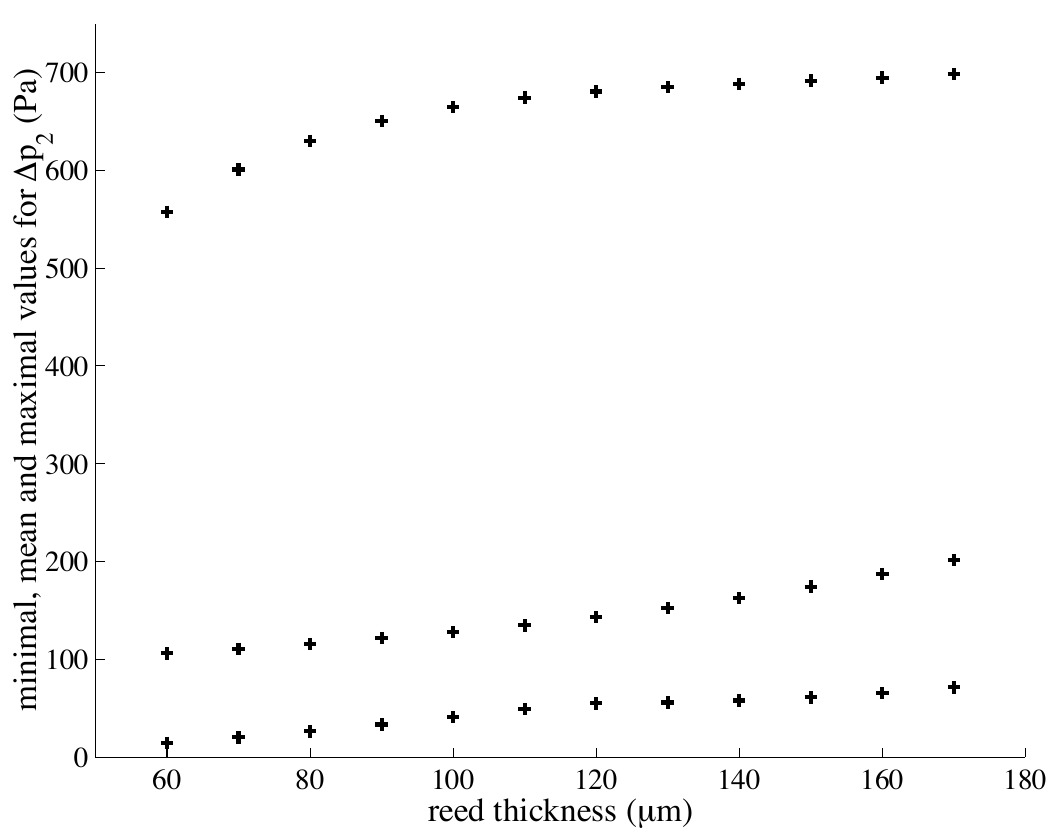}
    \caption{Influence of the variation of $e_r$, and $L_r$, over the over-pressure $\Delta p_2$ for a (-,+) reed (left) and a (+,-) reed. The values for the minimum, mean and maximum over-pressure are systematically plotted. The influence on both reed seems quite similar as all the values increase with the reed thickness. But the (+,-) reed present weaker reductions of the magnitude.}
    \label{f25}
\end{figure}

\paragraph{}Figure \ref{f25} presents the evolution of the minimum, mean and maximum over-pressure $\Delta p_2$. These two plots show that the magnitude of the over-pressure decreases as the reed thickness increases, for both reeds. Indeed, the minimum over-pressure increases faster than the maximum over-pressure for both reeds. The average over-pressure also increases but still remains quite close to the minimum over-pressure which is related to the nature of the signals: fast succession of sharp peaks. This could be considered as a reason to think that the mean over-pressure has only a mathematical sense as the signal is seldom closed to its mean value. This is an argument to consider an acoustical flow rather than a mean flow with acoustical perturbations.
 
\begin{figure}[h!tbp]
    \centering
    \includegraphics[width=7.5cm]{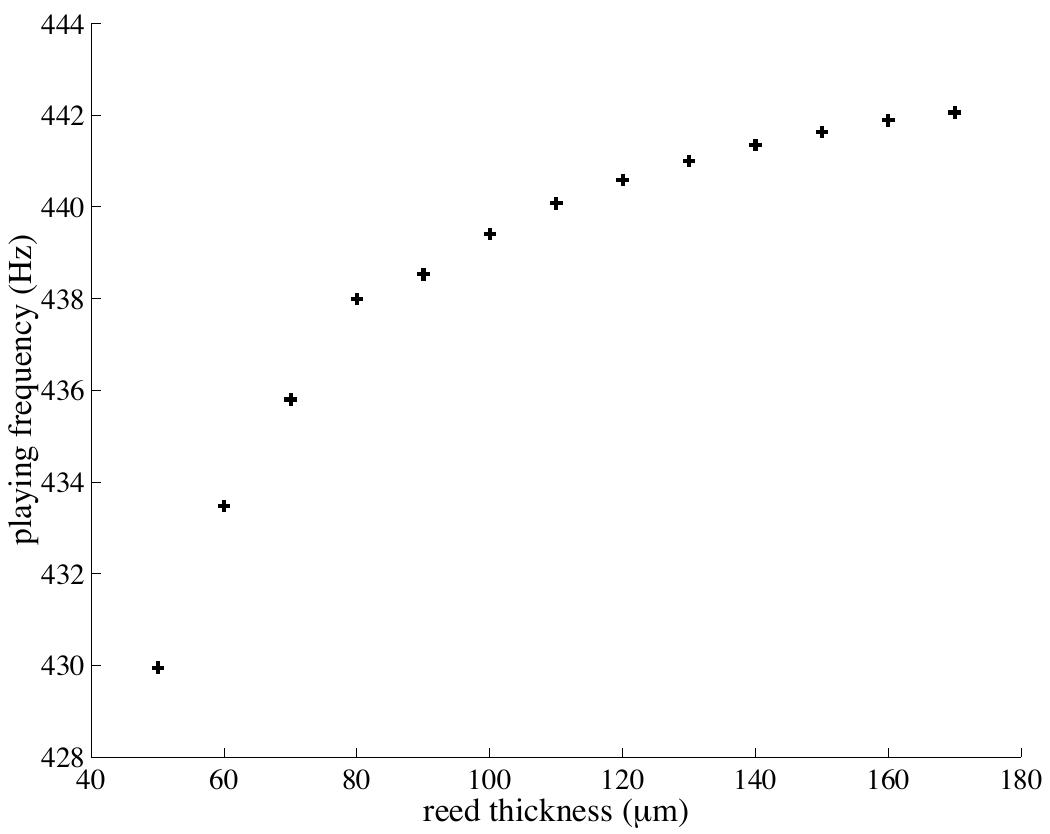}
   \includegraphics[width=7.5cm]{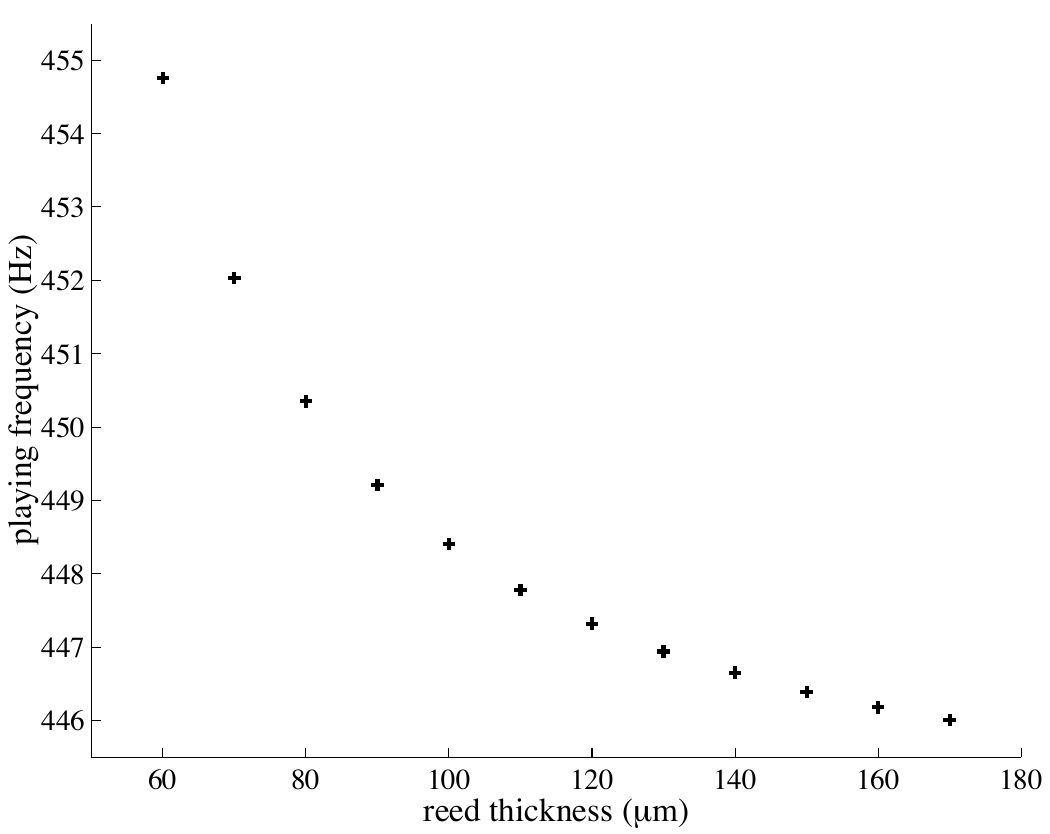}
    \caption{Influence of the variation of $e_r$, and $L_r$, over the playing frequency  for a (-,+) reed (left) and a (+,-) reed. For a (-,+) reed the playing frequency increases when the reed thickness increases while the playing frequency decreases for a (+,-) reed.}
    \label{f26}
\end{figure}

\paragraph{}Influence of the reed thickness over the playing frequency is illustrated in Figure \ref{f26}. One can see that the evolutions of the playing frequency are different according to the kind of reed we consider. Indeed, the playing frequency increases with the reed thickness for a (-,+) reed but  still remains lower than the reed eigenfrequency (444 Hz). For the (+,-) reed we have an inverse result as the playing frequency decreases  when the reed thickness increases. But, even if the playing frequency decreases, it is still higher than the reed eigenfrequency. It is also interesting to point out the fact that the variation of the playing frequency is a little larger for the (-,+) reed while we had the inverse result when we studied the influence of the volume $V_1$ or of the excitation velocity $v_0$ (even if in this case the variations were  tiny). So varying the reed thickness have a strong influence over the note to be played.

\begin{figure}[h!tbp]
    \centering
    \includegraphics[width=7.5cm]{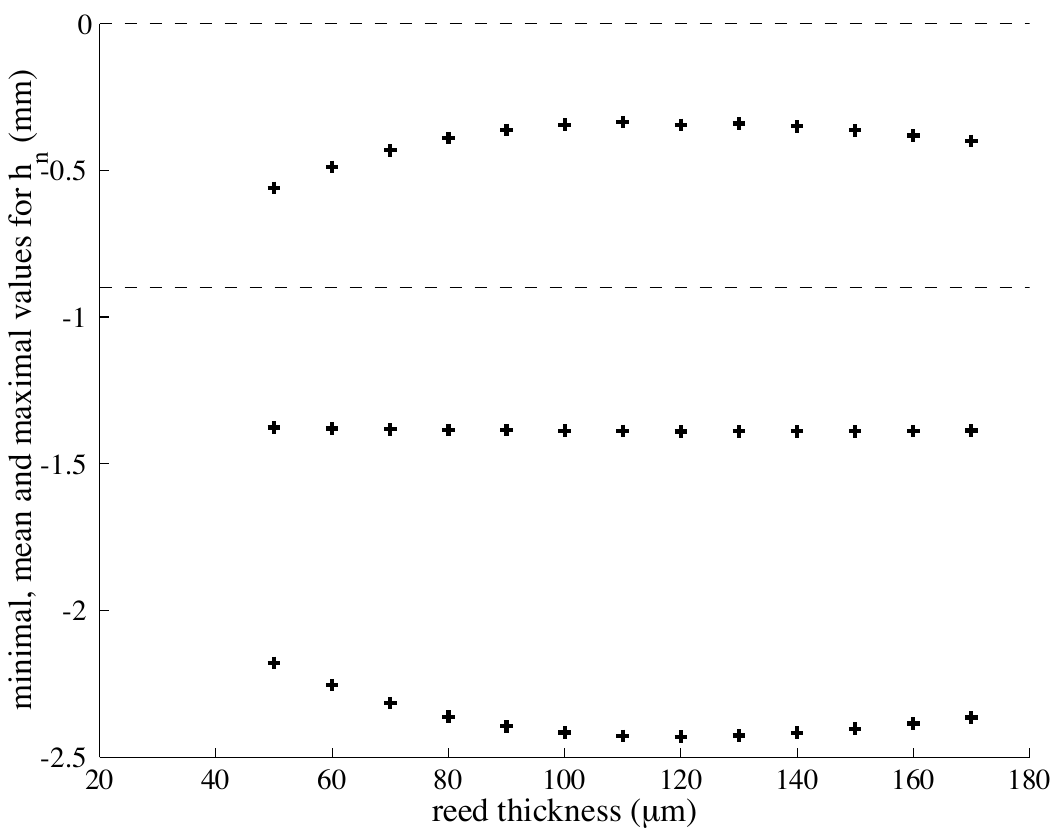}
   \includegraphics[width=7.5cm]{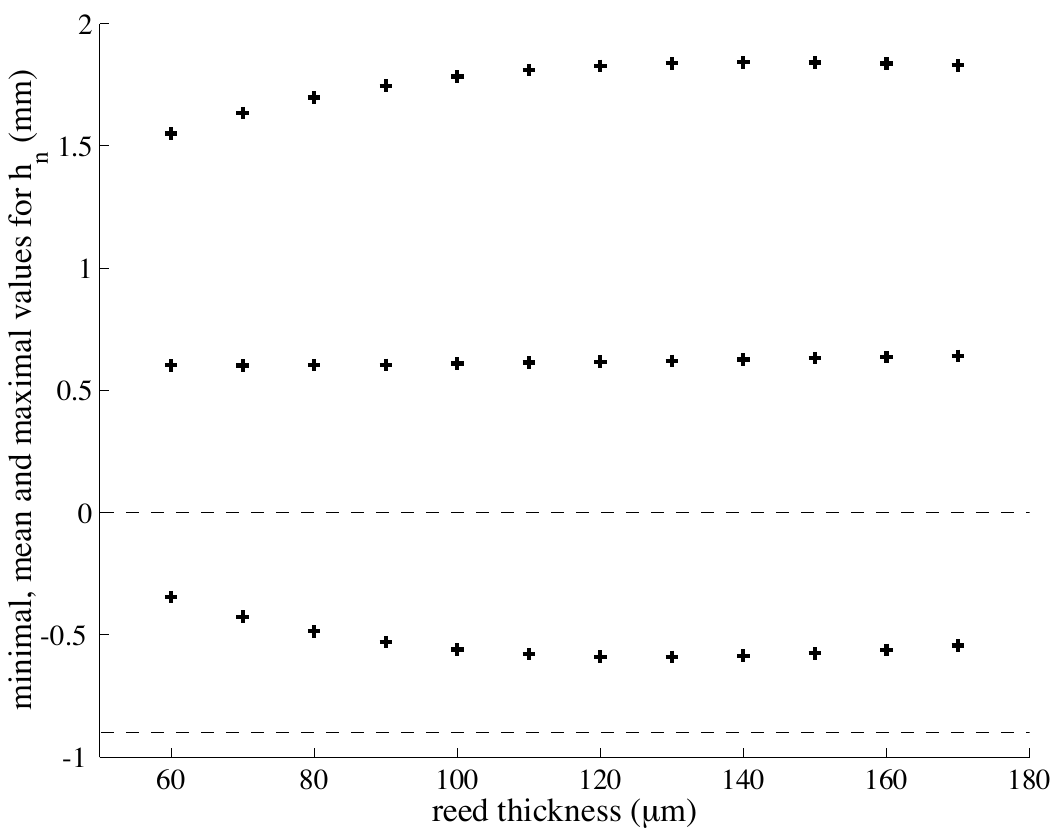}
    \caption{Influence of the variation of $e_r$, and $L_r$, over the tip reed opening $h_n$ for a (-,+) reed (left) and a (+,-) reed. The values for the minimum, mean and maximum over-pressure are systematically plotted.}
    \label{f27}
\end{figure}

\paragraph{}Figure \ref{f27} present the variations of the mean, minimum and maximum value of the free tip opening $h_n$. The mean opening presents strong variations while minimum and maximum present significant ones. The evolutions are almost similar for both reeds and we can notice that there is something like an optimum for the reed opening magnitude where the maximum opening is the biggest and the minimal opening is the tiniest. And this "optimum" thickness configuration corresponds to original reed thickness. We can point out that the magnitude of the reed displacement is about 2~mm which may still correspond to the assumption of a beam with reasonable bending. It is also interesting to note that the mean value is associated with the mean of the minimum and maximum openings which seems normal as the reed motion is quasi sinusoidal.

\subsubsection{Case of varying reed width $W_r$}

\begin{figure}[h!tbp]
    \centering
    \includegraphics[width=7.5cm]{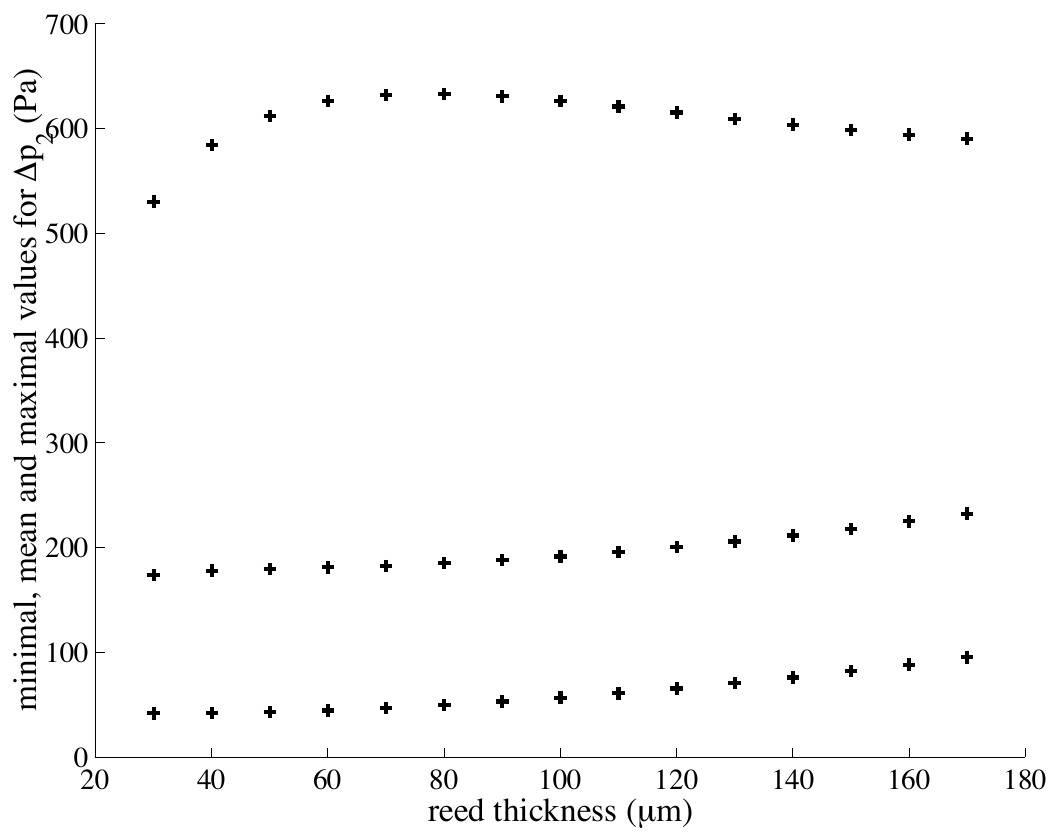}
   \includegraphics[width=7.5cm]{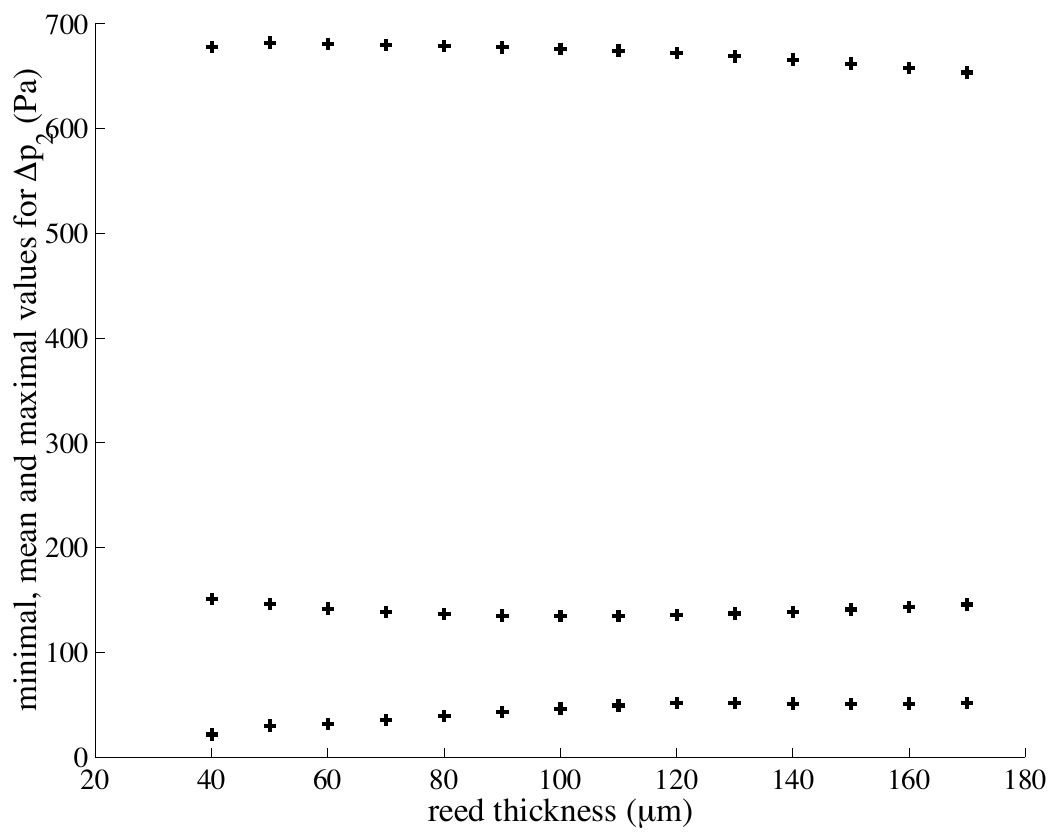}
    \caption{Influence of the variation of $e_r$, and $W_r$, over the over-pressure $\Delta p_2$ for a (-,+) reed (left) and a (+,-) reed. The values for the minimum, mean and maximum over-pressure are systematically plotted. }
    \label{f28}
\end{figure}

\paragraph{}On figure \ref{f28} we plot the evolution of the mean, minimum and maximum over-pressure as a function of the reed thickness but, now, with a variable reed width to keep the reed mass constant. Compared to the reed length variable case, while we find a similar plot for the (-,+) reed, the characteristics for the (+,-) have changed: the maximum over-pressure decreases with the increase of the reed thickness; the minimum over-pressure grows and then decreases a little, the mean over-pressure 
decreases up to a minimum and then increases. But we still find the fact that the mean over-pressure is quite closed to the minimum one so that the mean over-pressure is not representative.

\begin{figure}[h!tbp]
    \centering
    \includegraphics[width=7.5cm]{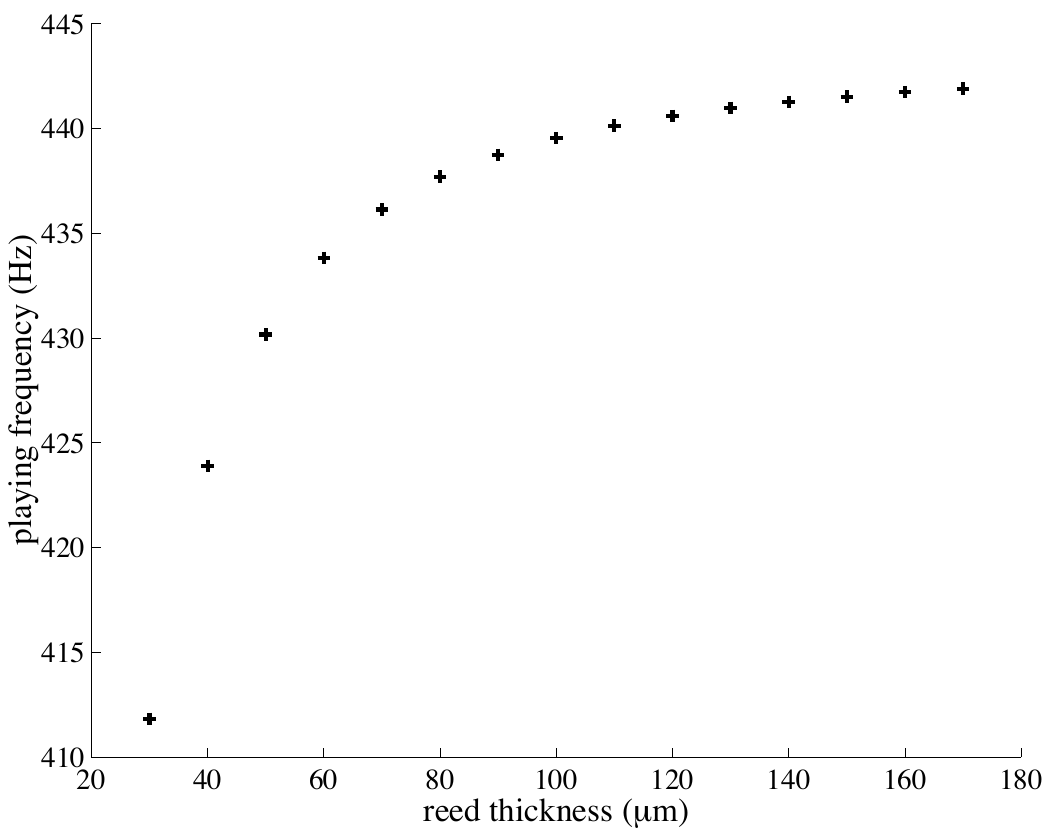}
   \includegraphics[width=7.5cm]{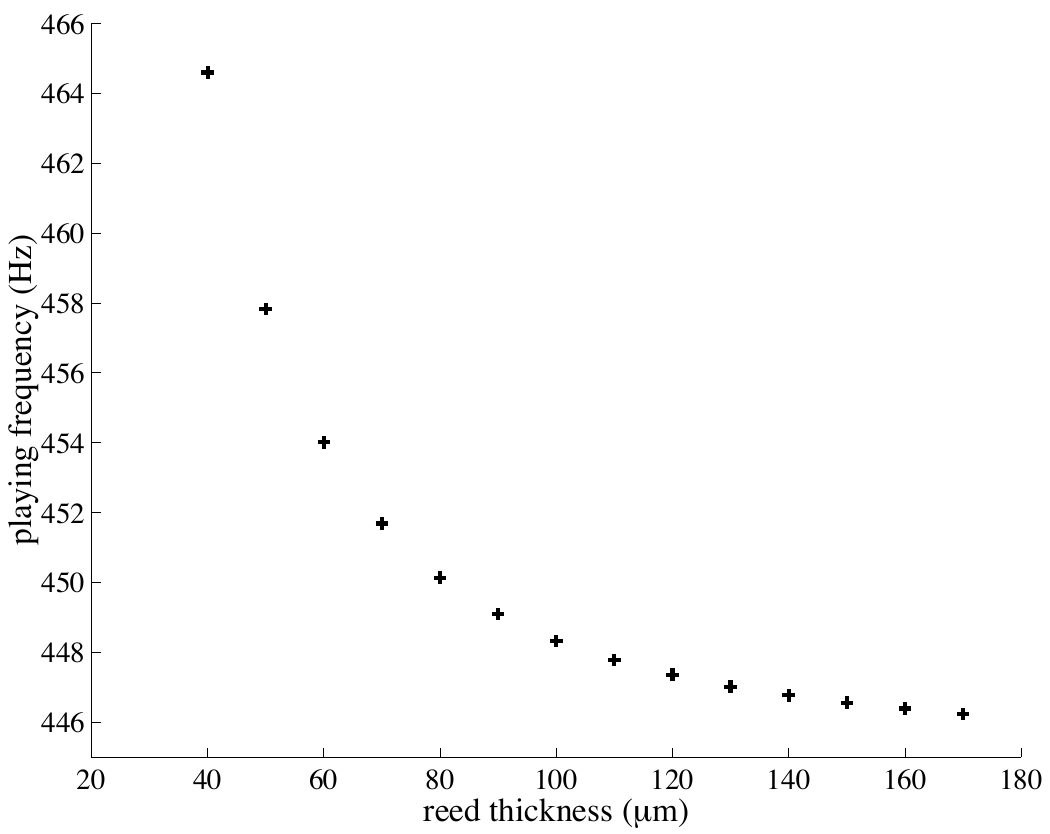}
    \caption{Influence of the variation of $e_r$, and $W_r$, over the playing frequency  for a (-,+) reed (left) and a (+,-) reed.}
    \label{f29}
\end{figure}

\paragraph{}With Figure \ref{f29} we access the characteristic of the playing frequency. We can then notice that the global behavior for both reeds is quite similar to the one found for the case where the reed length is variable. We still have an increasing playing frequency for the (-,+) reed, a decreasing playing frequency for the (+,-) reed and a magnitude of variation larger for the (-,+) reed than for the (+,-) reed. Moreover, the playing frequency still remains lower than the reed eigenfrequency for the (-,+) reed and higher for the (+,-) reed.

\begin{figure}[h!tbp]
    \centering
    \includegraphics[width=7.5cm]{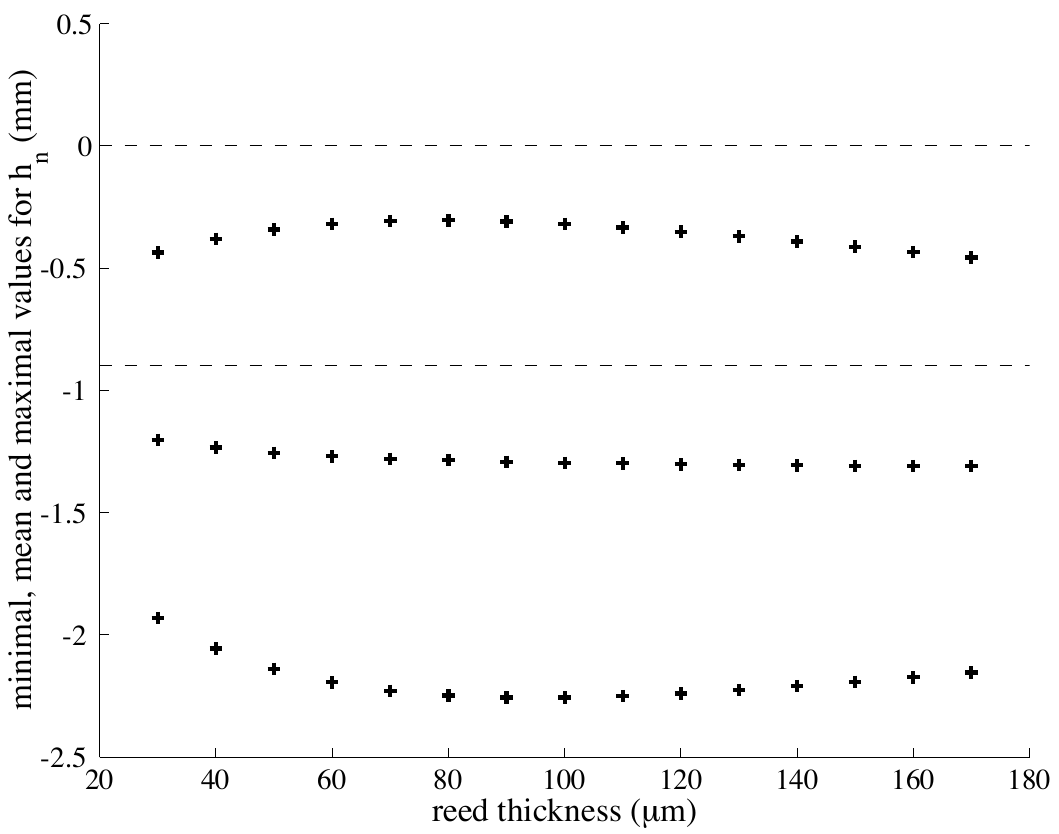}
   \includegraphics[width=7.5cm]{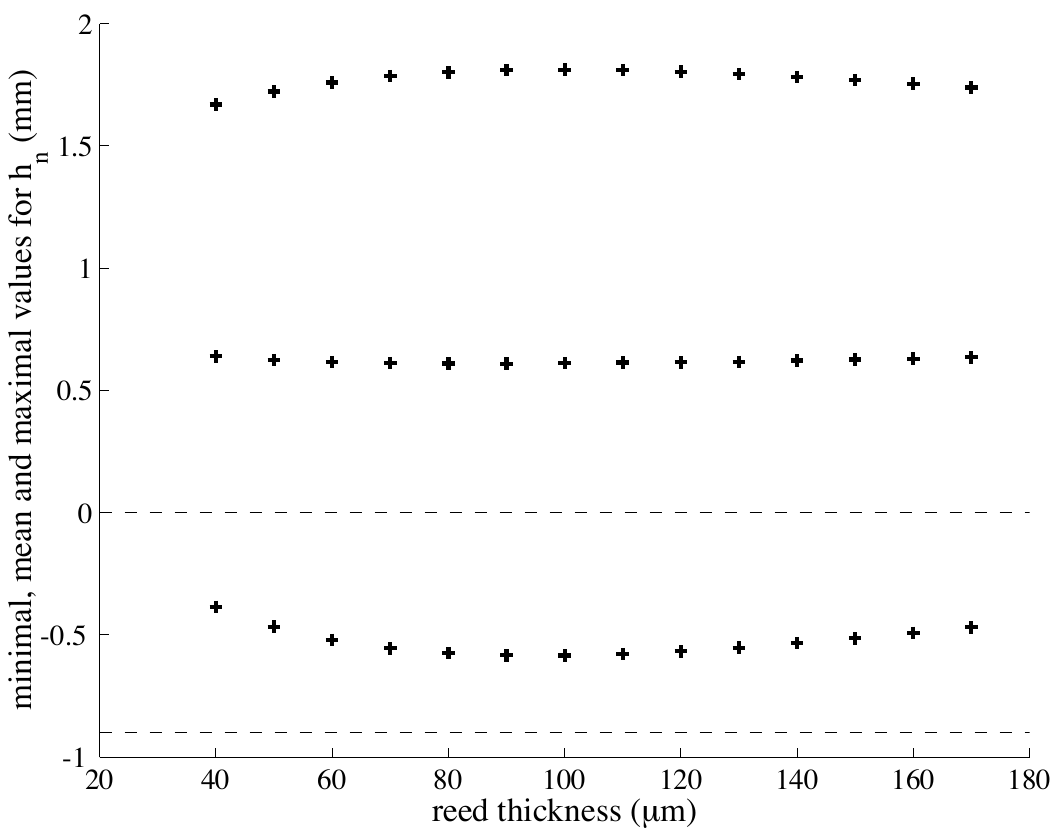}
    \caption{Influence of the variation of $e_r$, and $W_r$, over the tip reed opening $h_n$ for a (-,+) reed (left) and a (+,-) reed. The values for the minimum, mean and maximum over-pressure are systematically plotted.}
    \label{f30}
\end{figure}

\paragraph{}Plots of the mean, minimum and maximum free tip openings $h_n$ on Figure \ref{f30} exhibit the same behavior as the one found for the variable reed length: optimum for the opening magnitude closed to the original value of the reed thickness, small variations of the mean opening, mean opening corresponding to the mean of the maximum and minimum openings.
 
\paragraph{}With the study of the case where either the reed length or the reed width is variable, we can point out that the reed thickness can have a great influence on the behavior of the reed. This justifies the need to take into account of the reed thickness in the model of the useful section and to distinguish 
the neutral, downstream and upstream sections.

\section{Conclusion}
\paragraph{}In this article, we have tried to propose a minimal model for free reeds which may be used to build efficient models for free reed instruments.   

\paragraph{}We have proposed to introduce corrections into the airflow and above all into the description of the section through which the airflow passes.  It seems necessary to take account of the lateral contribution and to define precisely the front contribution. These definitions also need a precise definition of the frontiers of the escape. The reed description and the useful section explicitly introduce the reed thickness. Even if one may think that the reed thickness is too tiny to be taken into account, the study of its influence over the behavior for both (-,+) and (+,-) reeds shows that its influence is quite important and that we cannot neglect it. 

\paragraph{}The comparison of numerical simulations derived with three different  models of the useful section (Hikichi, Debut and our model) has shown that our model has not the same dynamic behavior: the reed response is faster and stronger. This result has been found for an artificial triplet excitation and also for a real time-varying one. We also have discussed the needed equations and the validity of their assumptions. 

\paragraph{}To be able to get some instabilities of both kinds of reeds, we have found that we must introduce a volume and a pipe at the upstream of the reed, on the basis of a linear analysis of the instability conditions. The temporal simulations seem to confirm the qualitative insight given by the instability conditions: reducing the upstream volume helps the oscillations of a (+,-) reed but increasing it is necessary for a (-,+) reed. It has also been shown that the playing frequency of both kinds of reeds is nearly independent of the excitation level. The playing frequency of a (-,+) reed is nearly independent of the size of the upstream volume while, for a (+,-) reed,  the playing frequency increases when the upstream volume is reduced and the variations of the playing frequency are important. The increase of the magnitude of the over-pressure is a consequence of the increase of the excitation. 

\paragraph{}Influence of the excitation velocity $v_0$ over the over-pressure makes us think that free reeds may present inverse bifurcations because we observe a violent departure from rest and we are able to maintain sound production for excitations lower than the threshold one.  But, some experiments should be done to confirm or refute this hypothesis. 

\paragraph{}We think it is now possible to use this minimal model in the case of free reed instruments such as harmonica, accordeon, harmonium or sho. The application to the case of the harmonica is on the way and the first results of the temporal simulations, to be present elsewhere, are quite promising.  But, there is a great need for experiments with artificial players for all free reeds instruments to be able to validate the models and the associated temporal simulations. It would also be interesting to study the nature of the airflow through the reeds with some more visualisations to improve its description.  A discussion  on the necessity to use a 3D model of the airflow may also be useful to decide if our model is sufficient not only for high quality synthesis but to study the real behavior of the instruments and to see if numerical simulations can have the same status as real experiments.

\section{Appendix}
\subsection{Description of the motion of the sections}
\begin{figure}[h!tbp]
    \centering
    \includegraphics[width=7.5cm]{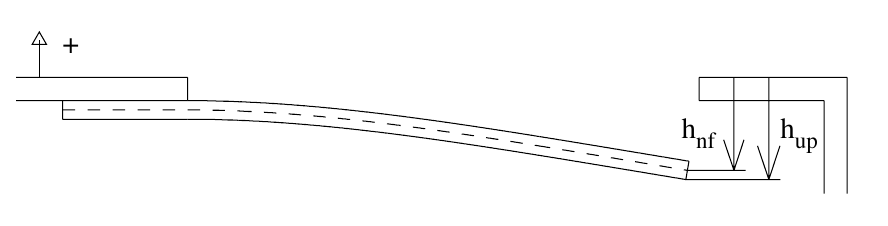}
    \includegraphics[width=7.5cm]{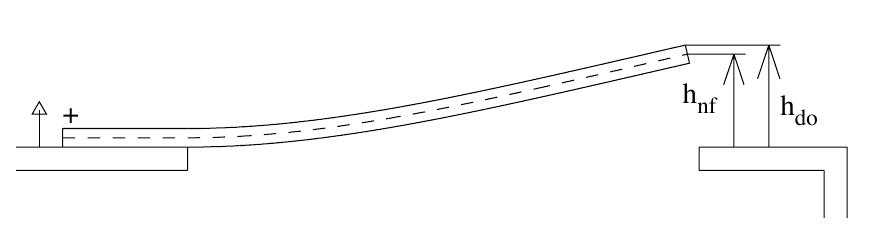}
    \caption{(left) Illustration of the convention chosen for the transverse displacements $h_{n}(s)$  and $h_{up}(s)$ of the neutral and downstream sections in the case of a (-,+) reed. (right) Illustration of the convention for the transverse displacement $h_{do}(s)$ of the downstream section in the case of a (+,-) reed. These displacements are used to calculate the useful sections.}
    \label{fa1}
\end{figure}

\paragraph{}In the following, we note $M_{n}(s)$, $M_{up}(s)$ and $M_{do}(s)$ respectively the points of the reed which are on the neutral section, on the upstream and downstream sections of the reed at a distance $s.L_r$ ($s\in[0,1]$, $L_r$ is the reed length) from the clamped end of the reed. We recall that the upstream and downstream sections respectively face the inside and the outside of the instrument. The reference point $O(0,0,0)$ is taken at the clamped end, on the middle of the width and on the downstream side of the support (outside of the instrument) as illustrated by figure \ref{fa1} for a (-,+) and a (+,-) reed.

\paragraph{}As we assume the reed to vibrate on its first transverse eigenmode and we note $h_{n}$ and $h_{n,000}$ the current and plane transverse position of the neutral section, the local neutral section point $M_{n}(s)$  coordinates are $(x_{n}(s), 0, h_{n}(s))$ (point situated on the middle of the width of the reed) where:
$$x_{n}(s)=s.Lr$$
and 
$$h_{n}(s)=h_{n,000}+(h_{n}-h_{n,000})\psi(s),$$
where $\psi(s)$ is the normalized modal reed factor for the first transverse eigenmode.

\paragraph{}Considering the figure \ref{fa1}, we can observe that when the reed is plane, the transverse positions $h_{n,000}$ of the (-,+) and (+,-) reeds are different:  
\begin{description}
    \item[$\bullet$] $\di h_{n,000}=-e_s-\frac{e_r}{2}$ for a (-,+) reed;
    
    \item[$\bullet$] $\di h_{n,000}=\frac{e_r}{2}$ for a (+,-) reed;
\end{description}
where $e_r$ and $e_s$ are the reed and support thickness.
 
\paragraph{}In these conditions, the coordinates of the normal $\overrightarrow{n(s)\;}$ to neutral section, at distance $s.L_r$ from the clamped end, are:
$$\overrightarrow{n(s)\;}=\left( \begin{array}{c}
\di \frac{-(h_{n}-h_{n,000})\psi'(s)}{\sqrt{L^2_r+(h_{n}-h_{n,000})^2\psi'(s)^2}}
\\
\\
0\\
\\
\di \frac{L_r}{\sqrt{L^2_r+(h_{n}-h_{n,000})^2\psi'(s)^2}}\\
\end{array}
\right).
$$  

\paragraph{}The points $M_{up}(s)$ and $M_{do}(s)$ of the upstream and downstream sections associated with the point $M_{n}(s)$ of the neutral section are then given by:
$$ \overrightarrow{OM_{up}(s)\;}=\overrightarrow{OM_{n}(s)\;}-\frac{e_r}{2}\overrightarrow{n(s)\;}$$
and   
$$ \overrightarrow{OM_{do}(s)\;}=\overrightarrow{OM_{n}(s)\;}+\frac{e_r}{2}\overrightarrow{n(s)\;}.$$

\paragraph{}So we have, for $x_{up}(s)$, $h_{up}(s)$, $x_{do}(s)$ and $h_{do}(s)$:
\begin{description}
    \item[$\bullet$] $\di x_{up}(s)=s.Lr +\frac{e_r}{2}\frac{(h_{n}-h_{n,000})\psi'(s)}
    {\sqrt{\di L^2_r+(h_{n}-h_{n,000})^2\psi'^2(s)}}$;
    
    \item[$\bullet$] $\di h_{up}(s)=h_{n}(s)-\frac{e_r}{2}\frac{L_r}{\sqrt{\di L^2_r+(h_{n}-h_{n,000})^2\psi'^2(s)}}$; 

    \item[$\bullet$] $\di x_{do}(s)=s.Lr -\frac{e_r}{2}\frac{(h_{n}-h_{n,000})\psi'(s)}
    {\sqrt{\di L^2_r+(h_{n}-h_{n,000})^2\psi'^2(s)}}$;
    
    \item[$\bullet$] $\di h_{do}(s)=h_{n}(s)+\frac{e_r}{2}\frac{L_r}{\sqrt{\di L^2_r+(h_{n}-h_{n,000})^2\psi'^2(s)}}$.
\end{description}

\paragraph{}We introduce the following quantities to simplify the expressions:
\begin{description}
    \item[$\bullet$] $\Delta x_{up}(s)= x_{up}(s)-s.L_r$ which corresponds to the longitudinal displacement of the upstream section;
    
    \item[$\bullet$] $\Delta x_{do}(s)= x_{do}(s)-s.L_r$ which corresponds to the longitudinal displacement of the downstream section.    
\end{description}

\subsection{Reed motion equation and pumped flow}
\paragraph{}In the case of free reeds, we have already seen that the reed motion corresponds to the first transverse eigenmode of a clamped-free beam so we can write the local transverse deformation $\xi(s,t)=\psi(s).\zeta(t)$ (of the neutral section at a distance $s.L_r$ from the clamped end) with $\zeta(t)=h_{n}(t)-h_{n,00}$ where $h_{n,00}$ is the unblown tip opening of the neutral section of the reed. 

\paragraph{}For the neutral section, the reed motion equation is classically given by 
\begin{equation}
\frac{d^2 \zeta}{dt^2}+Q^{-1}\omega_0 \frac{d \zeta}{dt} + \omega^2_0 \zeta= \mu \Delta p,
\end{equation}
where :
\begin{description}
\item[$\bullet$] $Q$ and $\omega_0$ are directly determined by calibration ;

\item[$\bullet$] $\di \mu=\frac{S_r}{M}$ with $\di S_r=W_r\int_0^{L_r}\psi(s)ds$ and $M$ given by $\di M=\frac{K}{\omega^2_0}$ as calibration gives the stiffness $K$ rather than $M$. 

\end{description}

\paragraph{}We also have to introduce the equation of the pumped flow. This flow can be defined by an equivalent expression $\di u_p=S_p \frac{d \zeta}{dt}$ or by a local expression $\di u_p= W_r\int_0^{L_r} \frac{\partial \xi(s,t)}{\partial t}ds$. As we have $\xi(s,t)=\psi(s).\zeta(t)$, we find after some calculation that:
\begin{equation}
S_p=S_r=W_r\int_0^{L_r} \psi(s)ds.
\end{equation}

\paragraph{}We do not distinguish $S_p$ and $S_r$ in practice and only use $S_r$ in case of free reeds instruments. 

\clearpage
\subsection{Derivation of the conditions for instabilities}
\paragraph{}Given the symmetry of the  laws for the useful section,  $S_u=S_{u00}+a_1\zeta^2$ is a  fair approximation of the useful section near the mean opening. In the following, we note $\zeta_0=h_{n,0}-h_{n,000}$, $\zeta_{00}=h_{n,00}-h_{n,000}$ and $S_{u0}=S_{u00}+a1(\zeta_0-\zeta_{00})^2$. We also use the prime suffix for the variable or acoustical part of quantities  used in the calculation: $\zeta'$ is for instance the acoustical part of $\zeta$ while $\zeta_0$ is the mean value of $\zeta$. 

\paragraph{}The Fourier transforms are noted using capital letters and are associated with the angular frequency $\omega$: $p'_2$ has $P'_2(\omega)$ for Fourier transform ; $\zeta'$ is an exception because its Fourier transform is written $\zeta'(\omega)$. We also need to introduce the useful quantity  $A=2a_1(\zeta_0-\zeta_{00})v_{j0}$.

\paragraph{}To derive the condition for the instabilities of both (-,+) and (+,-) reeds, we first need to linearize the equations and then calculate the Fourier transform of the linearized equations. We then get the following set of equations :
\begin{description}
	\item[$\bullet$] $\di \frac{V_1}{c^2_0} \frac{d (p_1-p_{atm})}{dt} = \rho_0 (u_0 - u)$ gives 
	$\di \ji \omega \frac{V_1}{c^2_0} P'_1(\omega) = -\rho_0.U'(\omega)$ (L1);
	
	\item[$\bullet$] $\di p_1 =p_2 + \rho_0 \frac{L_2}{S_2} \frac{du}{dt}$ becomes 
	$\di P'_1(\omega)=P'_2(\omega)+\ji \omega \rho_0 \frac{ L_2 }{ S_2 } U'(\omega)$ (L2);

	\item[$\bullet$] $\di p_2 = p_{atm} + \frac{1}{2} \rho_0 v^2_j$ gives $\di V_j'(\omega)=\frac{ P'_2(\omega) }{ \rho_0.v_{j0} }$ (L3) ;

	\item[$\bullet$] $\di \frac{d^2 \zeta}{dt^2} 
					+ Q^{-1}\omega_0 \frac{d \zeta}{dt} 
					+ \omega^2_0 \zeta
					= 
					\mu (p_2 - p_{atm})$ 
	becomes $\di ( -\omega^2+\ji \omega Q^{-1}\omega_0 +\omega^2_0 ) \zeta'(\omega)
				= \mu P'_2(\omega)$ (L4) ;

	\item[$\bullet$] $\di u=S_r\frac{d \zeta}{dt}+\alpha S_uv_j$ gives 
	$\di U'(\omega) = \ji \omega S_r \zeta'(\omega) 
	+ \alpha S_{u0} V'_j(\omega)
	+ \alpha A \zeta'(\omega)$ (L5).	
	\end{description}
	
\paragraph{}We can then introduce the expression of $P'_1(\omega)$ using equation (L2) and of $U'(\omega)$ using equation (L5) in the mass conservation (L1) which gives, when replacing $V'_j(\omega)$ by its expression as a function of $P'_2(\omega)$ using (L3):

\begin{multline*}
\ji \omega \frac{V_1}{c^2_0} \Big[
P'_2(\omega) + \ji \omega \rho_0 \frac{ L_2 }{ S_2 }
\Big(
\ji \omega S_r \zeta'(\omega) 
+ \frac{ \alpha S_{u0} }{ \rho_0 v_{j0} } P'_2(\omega)
+ \alpha A \zeta'(\omega)
\Big)
\Big]\\
=
-\rho_0 \Big[
\ji \omega S_r \zeta'(\omega)
+ \frac{ \alpha S_{u0} }{ \rho_0 v_{j0} } P'_2(\omega)
+ \alpha A \zeta'(\omega)
\Big],
\end{multline*}

which can also be written, after some calculation, as 

\begin{multline*}
P'_2(\omega) =
\frac{ \di - \rho_0 \Big( 1 - \frac{V_1 L_2}{c^2_0 S_2} \omega^2 \Big) }
{\di 
\Big( \frac{ \alpha S_{u0} }{ v_{j0} } \Big)^2 . \Big( 1 - \frac{V_1 L_2}{c^2_0 S_2} \omega^2 \Big)^2
+ \Big( \frac{V_1}{c^2_0}\omega  \Big)^2 
}
.
\bigg[
 \alpha A \frac{ \alpha S_{u0} }{ v_{j0} } \Big( 1- \frac{V_1 L_2}{c^2_0 S_2} \omega^2 \Big) 
 + \omega^2 S_r \frac{V_1}{ c^2_0}\\
+ \ji \omega \bigg( S_r \frac{ \alpha S_{u0} }{ v_{j0} } \Big( 1- \frac{V_1 L_2}{c^2_0 S_2} \omega^2 \Big)
- \alpha A \frac{V_1}{ c^2_0} \bigg) 
\bigg]
.\zeta'(\omega).
\end{multline*}

\paragraph{}The instability of the reed should be possible if the damping term in (L4), $\mathrm{j}\omega Q^{-1}\omega_0\zeta'(\omega)$, is at least compensated by the part of $P'_2(\omega)$ associated with $\ji \omega P'_2(\omega)$. In such condition, we must verify :
$$
Q^{-1}\omega_0 
<
-\mu \rho_0
\Big( 1- \frac{V_1 L_2}{c^2_0 S_2} \omega^2 \Big)
\frac{\di S_r \frac{ \alpha S_{u0} }{v_{j0} } \Big( 1 - \frac{V_1L_2}{c^2_0S} \omega^2 \Big) 
		- \alpha A \frac{V_2}{ c^2_0}}
{\di 
\Big(\frac{\alpha S_{u0} }{ v_{j0} }\Big)^2 \Big( 1 - \frac{V_1L_2}{c^2_0S_2} \omega^2 \Big)^2
	+ \Big(\frac{V_1}{c^2_0}\omega  \Big)^2
},
$$
which can only be fulfilled if we have 
$$\Big( 1- \frac{V_1L_2}{c^2_0S_2} \omega^2 \Big)
.\Big[ 
S_r \frac{ \alpha S_{u0} }{v_{j0} } \Big( 1- \frac{V_1L_2}{c^2_0S} \omega^2 \Big) - \alpha A \frac{V_2}{ c^2_0}
\Big]
< 0.
$$

\paragraph{}As $A>0$ for a (+,-) reed,  $\di 1- \frac{V_1L_2}{c^2_0S_2} \omega^2>0$ must be fullfilled to get potential instabilities while, as $A<0$ for a (-,+) reed, it is $\di 1- \frac{V_1L_2}{c^2_0S_2} \omega^2<0$ which must be verified.

\subsection{Keypoints of the numerical simulations method}
\paragraph{}To perform numerical simulations with these models of free reeds, we have followed the method proposed by Gazengel \textit{et al} in \cite{Gazengel:95} to derive the numerical equivalents for the models so we invite the reader to consult the article of Gazengel for full details: we only give the keypoints of the numerical algorithm.

\paragraph{}Except for the reed motion equation, the numerical equivalents are derived with replacement of derivatives using the backward Euler numerical scheme. 

\paragraph{}The numerical equivalent for the reed motion equation is derived using a bilinear transform, with pre-dilatation of the frequency $\omega_0$ as proposed by Gazengel \cite{Gazengel:95} and the introduction of the pumped flow to replace a second-order differential equation by a system of two first-order differential equations.

\paragraph{}To calculate the signals for sample $n$ we assume that the over-pressure $\Delta p_2[n]$ equals the $\Delta p_2[n-1]$ and considering $\Delta p_2[n]$ we express all the other quantities ($\zeta_{n}[n]$, $v_j[n]$, $S_u[n]$, $u_p[n]$, $u_t[n]$, $u[n]$ and $\Delta p_1[n]$) as functions of $\Delta p_2[n]$. The main, non linear equation to solve corresponds to the numerical version of the mass conservation for volume $V_1$:
$$NL\big[\Delta p_2[n]\big]=\frac{V_1}{\rho_0 c^2_0}F_s\big( \Delta p_1[n]-\Delta p_1[n-1]\big)-u_0[n]-u[n],$$
where $F_s$ is the sampling frequency.

\paragraph{}Using the Newton-Raphson scheme,  we calculate the new value of $\Delta p_2[n]$, which would be closest to the wanted value, as :
$$\Delta p_2[n] \; \mapsto \Delta p_2[n] -\frac{1}{NL\big[\Delta p_2[n]\big]}.\frac{d }{d\Delta p_2[n]}
\bigg(NL\big[\Delta p_2[n]\big]\bigg).$$

\paragraph{}We keep on calculate new values of $\Delta p_2[n]$ until the variation is lower than the chosen precision. When this value of $\Delta p_2[n]$ is finally found we calculate the other quantities and begin the calculation of the next sample ($\Delta p_2[n+1]$).

\paragraph{}As proposed by Gazengel, we always verify if the corrections for the $\Delta p_2[n]$ make sense, that is to say if the correction does not correspond to oscillations around the wanted value. If we note that the algorithm is oscillating around the solution we then adopt a dichotomy scheme to calculate the good approximation of $\Delta p_2[n]$.

\paragraph{}As previously indicated, we pre-calculate tables for the useful section and its derivative $dS_u / d\Delta p_2$ to increase the velocity of the calculation.

\paragraph{}At the beginning of the calculation we know the whole excitation signal and suppose that the reed is at rest.

\paragraph{}We must emphasize that we calculate all the quantities with the assumption of an acoustical flow. Indeed, we do not distinguish a mean flow and acoustical perturbations as mean quantities do not seem to make sense for the configurations we have studied.

\clearpage
\bibliographystyle{unsrt}

\end{document}